\documentclass[prd,preprint,tightenlines,floatfix,showpacs,preprintnumbers,nofootinbib,eqsecnum]{revtex4}

 \usepackage[dvips,final]{graphicx}
  \usepackage{amssymb}
   \usepackage{amsmath}
    \usepackage{amsfonts}
     \usepackage{epsfig}
      \usepackage{bm}

\begin{document}

\title{Central exclusive quark-antiquark dijet \\
and Standard Model Higgs boson production\\
in proton-(anti)proton collisions}

\author{Rafa{\l} Maciu{\l}a}
\email{rafal.maciula@ifj.edu.pl} \affiliation{Institute of Nuclear
Physics PAN, PL-31-342 Cracow, Poland}

\author{Roman Pasechnik}
\email{roman.pasechnik@fysast.uu.se} \affiliation{Department of
Physics and Astronomy, Uppsala University, Box 516, SE-751 20
Uppsala, Sweden}

\author{Antoni Szczurek}
\email{antoni.szczurek@ifj.edu.pl} \affiliation{Institute of Nuclear
Physics PAN, PL-31-342 Cracow,
Poland and\\
University of Rzesz\'ow, PL-35-959 Rzesz\'ow, Poland}

\date{\today}

\begin{abstract}
We consider the central exclusive production of $q\bar{q}$ pairs and Higgs boson in
proton-proton collisions at LHC. The amplitude
for the process is derived within the $k_{\perp}$-factorization approach
and considered in different kinematical asymptotics, in particular,
in the important high quark transverse momenta and massless quark limits. Quark helicity
and spin-projection amplitudes in two different frames are shown in extenso. Rapidity distributions, quark jet
$p_{\perp}$ distributions, invariant $q\bar{q}$ mass distributions,
angular azimuthal correlations between outgoing protons and jets are
presented. Irreducible $b{\bar b}$ background to the central
exclusive Higgs boson production is analyzed in detail, in
particular how to impose cuts to maximize signal-to-background ratio.
\end{abstract}


\maketitle

\section{Introduction}

Exclusive double diffractive production (EDD) of the Higgs boson
has been suggested some time ago as alternative to inclusive measurements \cite{Nachtman,BL}.
Exclusive diffractive dijets production attracted recently a lot of
attention due to new data from CDF run II \cite{CDF-dijets}. The
standard approach for the calculation of central dijets production in
proton-(anti)proton collisions is based on the
Kaidalov-Khoze-Martin-Ryskin (KKMR) QCD mechanism which was
initially developed for the central exclusive Higgs production in
Refs.~\cite{KMR_Higgs} which is expected to provide a really robust signal due
to a clean environment and highly suppressed backgrounds (see, e.g.
Refs.~\cite{KMR_bbar_suppression,KMR_Higgs_bbbar_background,KMR-bb}).
For more details on the central exclusive processes and related
physics, we refer to the most recent reviews in Ref.~\cite{CEP-review}.

\begin{figure}[!h]
\begin{minipage}{0.39\textwidth}
 \centerline{\includegraphics[width=1.0\textwidth]{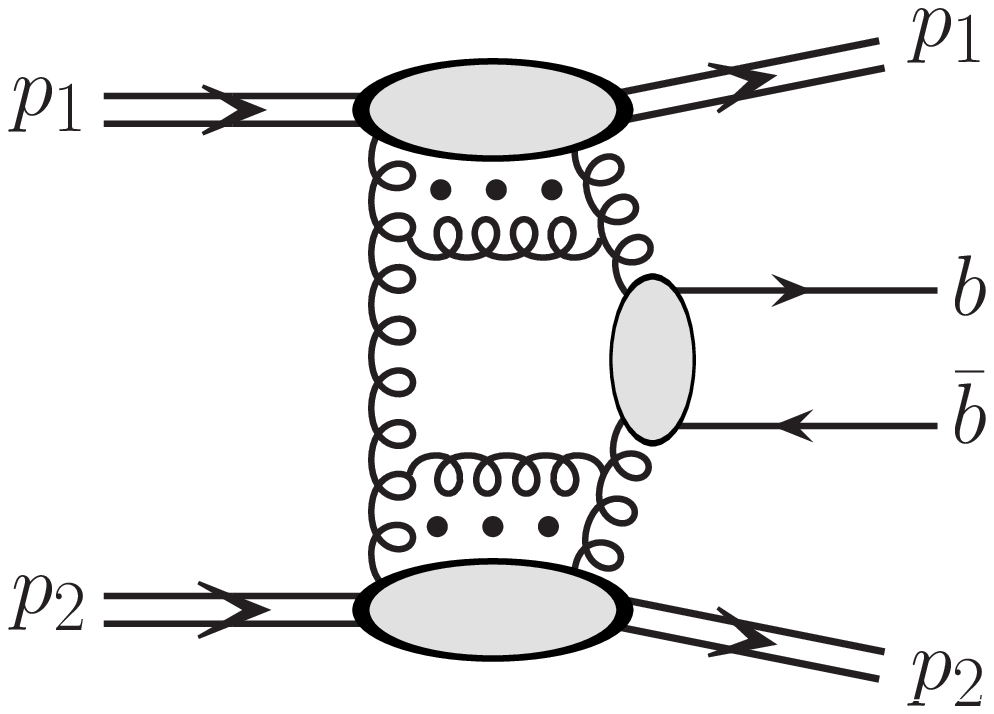}}
   \caption{
\small Direct central exclusive $b{\bar b}$ pair production in
$k_{\perp}$-factorization approach. It is considered to be the main
irreducible background for Higgs CEP.}
\label{fig:backgr}
\end{minipage} \hspace{0.5cm}
\begin{minipage}{0.45\textwidth}
 \centerline{\includegraphics[width=1.0\textwidth]{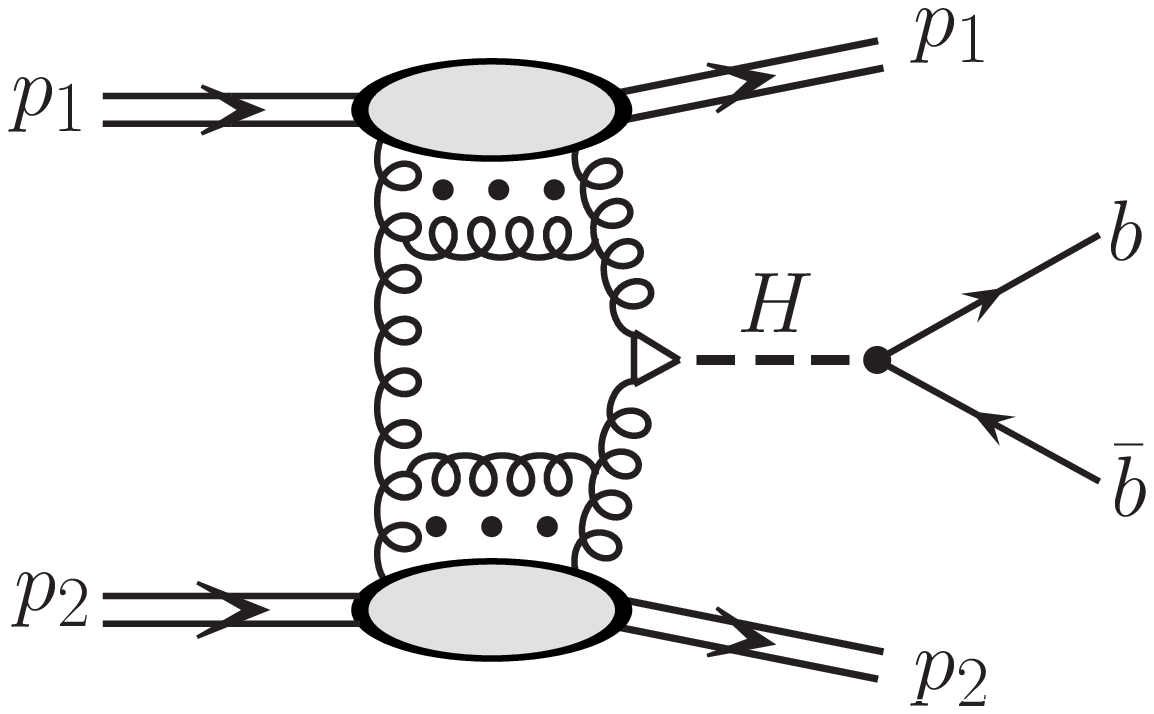}}
   \caption{
\small Central exclusive production of Higgs boson production with its subsequent decay into
$b{\bar b}$ pair, which competes with the direct $b{\bar b}$
pair production shown in Fig.~\ref{fig:backgr}.}
\label{fig:Higgs-dec}
\end{minipage}
\end{figure}

It is known, however, that the process $pp\to pp(q\bar q)$ is
dominated by the non-perturbative region of gluon transverse
momenta, and even perturbative ingredients like the Sudakov form
factor are not under full theoretical control \cite{Cudell:2008gv}.
Uncertainties on exclusive diffractive production of Higgs at the LHC
were discussed in Ref.~\cite{DKRS2011} together with uncertainties
on gluonic jet production which was measured by the CDF collaboration.
The problem becomes even more pronounced when considering the
irreducible backgrounds in central exclusive production of Higgs boson
originating from the direct exclusive $b{\bar b}$ pair production in
a fusion of two off-shell gluons. In particular, in
Ref.~\cite{our-bb} it was shown that the central exclusive
production (CEP) of $b{\bar b}$ jets at LHC (see
Fig.~\ref{fig:backgr}), may totally shadow the corresponding signal
of the Higgs boson in the $b{\bar b}$ channel (see
Fig.~\ref{fig:Higgs-dec}), which may lead to significant problems in
experimental identification.

Therefore, it becomes very important to investigate the
exclusive quark jets production in different kinematical domains and
quantify the related theoretical uncertainties. On the other hand,
the analysis of various differential distributions and experimental
cuts in considered four-body reaction $pp\rightarrow
p+\mathrm{``gap"}+(q\bar{q})+ \mathrm{``gap"}+p$ could help in a
reduction of the corresponding backgrounds.

Unpolarized exclusive $c$ and $\bar c$ jets production was
investigated numerically in Ref.~\cite{our-cc}. It was found that
the whole process is dominated by quark/antiquark production with
low transverse momentum $k_{\perp}$ in the very forward limit of
outgoing protons, whereas it is strongly suppressed at high
$k_{\perp}$'s, much stronger than in inclusive case. The same should
hold for $b$ and $\bar b$ jets CEP, which constitutes the major part
of the irreducible background for central exclusive Higgs
production\footnote{Other backgrounds although in principle irreducible
can be large \cite{Pilkington}.}. As was demonstrated numerically in Ref.~\cite{our-bb},
such a background turned out to be dominated by small gluon
transverse momenta $\sim q_{\perp}$ (which are the same for both
active and screening gluons in the forward limit) coming into hard
subprocess amplitude $g^*g^*\to q{\bar q}$ contracted with the gluon
transverse polarization vectors $\sim q_{\perp}^{\mu}/x\sqrt{s}$ and
integrated over in the diffractive amplitude. The presence of the
screening gluon in the loop, actually, violates the well known
$J_z=0$ selection rule \cite{Khoze:2000jm}, which was initially
established in one-step mechanisms, like $\gamma^*\gamma^*$ and $PP$
fusion processes. Such a violation manifests itself in the fact that
the hard subprocess amplitude is non-zeroth and proportional to
gluon transverse momentum, and it is indeed strongly suppressed only
in the high quark transverse momentum limit, whereas in the
low-$k_{\perp}$ limit it may lead to a significant contribution
\cite{our-cc,our-bb}.

These observations still suffer from a lack of solid theoretical
background. In order to predict observable signal from quark dijets
at LHC and to make a decisive conclusion about irreducible
background for Higgs CEP, it is worth to analyze carefully different
kinematical limits both numerically and analytically. The main goal
of this paper is to derive explicitly the hard subprocess amplitude
$g^*g^*\to q{\bar q}$ for any quark helicity states $\lambda_q$ and
$\lambda_{\bar q}$ in diffractive kinematics in two different
configurations of large $k_{\perp}\gg m_q$ and small $k_{\perp}\ll
m_q$ quark transverse momenta focusing primarily on large invariant
mass of $q{\bar q}$ dijets. This would give us an opportunity to
analyze different kinematical asymptotics of the diffractive
amplitude. Having such amplitudes we can then numerically evaluate
differential distributions in quark transverse momenta, rapidity,
relative angle between the quark jets and invariant mass of the
$q{\bar q}$ dijet. These theoretical elements are necessary for
upcoming Higgs searches and diffractive dijets measurements at LHC.

In the present paper we extend earlier studies related to
Higgs and jet production \cite{KMR_Higgs,Cudell:2010,Cudell:2008gv} to the production
of quark-antiquark jets. In our approach we use unintegrated
gluon distributions as proposed by the Durham group \cite{KMR_Higgs}.
Slightly different gluon distributions, fitted to the HERA data,
have been used in Ref.~\cite{Cudell:2008gv}.
The choice of gluon distributions brings uncertainties of a factor of about 2.
In our case we consistenly use the same
gluon distributions for the signal and background.
In the case of Higgs production we shall use off-shell matrix element
for $g^* g^* \to H$ compared to the on-shell matrix element used
before.

The paper is organized as follows. In Section II we consider the
general kinematics of the central exclusive dijet production.
Section III is devoted to a discussion of the diffractive amplitude,
in particular, its hard and soft constituents. Explicit derivation
of the helicity amplitudes for the hard subprocess part $g^*g^*\to
q{\bar q}$ in general kinematics and in some important limits is
given in Section IV. Electromagnetic $\gamma^*\gamma^*$ contribution
is discussed in Section V. In Section VI we reexamine the central
Higgs production taking into account gluon virtualities and explore
contributions of the exclusive $b{\bar b}$ pair and $Z$ production
as backgrounds for Higgs. Section VII contains discussion of numerical
results. Finally, some concluding remarks and outlook are given in
Section VIII.

\section{Kinematics of the central exclusive dijet production}

Inclusive heavy quark/antiquark pair production in the framework of the
$k_{\perp}$-factorization approach \cite{ktfac} was
considered in detail in Refs.~\cite{HKSST-qq,ktfac-qq,LSZ02}. In
particular, it was shown that the combination of the
$k_{\perp}$-factorization approach and the
next-to-leading-logarithmic-approximation (NLLA) BFKL vertex in
Quasi-multi-Regge kinematics (QMRK) \cite{FL96} together with the
concept of unintegrated gluon distribution functions (UGDFs) gives
quite good agreement with data on inclusive heavy $q{\bar q}$ pair
production.

\begin{figure}[!h]
\begin{minipage}{0.48\textwidth}
 \centerline{\includegraphics[width=0.6\textwidth]{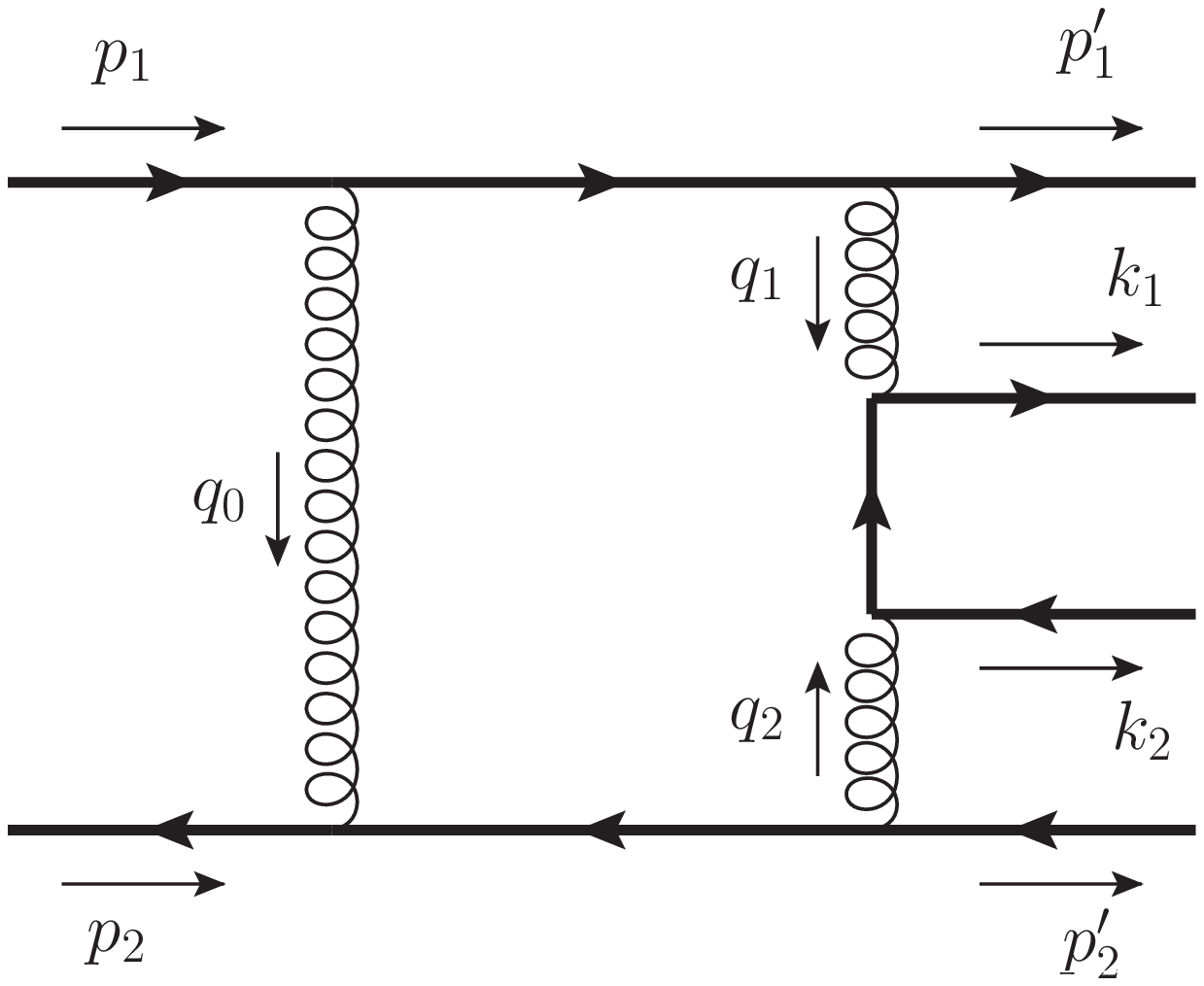}}
   \caption{
\small General kinematics of exclusive diffractive $q\bar{q}$
pair production in $pp$ collisions at the parton level.}
\label{fig:kinematics_qcd}
\end{minipage} \hspace{0.5cm}
\begin{minipage}{0.38\textwidth}
 \centerline{\includegraphics[width=1.1\textwidth]{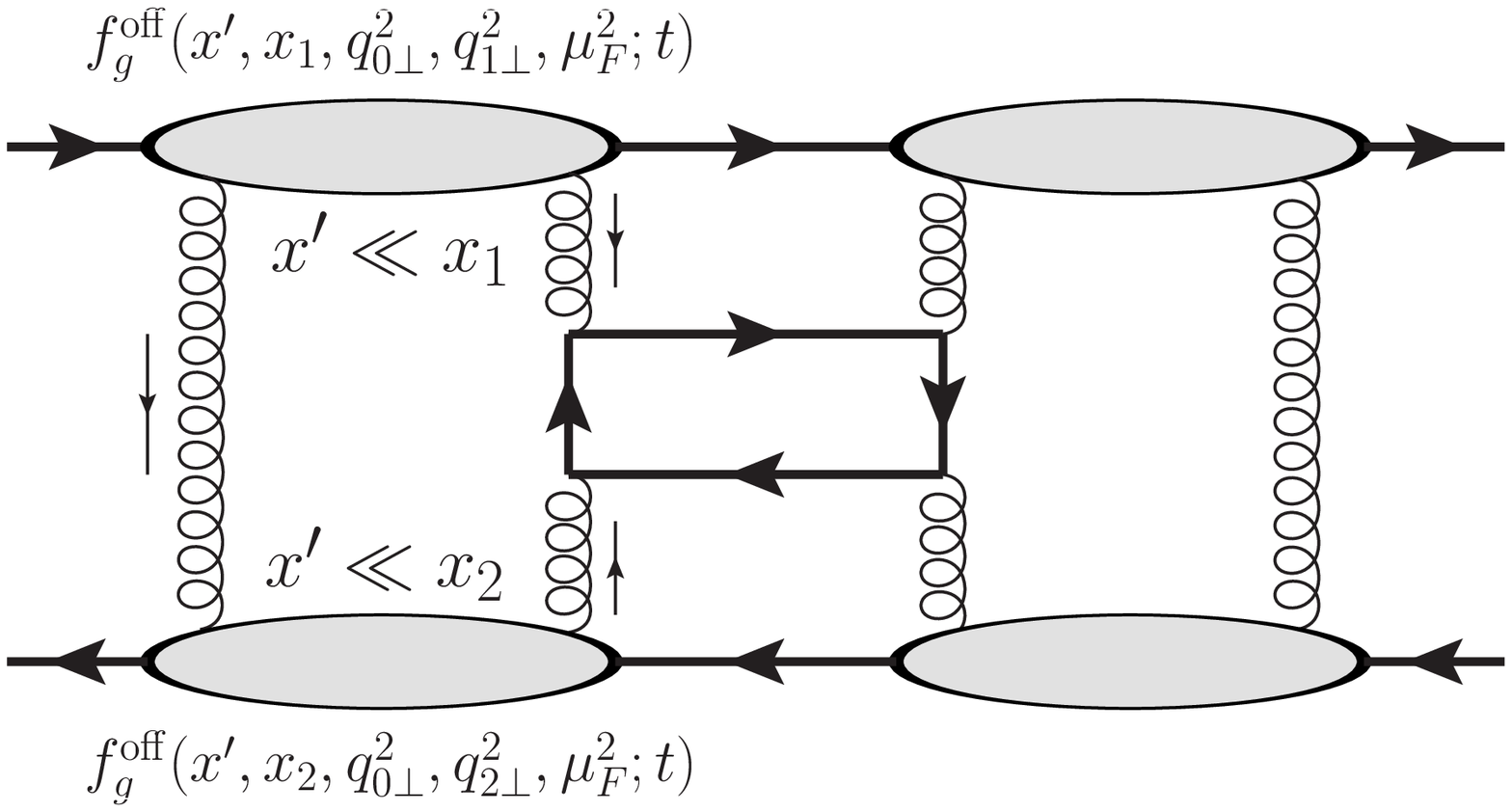}}
   \caption{
\small Cross section of $2\to4$ process of exclusive diffractive
$q\bar{q}$ pair production at the hadron level.} \label{fig:cross-sect}
\end{minipage}
\end{figure}

It looks quite natural to apply similar ideas to exclusive
diffractive $q\bar{q}$ production in proton-(anti)proton collisions
at different energies. In Figs.~\ref{fig:kinematics_qcd} and
\ref{fig:cross-sect} we show the general kinematics for the process
$pp\rightarrow p+\mathrm{``gap"}+(q\bar{q})+\mathrm{``gap"}+p$ under
consideration at the parton and the hadron levels, respectively.

The decomposition of gluon momenta into longitudinal and transverse
parts in the high energy limit in the c.m.s. frame is
\begin{eqnarray}\nonumber
&&q_1=x_1p_1+q_{1\perp},\qquad q_2=x_2p_2+q_{2\perp},\qquad 0<x_{1,2}<1, \\
&&q_0=x'_1p_1+x'_2p_2+q_{0\perp},\quad x'_1\sim x'_2=x'\ll x_{1,2},\quad
q_{0,1,2}^2\simeq q_{0/1/2\perp}^2. \label{dec}
\end{eqnarray}
Making use of conservation laws
\begin{eqnarray}
q_1=p_1-p'_1-q_0,\qquad q_2=p_2-p'_2+q_0,\qquad q_1+q_2=k_1+k_2\,,
\label{CL}
\end{eqnarray}
we write
\begin{eqnarray}\label{sx1x2}
s\,x_1x_2=M_{q\bar{q}}^2+|{\bf k}_{\perp}|^2\equiv
M_{q\bar{q}\perp}^2,\quad M_{q\bar{q}}^2=(k_1+k_2)^2,\quad
x_{1,2}=\frac{M_{q\bar{q}\perp}}{\sqrt{s}}e^{\pm y_{q\bar{q}}},
\end{eqnarray}
where $M_{q\bar{q}}$ and $y_{q\bar{q}}$ is the invariant mass and
rapidity of the $q\bar{q}$ pair, respectively, and
\begin{eqnarray*}
k_{\perp}=-(p'_{1\perp}+p'_{2\perp})=q_{1\perp}+q_{2\perp}=k_{1\perp}+k_{2\perp},
\end{eqnarray*}
is its transverse momentum, where $q_{1/2\perp}$ and $k_{1/2\perp}$ are
gluon and quark transverse momenta with respect to the c.m.s. beam
axis. In analogy with Eq.~(\ref{dec}), we can write
\begin{eqnarray}
p'_1=\xi_1p_1+p'_{1\perp},\qquad p'_2=\xi_2p_2+p'_{2\perp},\qquad
\xi_{1,2}=1-x_{1,2}\,, \label{decpp}
\end{eqnarray}
where ${p'}^2_{1/2\perp}=t_{1,2}$ in terms of the momentum transfers
along the proton lines $t_{1,2}$.

In the c.m.s. frame it is convenient to choose the basis with
$z$-axis collinear to the proton beam. Then the proton momenta are
\begin{eqnarray}\label{protons}
p_1=\frac{\sqrt{s}}{2}(1,\,0,\,0,\,1),\quad
p_2=\frac{\sqrt{s}}{2}(1,\,0,\,0,\,-1) \; .
\end{eqnarray}
Let us choose the $y$-axis in such a way that $q_1^y=-q_2^y\equiv q^y$. In
these coordinates the gluon transverse momenta are
\begin{eqnarray}\label{qt}
q_{1\perp}=(0,\,q_1^x,\,q^y,\,0),\quad
q_{2\perp}=(0,\,q_2^x,\,-q^y,\,0)\,.
\end{eqnarray}

Conservation laws provide us with the following relations between
components of gluon transverse momenta and covariant scalar products
\begin{eqnarray} \nonumber
&&q_1^x=-\frac{q_{1\perp}^2+(q_{1\perp}q_{2\perp})}{|{\bf k}_{\perp}|},\quad
q_2^x=-\frac{q_{2\perp}^2+(q_{1\perp}q_{2\perp})}{|{\bf k}_{\perp}|},\quad
q^y=\frac{\sqrt{q_{1\perp}^2q_{2\perp}^2-(q_{1\perp}q_{2\perp})^2}}{|{\bf
k}_{\perp}|}\,
\mathrm{sign}(q^y), \\
&&k_{\perp}^2=-|{\bf k}_{\perp}|^2=q_{1\perp}^2+q_{2\perp}^2+2(q_{1\perp}q_{2\perp}),\quad
q_{1/2\perp}^2=-|{\bf q}_{1/2\perp}|^2,\label{Qcomp}
\end{eqnarray}
where $|{\bf k}_{\perp}|$ is the $q\bar{q}$-pair transverse momentum
with respect to $z$-axis. On the other hand, momentum conservation
${\bf p}'_{1\perp}+{\bf p}'_{2\perp}={\bf k}_{\perp}$ leads to a
useful relation
\begin{eqnarray}
-t_1-t_2+2\sqrt{t_1t_2}\cos\Phi=|{\bf k}_{\perp}|\,,
\end{eqnarray}
where $\Phi$ is the relative angle between the outgoing protons.

The appearance of the factor $\mathrm{sign}(q^y)$ guarantees the
applicability of Eq.~(\ref{Qcomp}) for both positive and negative $q^y$.
Note that under permutations $q_{1\perp}\leftrightarrow q_{2\perp}$
implied by the Bose statistics the components interchange as
$q_1^x\leftrightarrow q_2^x$ and $q^y\leftrightarrow -q^y$.

In analogy to Eq.~(\ref{dec}) one can introduce the Sudakov
expansions for quark momenta as
\begin{eqnarray}
k_1=x_1^qp_1+x_2^qp_2+k_{1\perp},\quad k_2=x_1^{\bar q}p_1+x_2^{\bar
q}p_2+k_{2\perp}
\end{eqnarray}
leading to
\begin{eqnarray}\label{xqq}
x_{1,2}=x_{1,2}^q+x_{1,2}^{\bar q},\quad
x_{1,2}^q=\frac{m_{1\perp}}{\sqrt{s}}e^{\pm y_1},\quad x_{1,2}^{\bar
q}=\frac{m_{2\perp}}{\sqrt{s}}e^{\pm y_2},\quad
m_{1/2\perp}^2=m_q^2+|{\bf k}_{1/2\perp}|^2\,,
\end{eqnarray}
in terms of quark/antiquark rapidities $y_1$, $y_2$ and transverse
masses $m_{1\perp}$, $m_{2\perp}$. In the considered coordinates we
write in analogy to Eq.~(\ref{qt})
\begin{eqnarray} \label{kts}
k_{1\perp}=(0,\,k_1^x,\,k^y,\,0),\quad
k_{2\perp}=(0,\,k_2^x,\,-k^y,\,0),
\end{eqnarray}
with components satisfying the relation $k_1^x+k_2^x=q_1^x+q_2^x$.
By construction, in order to get the diffractive amplitude in the
covariant form useful in any coordinates, we should relate
components $k_{1/2}^x$ and $k^y$ with the scalar products in the
similar way as for gluon momentum components (see
Eq.~(\ref{Qcomp})):
\begin{eqnarray}\nonumber
k_1^x=-\frac{k_{1\perp}^2+(k_{1\perp}k_{2\perp})}{|{\bf k}_{\perp}|},\quad
k_2^x=-\frac{k_{2\perp}^2+(k_{1\perp}k_{2\perp})}{|{\bf k}_{\perp}|},\quad
k^y=\frac{\sqrt{k_{1\perp}^2k_{2\perp}^2-(k_{1\perp}k_{2\perp})^2}}{|{\bf
k}_{\perp}|}\, \mathrm{sign}(k^y)\,.\\
\label{Kcomp}
\end{eqnarray}

In subsequent calculations we will construct the $q\bar{q}$
diffractive amplitude in explicitly covariant form and analyze its
behavior in different regions of the 4-particle phase space.

\section{Diffractive amplitude}

Generally, in the case of the central exclusive production (CEP) with
the leading protons, the central system $X$ should necessarily be
produced in the color singlet state, such that the proton remnants
and the $X$ system are disconnected in the color space and their
hadronisation occurs independently giving rise to rapidity gaps
\cite{GI}. So, without the loss of generality we are concentrated on
the simplest case of $q\bar{q}$ pair produced in the color singlet
state.

According to the KKMR approach
\cite{KMR_Higgs,KMR_bbar_suppression,KMR_Higgs_bbbar_background,KMR-bb}
we write the amplitude of the exclusive diffractive $q\bar{q}$ pair
production $pp\to p(q\bar{q})p$ as
\begin{eqnarray}
{\cal
M}_{\lambda_q\lambda_{\bar{q}}}=s\cdot\pi^2\frac12\frac{\delta_{c_1c_2}}{N_c^2-1}\,
\Im\int d^2
q_{0\perp}V_{\lambda_q\lambda_{\bar{q}}}^{c_1c_2}\frac{f^{\mathrm{off}}_g(x',x_1,q_{0\perp}^2,
q_{1\perp}^2,t_1)f^{\mathrm{off}}_g(x',x_2,q_{0\perp}^2,q_{2\perp}^2,t_2)}
{q_{0\perp}^2\,q_{1\perp}^2\,q_{2\perp}^2} \, , \label{ampl}
\end{eqnarray}
where the transverse momenta and the longitudinal fractions of
gluons are defined in the previous Section,
$\lambda_q,\,\lambda_{\bar{q}}$ are the helicities of heavy $q$ and
$\bar{q}$, respectively, $f^{\mathrm{off}}_g$ is the unintegrated
gluon density function (UGDF) and $V_{\lambda_q\lambda_{\bar{q}}}$
is the hard subprocess $g^*g^*\to b{\bar b}$ amplitude. Averaging
over color indices $c_1,\,c_2$ of $t$-channel fusing gluons is made
explicitly. The normalization convention of this amplitude differs
from the KKMR one by a factor $s$. The amplitude is averaged over
the color indices and over the two transverse polarizations of the
incoming gluons. The bare amplitude above is subjected to absorption
corrections which depend on collision energy and typical proton
transverse momenta. We shall discuss this issue shortly when
presenting our results.

\subsection{Matrix element of the hard subprocess $g^*g^*\to
Q\bar{Q}$}

\begin{figure}[h!]
 \centerline{\epsfig{file=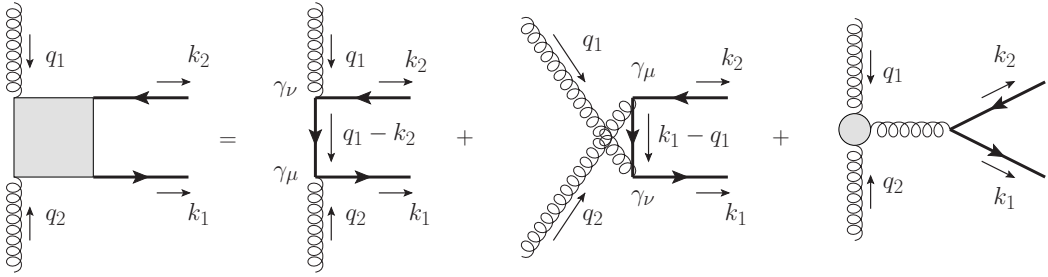,width=14cm}}
 \caption{Effective vertex in QMRK approach \cite{FL96}. Last diagram with effective
 3-gluon vertex drops out in projection to the color singlet final state.}
 \label{flvertex}
\end{figure}

Let us consider the subprocess amplitude for the $q\bar{q}$ pair
production via off-shell gluon-gluon fusion. The vertex factor
$V_{\lambda_q\lambda_{\bar{q}}}^{c_1c_2}=V_{\lambda_q\lambda_{\bar{q}}}^{c_1c_2}(k_1,k_2)$
in expression (\ref{ampl}) is the production amplitude of a pair of
massive quark $q$ and antiquark $\bar{q}$ with helicities
$\lambda_q$, $\lambda_{\bar{q}}$ and momenta $k_1$, $k_2$,
respectively. Within the QMRK approach \cite{FL96} we have
\begin{eqnarray}\label{qqamp}
&&V_{\lambda_q\lambda_{\bar{q}}}^{c_1c_2}(q_1,q_2)\equiv
n^+_{\mu}n^-_{\nu}V_{\lambda_q\lambda_{\bar{q}}}^{c_1c_2,\,\mu\nu}(q_1,q_2),\qquad
n_{\mu}^{\mp}=\frac{p_{1,2}^{\mu}}{E_{p,cms}},\\
&&V_{\lambda_q\lambda_{\bar{q}}}^{c_1c_2,\,\mu\nu}(q_1,q_2)=-g^2\sum_{i,k}\left\langle
3i,\bar{3}k|1\right\rangle\bar{u}_{\lambda_q}(k_1)
(t^{c_1}_{ij}t^{c_2}_{jk}b^{\mu\nu}(k_1,k_2)-
t^{c_2}_{kj}t^{c_1}_{ji}\bar{b}^{\mu\nu}(k_2,k_1))v_{\lambda_{\bar{q}}}(k_2),
\nonumber
\end{eqnarray}
where $E_{p,cms}=\sqrt{s}/2$ is the c.m.s. proton energy, $t^c$ are
the color group generators in the fundamental representation,
$u(k_1)$ and $v(k_2)$ are on-shell quark and antiquark spinors,
respectively, $b^{\mu\nu},\,\bar{b}^{\mu\nu}$ are the effective
vertices (\ref{bb}) arising from the Feynman rules in the QMRK
approach illustrated in Fig.~\ref{flvertex}:
\begin{eqnarray} \label{bb}
b^{\mu\nu}(k_1,k_2)=\gamma^{\nu}\frac{\hat{q}_{1}-\hat{k}_{1}-m_q}{(q_1-k_1)^2-m_q^2}
\gamma^{\mu}-\frac{\gamma_{\beta}\Gamma^{\mu\nu\beta}(q_1,q_2)}{(k_1+k_2)^2}
\; , \\
\bar{b}^{\mu\nu}(k_2,k_1)=\gamma^{\mu}\frac{\hat{q}_{1}-\hat{k}_{2}+m_q}{(q_1-k_2)^2-m_q^2}
\gamma^{\nu}-\frac{\gamma_{\beta}\Gamma^{\mu\nu\beta}(q_1,q_2)}{(k_1+k_2)^2}
\; , \nonumber
\end{eqnarray}
where by the Dirac convention $\hat{a}\equiv \gamma\cdot a$ for any
4-vector $a^{\mu}$ is adopted, $\Gamma^{\mu\nu\beta}(q_1,q_2)$ is
the effective three-gluon vertex. These effective vertices were
initially proposed for massless quarks in Refs.~\cite{FL96} and then
extended for massive case in Ref.~\cite{HKSST-qq,HKSST-charm}. The
effective $ggg$-vertices are canceled out when projecting the
$q\bar{q}$ production amplitude Eq.~(\ref{qqamp}) onto the color
singlet state, so only the first two diagrams in Fig.~\ref{flvertex}
contribute to the final result for the production amplitude. Since
we will adopt the definition of gluon polarization vectors
proportional to transverse momenta $q_{1/2\perp}$, i.e.
$\varepsilon_{1,2}\sim q_{1/2\perp}/x_{1,2}$ (see below), then we
take into account the longitudinal momenta in the numerators of
effective vertices (\ref{bb}).

The SU(3) Clebsch-Gordan coefficient $\left\langle
3i,\bar{3}k|1\right\rangle=\delta^{ik}/\sqrt{N_c}$ in
Eq.~(\ref{qqamp}) projects out the color quantum numbers of the
$q\bar{q}$ pair onto the color singlet state. Factor $1/\sqrt{N_c}$
provides the averaging of the matrix element squared over
intermediate color states of quarks.

Therefore, we have the following amplitude
\begin{eqnarray}\nonumber
&&{}V_{\lambda_q\lambda_{\bar{q}}}^{c_1c_2,\,\mu\nu}=-\frac{g^2}{2\sqrt{N_c}}\,
\delta^{c_1c_2}\,\bar{u}_{\lambda_q}(k_1)
\biggl(\gamma^{\nu}\frac{\hat{q}_{1}-\hat{k}_{1}-m_q}
{(q_1-k_1)^2-m_q^2}\gamma^{\mu}-\gamma^{\mu}\frac{\hat{q}_{1}-
\hat{k}_{2}+m_q}{(q_1-k_2)^2-m_q^2}\gamma^{\nu}\biggr)v_{\lambda_{\bar{q}}}(k_2).\\
\label{fin-vert-mn}
\end{eqnarray}
This amplitude can be simplified by using Dirac equations for
quark/antiquark spinors
\begin{eqnarray}\label{Dirac}
\bar{u}_{\lambda_q}(k_1)\hat{k}_1=m_q\bar{u}_{\lambda_q}(k_1),\qquad
\hat{k}_2v_{\lambda_{\bar{q}}}(k_2)=-m_qv_{\lambda_{\bar{q}}}(k_2),\quad
k_1^2=k_2^2=m_q^2.
\end{eqnarray}
Moving $\hat{k}_1$ to the left, and $\hat{k}_2$ to the right, until
they disappear upon acting on the spinor, we finally get
\begin{eqnarray}\nonumber
&&{}V_{\lambda_q\lambda_{\bar{q}},\,\mu\nu}^{c_1c_2}=
-\frac{g^2}{2\sqrt{N_c}}\,\delta^{c_1c_2}\,\bar{u}_{\lambda_q}(k_1)
\biggl(\frac{\gamma^{\nu}\hat{q}_{1}-2k^{\nu}_{1}}
{q_{1}^2-2(k_1q_1)}\,\gamma^{\mu}-\gamma^{\mu}\,\frac{\hat{q}_{1}\gamma^{\nu}-
2k^{\nu}_{2}}{q_{1}^2-2(k_2q_1)}\biggr)v_{\lambda_{\bar{q}}}(k_2).\\
\label{fin-vert-1mn}
\end{eqnarray}

Amplitude of fusion of two off-shell (reggeized) gluons $g^*g^*\to
q{\bar q}$ turns out to be explicitly gauge invariant. Indeed, by
direct calculation we see that the gauge invariance over the first
gluon line is satisfied:
\begin{eqnarray}\label{GI-1}
q_1^{\nu}V_{\lambda_q\lambda_{\bar{q}},\,\mu\nu}^{c_1c_2}=0.
\end{eqnarray}
Now due to momentum conservation $q_1+q_2=k_1+k_2$ we may rewrite
the amplitude (\ref{fin-vert-mn}) as follows
\begin{eqnarray}\nonumber
&&{}V_{\lambda_q\lambda_{\bar{q}},\,\mu\nu}^{c_1c_2}=
\frac{g^2}{2\sqrt{N_c}}\,\delta^{c_1c_2}\,\bar{u}_{\lambda_q}(k_1)
\biggl(\gamma^{\nu}\,\frac{\hat{q}_{2}\gamma^{\mu}-2k^{\mu}_{2}}
{q_{2}^2-2(k_2q_2)}-\frac{\gamma^{\mu}\hat{q}_{2}-
2k^{\mu}_{1}}{q_{2}^2-2(k_1q_2)}\,\gamma^{\nu}\biggr)v_{\lambda_{\bar{q}}}(k_2).\\
\label{fin-vert-2mn}
\end{eqnarray}
Thus, the gauge invariance over the second gluon line is also
satisfied:
\begin{eqnarray}\label{GI-2}
q_2^{\mu}V_{\lambda_q\lambda_{\bar{q}},\,\mu\nu}^{c_1c_2}=0.
\end{eqnarray}
Comparing Eqs.~(\ref{fin-vert-1mn}) and (\ref{fin-vert-2mn}) we see
that the amplitude is symmetric w.r.t. interchanges
$q_1\leftrightarrow q_2$ and $\mu\leftrightarrow\nu$ as it should
be.

Taking into account definition (\ref{qqamp}) and momentum
conservation (\ref{dec}) and using the gauge invariance properties
(\ref{GI-1}) and (\ref{GI-2}), we get the following projection to
the light cone vectors (so called ``Gribov's trick'')
\begin{eqnarray}
&&V_{\lambda_q\lambda_{\bar{q}}}^{c_1c_2}=
n^+_{\mu}n^-_{\nu}V_{\lambda_q\lambda_{\bar{q}},\,\mu\nu}^{c_1c_2}=
\frac{4}{s}\frac{q^{\nu}_1-q^{\nu}_{1\perp}}{x_1}\frac{q^{\mu}_2-q^{\mu}_{2\perp}}{x_2}
V^{c_1c_2}_{\lambda_q\lambda_{\bar{q}},\,\mu\nu}=
\frac{4}{s}\frac{q^{\nu}_{1\perp}}{x_1}\frac{q^{\mu}_{2\perp}}{x_2}
V^{c_1c_2}_{\lambda_q\lambda_{\bar{q}},\,\mu\nu}.
\label{decomp}
\end{eqnarray}
Last expression shows that an important consequence of the gauge
invariance is the vanishing of the matrix element of the effective
$ggq\bar{q}$-vertex between on-mass-shell quark and antiquark states
in the limit of small $q_{1\perp}$ and $q_{2\perp}$
\cite{HKSST-qq,HKSST-charm}
\begin{eqnarray}\label{GI}
V_{\lambda_q\lambda_{\bar{q}}}^{c_1c_2}\to0\quad\mathrm{for}\quad
q_{1\perp}\;\mathrm{or}\;q_{2\perp}\to0,
\end{eqnarray}

The normalization of polarization vectors coincides with that of
Ref.~\cite{Forshaw05}. Now using Eqs.~(\ref{sx1x2}), (\ref{decomp})
and (\ref{bb}) we finally get the following $q\bar{q}$ production
vertex
\begin{eqnarray}\nonumber
&&{}V_{\lambda_q\lambda_{\bar{q}}}^{c_1c_2}=-\frac{2g^2}{M_{q\bar{q}\perp}^2\sqrt{N_c}}\,
\delta^{c_1c_2}\,\bar{u}_{\lambda_q}(k_1)
\biggl(\frac{\hat{q}_{1\perp}\hat{q}_{1}-2(k_{1\perp}q_{1\perp})}
{q_{1\perp}^2-2(k_1q_1)}\,\hat{q}_{2\perp}-\hat{q}_{2\perp}\,\frac{\hat{q}_{1}\hat{q}_{1\perp}-
2(k_{2\perp}q_{1\perp})}{q_{1\perp}^2-2(k_2q_1)}\biggr)v_{\lambda_{\bar{q}}}(k_2).\\
\label{fin-vert-fin}
\end{eqnarray}
It is interesting to note that this vertex function would be equal to zero
if one substitutes $q_1\to q_{1\perp}$, i.e. when one neglects the
longitudinal components of gluon momenta $q_{1/2,l}$ putting
$x_1\to0$ or $x_2\to0$. So, it turns out that the longitudinal
momenta play a critical role in diffractive production of $q\bar{q}$
pair and cannot be neglected. At the same time, we keep the gluon
virtualities in the propagators in Eq.~(\ref{fin-vert-fin}) as they
apparently become important in the small quark masses and quark
transverse momenta.

It is worth to notice that the mass terms disappear when applying
the Dirac equations (\ref{Dirac}). The quark mass $m_q$ is present
in the spinors and in the scalar products only. So, we see that for
massless quarks the production amplitude (\ref{fin-vert-fin}) has
the same covariant form as for massive ones.

In both particular cases of ${\bf p}'_{1\perp}=-{\bf p}'_{2\perp}$ and
in the forward limit $|{\bf p}'_{1\perp}|=|{\bf p}'_{2\perp}|\to0$,
we have $q_{1\perp}=-q_{2\perp}\equiv q_{\perp}$ and, hence,
$k_{1\perp}=-k_{2\perp}\equiv k_{\perp}$. High-$k_{\perp}$ jets limit
corresponds to $m_q\ll k_{\perp}$ and $q_{\perp}\ll k_{\perp}$. Invariant mass of the
$q{\bar q}$ pair is then given by $M_{q\bar{q}}^2\simeq 4|{\bf
k}_{\perp}|^2$, and the calculation of the matrix element squared
$|V(q_1,q_2)|^2$ for the considered hard subprocess
(\ref{fin-vert-fin}) in this limit leads to
\begin{eqnarray}\label{V2}
\sum_{\lambda_q\lambda_{\bar
q}}|V_{\lambda_q\lambda_{\bar{q}}}|^2\simeq
\frac{8g^4}{N_c}\Big(\frac{q_{\perp}}{k_{\perp}}\Big)^4\sin^2(2\phi),
\end{eqnarray}
where $\phi$ is the relative angle between ${\bf k}_{\perp}$ and
${\bf q}_{\perp}$ vectors. We see now that the amplitude
(\ref{fin-vert-fin}) in the high-$k_{\perp}$ limit is not exactly
zero, but rather suppressed by a factor $\sim
q_{\perp}^2/k_{\perp}^2$. However, relation (\ref{V2}) cannot be
used for prediction of the corresponding high-$k_{\perp}$
asymptotics of the diffractive amplitude (\ref{ampl}) since it
contains the hard subprocess amplitude $V(q_1,q_2)$ in the first
power integrated over $q_{\perp}$. For this purpose we have to
consider initial expression (\ref{fin-vert-fin}) for particular
quark helicity configurations separately paying attention not only
at their asymptotical behavior, but also at symmetry w.r.t. ${\bf
q}_{\perp}\leftrightarrow -{\bf q}_{\perp}$. Large-$k_{\perp}$
behavior of the 4-particle phase space is important as well.

\subsection{$Q\bar{Q}$ center of mass helicity amplitudes for the hard subprocess $g^*g^*\to
Q\bar{Q}$}

Let us consider now separate quark/antiquark helicity contributions of
the off-shell gluon fusion (hard) subprocess $g^*g^*\to
Q_{\lambda_q}\bar{Q}_{\lambda_{\bar q}}$, given by the matrix
element in the general covariant form (\ref{fin-vert-fin}).

The most convenient way is to determine the quark/antiquark
helicities in the c.m.s. frame of the $Q\bar{Q}$ pair with $z$ axis
along the proton beam, so ${\bf k}_1=-{\bf k}_2$ and
$k_{1,2}^0=M_{q{\bar q}}/2$. For simplicity, we work in the limit of
forward scattering, so $p'_{1\perp}=p'_{2\perp}=0$, so ${\bf
q}_{1\perp}=-{\bf q}_{2\perp}={\bf q}_{0\perp}$. In this frame
momenta of protons and final-state quarks are
\begin{eqnarray}\nonumber
&&p_1^{\mu}=\frac{E_1}{\sqrt{2}}(1,\,0,\,0,\,1),\qquad
p_2^{\mu}=\frac{E_2}{\sqrt{2}}(1,\,0,\,0,\,-1),\\
&&k_1^{\mu}=E_q(1,\,\gamma\sin\theta_q\cos\varkappa,\,\gamma\sin\theta_q\sin\varkappa,\,\gamma\cos\theta_q),\quad \gamma=\frac{\sqrt{E_q^2-m_q^2}}{E_q}<1,\nonumber\\
&&k_2^{\nu}=E_q(1,\,-\gamma\sin\theta_q\cos\varkappa,\,-\gamma\sin\theta_q\sin\varkappa,\,-\gamma\cos\theta_q),\nonumber
\label{frame}
\end{eqnarray}
so that the proton and quark energies $E_{1,2},\,E_q$ and the polar
angle of a (anti)quark jet $\theta_q$ w.r.t. the $z$-axis are
defined as
\begin{eqnarray*}
&&E_q\equiv\frac{M_{q\bar
q}}{2}=\frac{E_1}{\sqrt{2}}(x_1^q+x_1^{\bar q})=
\frac{E_2}{\sqrt{2}}(x_2^q+x_2^{\bar q}),\\
&&\cos\theta_q=\frac{1}{\gamma}\frac{x_1^q-x_1^{\bar
q}}{x_1^q+x_1^{\bar q}},\quad
\sin\theta_q=\frac{1}{\gamma}\frac{\sqrt{\gamma^2(x_1^q+x_1^{\bar
q})^2-(x_1^q-x_1^{\bar q})^2}}{x_1^q+x_1^{\bar q}},
\end{eqnarray*}
where $x_{1,2}^{q,\bar q}$ are the Sudakov fractions defined in
Eq.~(\ref{xqq}). The gluon and quark transverse momenta (with
respect to the proton beam) in the polar coordinates are then
defined as
\begin{eqnarray}\nonumber
{\bf q}_{0\perp}=q_{\perp}(\cos\psi,\sin\psi),\qquad {\bf
k}_{1\perp}=-{\bf
k}_{2\perp}=k_{\perp}(\cos\varkappa,\sin\varkappa),\quad
\end{eqnarray}
respectively, and
\begin{eqnarray}\label{inv-kin}
k_{\perp}=E_q\frac{\sqrt{\gamma^2(x_1^q+x_1^{\bar
q})^2-(x_1^q-x_1^{\bar q})^2}}{x_1^q+x_1^{\bar q}},\quad
k_z=E_q\frac{x_1^q-x_1^{\bar q}}{x_1^q+x_1^{\bar q}},\quad |{\bf
k}|=\sqrt{k_{\perp}^2+k_z^2}=E_q\gamma\,.
\end{eqnarray}

Using these notations, the different helicity amplitudes
$V_{\lambda_q\lambda_{\bar{q}}}$ can be written as follows:
\begin{eqnarray}\nonumber
&&V_{+-}={\cal C}\,\frac{q_{\perp}^2}{|{\bf k}|}\,\bigg[2|{\bf
k}|q_{\perp}\Big(|{\bf
k}|\cos(\psi-\varkappa)-ik_z\sin(\psi-\varkappa)\Big)+
M_{q\bar{q}}k_{\perp}\Big(k_z\cos(2\psi-2\varkappa)-\\&&i|{\bf
k}|\sin(2\psi-2\varkappa)\Big)\bigg]/
\bigg[M_{q\bar{q}}^2(k_{\perp}^2+q_{\perp}^2+m_q^2)+4M_{q\bar{q}}k_{\perp}q_{\perp}k_z\cos(\psi-\varkappa)-\nonumber\\&&
2k_{\perp}^2q_{\perp}^2(1+\cos(2\psi-2\varkappa))+q_{\perp}^4\bigg],\label{Vpm}\\
&&V_{++}=-2{\cal C}\,e^{-i\varkappa}\frac{q_{\perp}^2m_q}{|{\bf
k}|}\,\bigg[k_{\perp}^2\cos(2\psi-2\varkappa)+|{\bf k}|^2\bigg]/
\bigg[M_{q\bar{q}}^2(k_{\perp}^2+q_{\perp}^2+m_q^2)+\nonumber\\&&4M_{q\bar{q}}k_{\perp}q_{\perp}k_z\cos(\psi-\varkappa)-
2k_{\perp}^2q_{\perp}^2(1+\cos(2\psi-2\varkappa))+q_{\perp}^4\bigg]\label{Vpp}
\end{eqnarray}
with normalisation ${\cal C}=2g^2\delta^{c_1c_2}/\sqrt{N_c}$. Up to
a phase factor, the helicity amplitudes are dependent on the
difference $\psi-\varkappa$ (coming from the scalar product $({\bf
k}_{\perp}{\bf q}_{0\perp})$) and thus explicitly invariant with
respect to rotations of ${\bf k}_{\perp}$ and ${\bf q}_{0\perp}$ in
the transverse plane or shifts of the angles
$\psi,\,\varkappa\to\psi+\delta,\,\varkappa+\delta$.

Covariant relations (\ref{inv-kin}) allow us to turn to any
desirable frame of reference, in particular, to the overall
c.m.s. frame, where the (anti)quark longitudinal momentum fractions
$x_{1,2}^{q,\bar q}$ are define through their rapidities $y_1,\,y_2$
in Eq.~(\ref{xqq}) and the $q\bar q$ invariant mass is given by
\begin{eqnarray}\label{Mqq}
M_{q\bar{q}}=\sqrt{2m_{\perp}^2(1+\cosh(y_1-y_2))}\,.
\end{eqnarray}
Let us now investigate the hard subprocess $g^*g^*\to q{\bar q}$
matrix elements (\ref{Vpm}) and (\ref{Vpp}) in two limits of high
and low-$k_{\perp}$ jets separately.

Physically interesting case is for
$q{\bar q}$ dijets with very high invariant mass $M_{q\bar{q}}\gg
m_q$, where the KKMR QCD mechanism \cite{KMR_Higgs} based on the
$k_{\perp}$-factorization and Sudakov evolution is strictly
justified. If one looks at centrally produced jets $y_{1,2}\to 0$,
then according to Eq.~(\ref{Mqq}) the only way to produce the large
invariant mass $M_{q\bar{q}}$ is to consider high-$k_{\perp}$ jets
limit $|k_{\perp}|\gg m_q,|q_{\perp}|$. As follows from
Eq.~(\ref{inv-kin}), such a limit corresponds to the quark and
antiquark at central rapidities $y_{1,2}\sim 0,\,x_1^q\sim x_1^{\bar
q}$ and the $q\bar q$ invariant mass $M_{q\bar q}\simeq 2k_{\perp}$.

We see from Eqs.~(\ref{Vpp}) that $V_{++}\to 0$ in the quark
massless limit $m_q\to 0$ as it should be, so it is generally
suppressed with respect to $V_{+-}$ in the limit of large invariant
mass $M_{q\bar q}$ and high-$k_{\perp}$ quarks, i.e. for
$M_{q\bar{q}},\,k_{\perp}\gg m_q,\,q_{\perp}$. Indeed, in
high-$k_{\perp}$ quarks limit, the helicity amplitudes are
\begin{eqnarray}\label{hels}
V_{+-}\simeq-i{\cal
C}\frac{q_{\perp}^2}{2k_{\perp}^2}\sin(2\psi-2\varkappa),\quad
V_{++}\simeq -{\cal
C}\,e^{-i\varkappa}\frac{q_{\perp}^2m_q}{2k_{\perp}^3}\,(1+\cos(2\psi-2\varkappa)).
\end{eqnarray}
Then summing up the non-zeroth contributions
$|V_{\lambda_q\lambda_{\bar q}}|^2$ we easily recover Eq.~(\ref{V2})
for unpolarized hard matrix element squared in the considered limit.
However, $V_{+-}$ actually drops out in the integration over $\psi$
in the diffractive amplitude due to antisymmetry with respect to the
interchange $\psi\leftrightarrow-\psi$. So the diffractive cross
section for the $q\bar q$ pair production is given by $V_{++}$
contribution only and gets significantly suppressed in the
high-$k_{\perp}$ limit by a factor $\sim
q_{\perp}^2m_q/k_{\perp}^3$. So, in this case quark masses $m_q$ and
off-forward corrections become important. This is in agreement with
the common belief that the Higgs CEP background is supposed to be
small in very forward and quark massless limits for centrally
produced $b{\bar b}$ jets (with small rapidities), and agrees well
with the $J_z=0$ selection rule \cite{Khoze:2000jm}.

However, the particular high-$k_{\perp}$ limit does not explain the whole story.
Observation we have made above means only that quark high-$k_{\perp}$
contributions in the central rapidity region may
be strongly suppressed with respect to low-$k_{\perp}$ jets, where
the gluon transverse momenta $q_{\perp}$ and quark masses are
significant, but it does not tell us that the whole $b{\bar b}$
background for Higgs production is suppressed.

In our previous analysis of the exclusive open charm production in
Ref.~\cite{our-cc} it was shown that the dominant contribution to
the $c{\bar c}$ dijet cross section comes from relatively small
quark transverse momenta $|k_{\perp}|\simeq 1$ GeV. The same should
hold for $b{\bar b}$ CEP relevant for the Higgs background. Indeed,
resolving relation (\ref{Mqq}) with respect to typical rapidity
difference $|y_1-y_2|=\Delta y$, neglecting quark transverse momenta
$|k_{\perp}|\ll m_q$ and keeping only the $b$-mass contributions
$m_b\simeq 4.5$ GeV at fixed $M_{q\bar{q}}\sim M_H=120$ GeV, we get
$\Delta y\simeq 6.6$, $x_1^q\gg x_1^{\bar q}$. So the
low-$k_{\perp}$ quark jets can provide a contribution to irreducible
background for Higgs CEP, if their rapidities are $y_{1,2}\simeq \pm
3.3$. Then the invariant mass of the quark/antiquark pair is given
by their longitudinal fractions (or rapidities) only
\begin{eqnarray}\label{Mqq-lowkt}
M_{q\bar{q}}\simeq 2m_q\sqrt{\cosh(y_1)\cosh(y_2)},\qquad
|y_{1,2}|\gg1\,.
\end{eqnarray}
In the last kinematical situation the helicity amplitudes
(\ref{Vpm}), (\ref{Vpp}) are not suppressed by a large denominator,
and significant contributions can be obtained. Indeed,
\begin{eqnarray}\label{hels-lowkt}
V_{+-}\simeq\frac{{\cal
C}\,e^{i(\varkappa-\psi)}}{2\sqrt{\cosh(y_1)\cosh(y_2)}}\frac{q_{\perp}^3}{m_q(m_q^2+q_{\perp}^2)},\quad
V_{++}\simeq \frac{-{\cal
C}\,e^{-i\varkappa}}{2\sqrt{\cosh(y_1)\cosh(y_2)}}\frac{q_{\perp}^2}{m_q^2+q_{\perp}^2}
\end{eqnarray}
which in the considered limit $q_{\perp}\sim k_{\perp}\ll m_q$ up to
a common phase factor coincide with our previously published result
in Ref.~\cite{our-bb}. Again, analogously to the previous case
$V_{+-}$ drops out in the integration over $\psi$ in the diffractive
amplitude. The leading symmetric w.r.t. $\psi\leftrightarrow -\psi$
contribution to $V_{+-}$
\begin{eqnarray*}
V^{sym}_{+-}\sim \frac{q_{\perp}^4k_{\perp}}{M_{q\bar q}m_q^4}\ll
V_{++}
\end{eqnarray*}
is again extremely suppressed w.r.t. $V_{++}$ in the considered
low-$k_{\perp}$ asymptotics.

Numerically, dominant low-$k_{\perp}$ contribution coming from
$V_{++}$ amplitude (of course, at not extremely large $y_{1,2}$) in
the case of $b$-jets can lead to a dominant contribution to the
exclusive background for the Higgs CEP. In this asymptotics, the quark
mass $m_q$ plays an important role since it comes into the
denominator in Eq.~(\ref{hels-lowkt}). Precise evaluation of the
corresponding signal, however, demands employing the formulae for
the hard amplitudes in the general form given by Eqs.~(\ref{Vpm})
and (\ref{Vpp}). Detailed numerical investigation of contributions
from different parts of the phase space will be presented below in
the Results section.

\subsection{Unintegrated gluon distributions}

In our approach we use unintegrated gluon distributions
as proposed by Khoze, Martin and Ryskin (see e.g.~\cite{KMR_Higgs}).
In Ref.~\cite{Cudell:2008gv} slightly different unintegrated
gluon distributions taken from the analysis of
Ivanov and Nikolaev \cite{IN2002} were used.
These gluon distributions have been adjusted
to the deep-inelastic HERA data.
In addition the authors have shown that their off-diagonal
UGDFs provide a good description of the Tevatron data
on exclusive dijet production \cite{Cudell:2008gv}.
The KMR UGDFs discussed in the present paper also reasonably
well describe the dijet data \cite{DKRS2011,MPS2011}.

Let us now consider in detail the couplings of gluons to protons. At the
parton level, we assume that hard active gluons (carrying the
momentum fractions $x_{1,2}$) and screening gluons (carrying the
momentum fraction $x'\ll x_{1,2}$) couple to a quark line in the
proton in the normal way. In order to turn to the hadron level, the
factor $C_F\alpha_s(\mu^2_{\text{soft}})/\pi$ \cite{Forshaw05} is
absorbed into the off-diagonal unintegrated gluon distribution
function (UGDF)
$f^{\mathrm{off}}_g(x',x_{1,2},q_{1/2\perp}^2,q_{0\perp}^2,\mu_F^2;t)$.
The absorbed coupling $\alpha_s(\mu^2_{\text{soft}})$ corresponds to
the coupling of the screening gluon with virtuality
$\mu_{\text{soft}}^2\sim q_{0\perp}^2$ to a quark in the proton,
whereas the coupling of the active gluons to the $q{\bar q}$ central
system or to a quark in the proton is purely perturbative (given at
the hard scale $\mu_F \sim M_{q{\bar q}\perp}$) and enters to the
hard subprocess amplitude.

In the forward limit the following factorization is assumed
\begin{equation}
f^{\mathrm{off}}_g(x',x_{1,2},q_{1/2\perp}^2,q_{0\perp}^2,\mu_F^2;t)=
f^{\mathrm{off}}_g(x',x_{1,2},q_{1/2\perp}^2,q_{0\perp}^2,\mu_F^2)\,{\rm
exp}(bt/2) \; ,
\end{equation}
with the slope parameter $b\simeq 4\,{\rm GeV}^{-2}$ \cite{KMR00}.
In the $x'\ll x_{1,2}$ limit the off-diagonal UGDFs can be written as
\cite{Kimber:2001sc,MR}
 \begin{equation}
 f^{\mathrm{off}}_g(x',x_{1,2},q_{1/2\perp}^2,q_{0\perp}^2,\mu_F^2)\simeq
 R_g(x')\cdot f_g(x_{1,2},q_{1/2\perp}^2,\mu_F^2),
 \label{rg}
 \end{equation}
where the skewedness parameter $R_g\simeq 1.2-1.3$ is roughly
constant at LHC energies, which accounts for the single $\log Q^2$
skewed effect \cite{Shuvaev:1999ce} and gives only a small
contribution to an overall normalization uncertainty.

Another more symmetrical prescription for skewed UGDFs was
introduced in Refs.~\cite{our-eta,PST_chic0}. It is inspired by the
positivity constraints for the collinear Generalized Parton
Distributions \cite{posit}, and can be considered as a saturation of
the Cauchy-Schwarz inequality for the density matrix
\cite{PhysRept}. It allows us to incorporate the actual dependence
of the off-diagonal UGDFs on longitudinal momentum fraction of the
soft screening gluon and its transverse momentum in explicitly
symmetric way:
 \begin{equation}
 f^{\mathrm{off}}_{1/2\,g}\simeq
 \sqrt{ f_g(x_{1,2},q_{1/2\perp}^2,\mu_F^2)\cdot
 f_g(x',q_{0\perp}^2,\mu_{\text{soft}}^2)}\,, \quad x'\sim \frac{q_{0\perp}}{\sqrt{s}}\,.  \label{sqrt}
 \end{equation}
As we see it explicitly depends on $x'\sim q_{0\perp}^2/s$. It works
well in the description of the recent CDF data on the central
exclusive charmonia production \cite{PST_chic0,PST_chic12} and the
precise HERA data on the diffractive structure function
\cite{PEI10}. Model (\ref{sqrt}) implies the factorization of the
generalized UGDF into the hard part depending on a hard scale
$\mu_F$ and $x_{1,2}$ describing the hard gluon coupling to the
proton, and the soft part defined at some soft scale
$\mu_{\text{soft}}$ and small $x'\ll x_{1,2}$. Together with the
factorization in transverse momentum space, model (\ref{sqrt})
provides the QCD factorization of the diffractive amplitude in the
full momentum space.
\begin{figure*}[tbh]
 \centerline{
\includegraphics[width=0.33\textwidth]{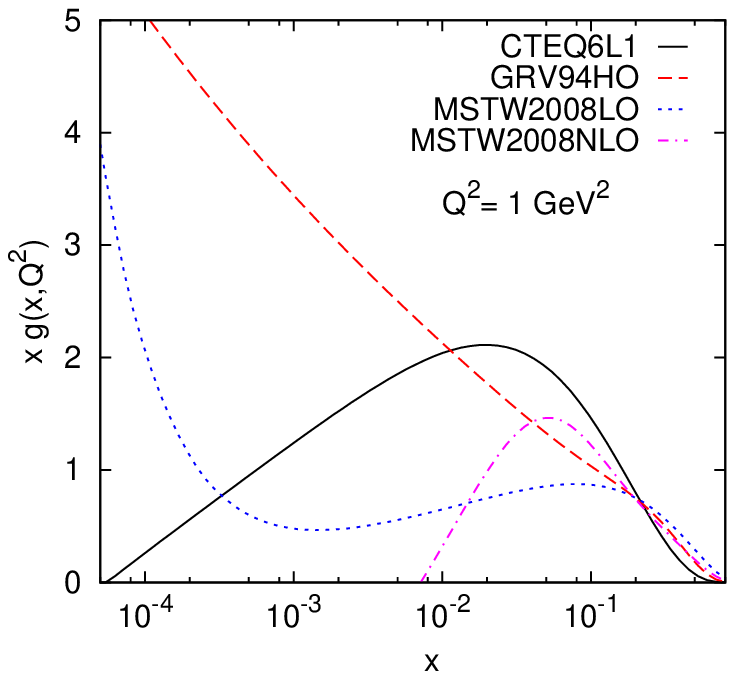}
\includegraphics[width=0.33\textwidth]{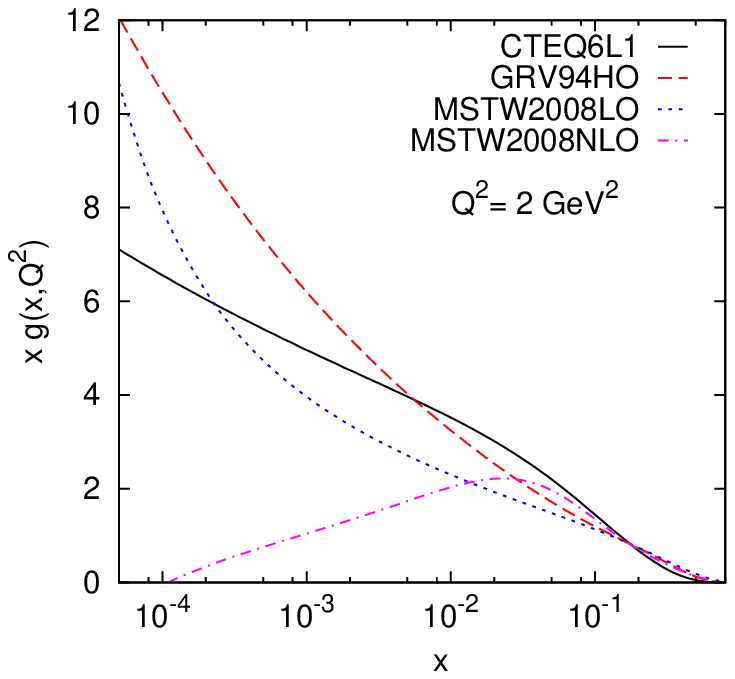}
\includegraphics[width=0.33\textwidth]{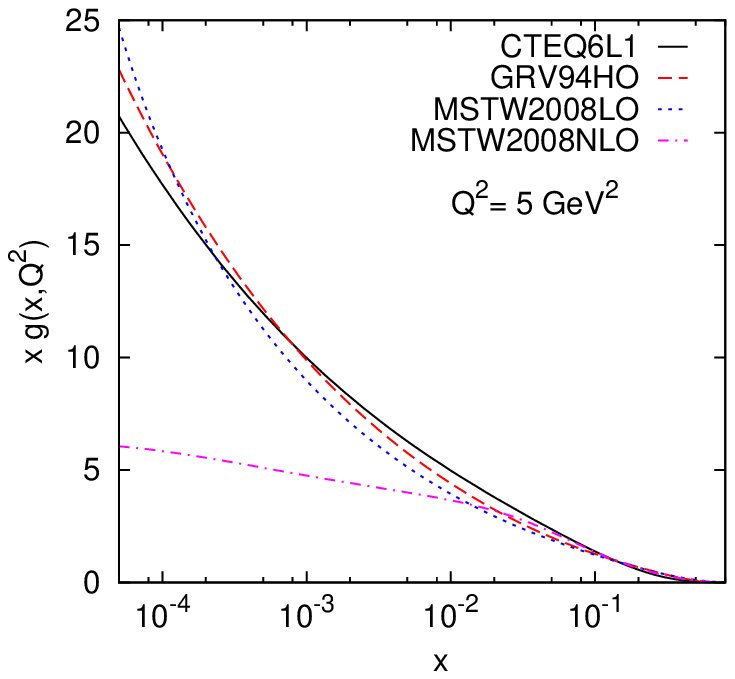}
}
   \caption{\label{fig:pdfs}
   Gluon densities as a function of longitudinal momentum fraction $x$
   at the scales $Q^2=1$, 2 and 5 GeV$^2$ given by
   the global parameterizations
   CTEQ6L1~\cite{CTEQ}, GRV94HO~\cite{GRV94},
   MSTW2008LO and NLO~\cite{MSTW}.}
\end{figure*}

In the considered kinematics the diagonal unintegrated densities can
be written in terms of the conventional (integrated) densities
$xg(x,q_{\perp}^2)$ as~\cite{MR}
\begin{equation}\label{ugdfkmr}
f_g(x,q_{\perp}^2,\mu^2)=\frac{\partial}{\partial\ln q_{\perp}^2}
\big[xg(x,q_{\perp}^2)\sqrt{T_g(q_{\perp}^2,\mu^2)}\big] \; ,
\end{equation}
where $T_g$ is the Sudakov form factor which suppresses real
emissions from the active gluon during the evolution, so that the
rapidity gaps survive. It is given by
\begin{equation}\label{Sudak}
T_g(q_\perp^2,\mu^2)={\rm exp}
\bigg(\!\!-\!\!\int_{q_\perp^2}^{\mu^2} \frac{d {\bf
k}_\perp^2}{{\bf k}_\perp^2}\frac{\alpha_s(k_\perp^2)}{2\pi}
\int_{0}^{1-\Delta} \!\bigg[ z P_{gg}(z) + \sum_{q} P_{qg}(z)
\bigg]dz \!\bigg) ,
\end{equation}
where the upper limit is taken to be
\begin{equation}\label{delta}
\Delta=\frac{k_\perp}{k_\perp+ a M_{q{\bar q}}} \; .
\end{equation}
The KMR group used $a = 0.62$ \cite{KKMR-spin}. It was argued
recently that $a = 1$ should be used instead \cite{CF09},
so in numerical calculations below we adopt the last
choice (for these two choices the final results for the Higgs CEP
cross section differ by about a factor of 2, which is not
negligible). In addition, following Ref.~\cite{CF09} we use the
factorization scale $\mu_F=M_{q{\bar q}}$ as compared to the KKMR
choice $\mu_F^{\text{KMR}}=M_{q{\bar q}}/2$. We will discuss the
sensitivity to the factorization scale choice below when presenting
numerical results.

In general, employing diagonal UGDF in the form (\ref{ugdfkmr}) we
encounter a problem of poorly known gluon PDFs at rather low
$x_{1,2}$ and especially small gluon virtualities $q_{\perp}^2$. For
an illustration of the corresponding uncertainties, in
Fig.~\ref{fig:pdfs} we show several gluon PDFs as functions of
fraction $x$ at evolution scale $\sim q_{\perp}^2$ fixed at values
1, 2 and 5 GeV$^2$ characteristic for the exclusive production of
Higgs boson. We see that at $x\lesssim 10^{-3}$ the PDF
uncertainties may strongly affect predictions for not sufficiently
large gluon transverse momenta. In this sense, the precise data on
the diffractive and central exclusive production could be used for
making constraints on the PDF parameterizations \cite{PEI10}.

Testing other models of UGDFs, different from Eq.~(\ref{ugdfkmr}),
may be important for estimation of an overall theoretical
uncertainty of our predictions and their stability.

\section{Electromagnetic $\gamma^*\gamma^*$-fusion process}

\begin{figure}[h!]
 \centerline{\epsfig{file=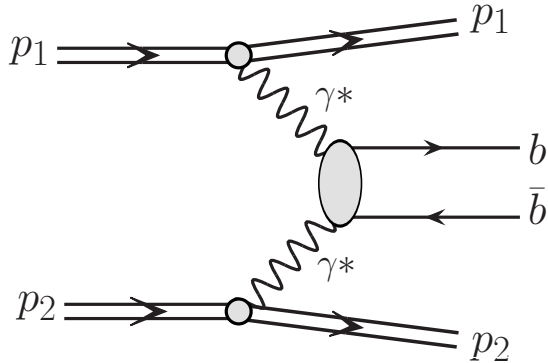,width=8cm}}
 \caption{
The QED $\gamma^* \gamma^*$ fusion mechanism of the exclusive $q
\bar q$ production.}
 \label{fig:QED}
\end{figure}

It is instructive to estimate the QED contribution to the central
exclusive $b{\bar b}$ production illustrated in Fig.~\ref{fig:QED}.

In the forward limit of small momentum transfers $t_{1,2}$
($|t_{1,2}| \ll 4 m_N^2$) the matrix element for $pp \to pp\,q{\bar
q}$ reaction via $\gamma^*\gamma^*$-fusion can be written as
\cite{our-eta}
\begin{eqnarray}
{\cal M}^{\gamma^* \gamma^*} \approx e F_1(t_1) \frac{(p_1 +
p_1')^{\nu}}{t_1}\, V_{\mu\nu}^{\gamma^* \gamma^*}(q_1,q_2)\,
\frac{(p_2 + p_2')^{\mu}}{t_2}\, e F_1(t_2)\,, \label{ggamp}
\end{eqnarray}
where $F_1(t_1)$ and $F_1(t_2)$ are Dirac proton electromagnetic
form factors, and the $\gamma^*\gamma^* \to q{\bar q}$ vertex
\cite{PST_chic0} has analogous form as (\ref{fin-vert-mn}), i.e.
\begin{eqnarray}\nonumber
&&{}V_{\lambda_q\lambda_{\bar{q}},\mu\nu}^{\gamma^*
\gamma^*}=\left(e_{q}e\right)^2\,\bar{u}_{\lambda_q}(k_1)
\biggl(\gamma^{\nu}\frac{\hat{q}_{1}-\hat{k}_{1}-m_q}
{(q_1-k_1)^2-m_q^2}\gamma^{\mu}-\gamma^{\mu}\frac{\hat{q}_{1}-
\hat{k}_{2}+m_q}{(q_1-k_2)^2-m_q^2}\gamma^{\nu}\biggr)v_{\lambda_{\bar{q}}}(k_2).\\
\label{ggvert}
\end{eqnarray}
Momentum conservation dictates us the following decompositions of
the photon momenta into the longitudinal and transverse parts w.r.t.
c.m.s. direction \cite{PST_chic0}:
\begin{eqnarray}\label{qgg}
q_1=x_1p_1+\frac{t_1}{s}\,p_2+q_{1\perp},\quad
q_2=x_2p_2+\frac{t_2}{s}\,p_1+q_{2\perp},\quad q_{1/2\perp}^2 \simeq
t_{1,2}(1-x_{1,2}) \; ,
\end{eqnarray}
where $t_{1,2}\equiv q_{1,2}^2$. Due to gauge invariance we have
(similarly to (\ref{decomp}))
\begin{eqnarray}
V^{\gamma^*\gamma^*}(q_1,q_2)=(p_1+p_1')^{\nu}
V^{\gamma^*\gamma^*}_{\mu\nu}(q_1,q_2)(p_2+p_2')^{\mu}=
4p_1^{\nu}p_2^{\mu}V_{\mu\nu}(q_1,q_2).
\end{eqnarray}
so the matrix element squared $|V^{\gamma^*\gamma^*}|^2$ is
proportional to the gluonic one found in Eq.~(\ref{V2}). The photon
virtualities in the relevant limit disappear $t_{1,2}\to 0$, so
$q_{\perp}\to 0$ leading to vanishing of the $\gamma^*\gamma^*\to
q\bar{q}$ amplitude in the high-$k_{\perp}$ limit and massless
quarks (see Eq.~(\ref{V2})). Thus, to estimate $\gamma^*\gamma^*$
contribution to exclusive production of quark jets we have to take into
account subleading corrections in the $m_q/k_{\perp}$-expansion. This
means that the electromagnetic mechanism may give some contribution
for moderate and large quark masses, which will be
evaluated numerically in the Results section.

\section{Off-shell effects in central exclusive Higgs production}

As one can see from Eq.~(\ref{decomp}), the subprocess vertex
$V_{\lambda_q\lambda_{\bar q}}$ of $g^*g^*\to b{\bar b}$ is
proportional to gluon transverse momenta squared due to projection
$\sim q_{1\perp}^{\mu}q_{2\perp}^{\nu}$. When considering the
irreducible $b{\bar b}$ background for Higgs boson production in the
amplitude $V_{\lambda_q\lambda_{\bar q}}^{\mu\nu}$ we can neglect,
in principle, the gluon virtualities in comparison with quark
transverse momenta $k_{\perp}$ and $b$-quark mass $m_q$ as it was
done in e.g. Ref.~\cite{KMR_bbar_suppression}.

However, as demanded by $k_{\perp}$-factorization framework, it may
be instructive to analyze in which region of the phase space the
gluon off-shell effects may play a significant role (if any), and
whether it is possible to see such effects in experiment. A complete
calculation of the off-shell effects in inclusive Higgs boson
production was performed in Ref.~\cite{incl-Higgs} (for the Higgs
boson production in $k_{\perp}$-factorisation, see also
Ref.~\cite{Hautmann}). It was shown there that the off-shell effects
can significantly affect the distribution of Higgs boson cross
section in azimuthal angle between fusing gluons $\phi$ in a very
close vicinity of $\phi=\pi/2$. The calculations of the central
exclusive Higgs production rates in the on-shell gluon approximation
are well-known \cite{KMR_Higgs}. Let us now investigate off-shell
effects in Higgs CEP.

\subsection{Off-shell effects in the hard vertex $g^*g^*\to H$}

Tensor decomposition of the hard subprocess amplitude $g^*g^*\to H$
can be written in the following general form \cite{incl-Higgs}
\begin{eqnarray}
T^{ab}_{\mu \nu}(q_1,q_2) &=&
i\delta^{ab}\frac{\alpha_{s}}{2\pi}\frac{1}{v}\biggl([(q_1q_2)g_{\mu
\nu}-q_{1,\nu}q_{2,\mu}]G_1+
\nonumber
\\
&+&\left[q_{1,\mu}q_{2,\nu}-\frac{q_1^2}{(q_1q_2)}q_{2,\mu}q_{2,\nu}-
\frac{q_2^2}{(q_1q_2)}q_{1,\mu}q_{1,\nu}+
\frac{q_1^2q_2^2}{(q_1q_2)^2}q_{1,\nu}q_{2,\mu}\right]G_2\biggr),
\label{ampmain}
\end{eqnarray}
where $\alpha_{s}$ is the strong coupling constant,
$v=(G_{F}\sqrt{2})^{-1/2}$ is the electroweak parameter of the
Standard Model, $a,\,b$ are the colour indices of two virtual gluons
with momenta $q_1,\,q_2$.

Let us introduce the dimensionless parameters
\begin{eqnarray*}
\chi=\frac{M_H^2}{4m^{2}_{f}}>0,\qquad
\chi_1=\frac{q_1^2}{4m^{2}_{f}}<0,\qquad
\chi_2=\frac{q_2^2}{4m^{2}_{f}}<0,
\end{eqnarray*}
so the heavy quark limit corresponds to
$\chi,\,\chi_1,\,\chi_2\rightarrow 0$. In the case of heavy Higgs
production we have $M_H^2\gg |{\bf q}_{1\perp}|^2,\,|{\bf
q}_{2\perp}|^2$, so in the expansion of the form factors we have to take
into account powers of $\chi$ higher than powers of gluon
virtualities $\chi_1,\,\chi_2$ \cite{incl-Higgs}
\begin{eqnarray}
 G_1(\chi,\chi_1,\chi_2)&=&\frac23
 \left[1+\frac{7}{30}\chi+\frac{2}{21}\chi^{2}+
 \frac{11}{30}(\chi_1+\chi_2)+...\right], \nonumber
\\
 G_{2}(\chi,\chi_1,\chi_2)&=&-\frac{1}{45}(\chi-\chi_1-\chi_2)-
 \frac{4}{315}\chi^{2}+... \; . \label{G1G2}
\end{eqnarray}
These expansions will be sufficient for our present calculations.

Let us turn now to the discussion of the exclusive diffractive Higgs
production within the KKMR double diffractive mechanism
\cite{KMR_Higgs}. The hard subprocess vertex entering the
diffractive amplitude (\ref{ampl}) $V\equiv
V(q_{1\perp}^2\,q_{2\perp}^2,P_{\perp}^2)$ with explicit taking into
account gluon virtualities reads
\begin{eqnarray}
V^{ab}_{g^*g^*\to
H}(q_{1\perp}^2\,q_{2\perp}^2,P_{\perp}^2)=n^+_{\mu}n^-_{\nu}T_{\mu\nu}^{ab}(q_1,q_2)=
\frac{4}{s}\frac{q^{\mu}_{1\perp}}{x_1}\frac{q^{\nu}_{2\perp}}{x_2}
T^{ab}_{\mu\nu}(q_1,q_2),\quad
q_1^{\mu}T^{ab}_{\mu\nu}=q_2^{\nu}T^{ab}_{\mu\nu}=0, \label{V-higgs}
\end{eqnarray}
where the amplitude of the gluon fusion $T_{\mu\nu}^{ab}(q_1,q_2)$
is defined in Eq.~(\ref{ampmain}). Contracting indices and
introducing the transverse Higgs mass,
\begin{eqnarray}
sx_1x_2=M_H^2-P_{\perp}^2\equiv M_{H\perp}^2,\qquad
P_{\perp}^2=-|{\bf P}_{\perp}|^2=-({\bf q}_{1\perp}^2+{\bf
q}_{2\perp}^2 +2|{\bf q}_{1\perp}||{\bf q}_{2\perp}|\cos{\phi}),
\end{eqnarray}
we get in terms of form factors (\ref{G1G2})
\begin{eqnarray}\label{proj-excl}
V^{ab}_{g^*g^*\to H}=-i\delta^{ab}\frac{\alpha_s}{\pi}\frac{|{\bf
q}_{1\perp}||{\bf q}_{2\perp}|}{v}\biggl[\cos{\phi}\;G_1-
\frac{2M_{H\perp}^2|{\bf q}_{1\perp}||{\bf q}_{2\perp}|}{(M_H^2+{\bf
q}_{1\perp}^2+{\bf q}_{2\perp}^2)^{2}}\;G_2\biggr]\,.
\end{eqnarray}

In the limit of real gluons for not extremely heavy Higgs we have
asymptotically
\begin{eqnarray}
V^{ab}_{gg\to
H}\simeq-i\delta^{ab}\frac{\alpha_{s}}{\pi}\frac{1}{v}({\bf
q}_{1\perp}{\bf q}_{2\perp})\cdot \frac23, \qquad ({\bf
q}_{1\perp}{\bf q}_{2\perp})=|{\bf q}_{1\perp}||{\bf
q}_{2\perp}|\cos{\phi}. \label{Vhiggsre}
\end{eqnarray}
Substituting this into the amplitude (\ref{ampl}), we get after
summation over colour indices
\begin{eqnarray*}
{\cal M}^{\mathrm{on-shell}}_{excl}= \frac{\pi\alpha_{s}}{v}\cdot
\frac{s}{3}\int d^2 {\bf q}_{0\perp}\frac{({\bf q}_{1\perp}{\bf
q}_{2\perp})}{{\bf q}_{0\perp}^2\,{\bf q}_{1\perp}^2\,{\bf
q}_{2\perp}^2}\,f^{\mathrm{off}}_{g,1}(x_1,x',q_{0\perp}^2,
q_{1\perp}^2,t)f^{\mathrm{off}}_{g,2}(x_2,x',q_{0\perp}^2,q_{2\perp}^2,t_2),
\end{eqnarray*}

Next-to-leading order contribution in the cross section of the hard
subprocess $\hat{\sigma}(gg\to H)$ can be accounted for by a factor
$K_{NLO}\simeq1.5$ assuming that the NLO corrections factor
$K_{\mathrm{NLO}}$ in the $g^*g^*\to H$ vertex is the same as in the
$H\to gg$ width provided that $|V_{gg\to H}|^2\sim\Gamma(H\to gg)$
\cite{KMR-Knlo}. As for the irreducible Higgs background, the
one-loop corrections to $gg\to b{\bar b}$ were calculated in
Ref.~\cite{KMR-bb}. For simplicity, in the current analysis we do
not take into account the NLO effects and concentrate only on the
leading order contributions in the hard subprocess part.

At high energies the cross section for the exclusive Higgs boson
production can be expressed as:
\begin{eqnarray*}
d\sigma_{pp\to pHp}=\frac{1}{2s}\,|{\cal M}|^2\cdot d^{\,3}PS,\quad
d^3 PS = \frac{1}{2^8 \pi^4\,s} dt_1 dt_2 dy_H d \Phi\,.
\end{eqnarray*}

\subsection{$b \bar b$ signal of the Higgs decay}

In order to estimate the observable signal from the central
exclusive production of Higgs in the $b\bar{b}$ decay channel, one
can multiply the matrix element squared $|{\cal M}_{excl}|^2$ by
the relativistic Breit-Wigner distribution over the invariant mass
$\mu\equiv M_{b\bar{b}}$ of $b\bar{b}$ pair with a proper
normalization
\begin{eqnarray}\label{BW}
 \rho_{b\bar{b}}(\mu)=\frac{1}{\pi}\,\frac{\mu\Gamma_H^{b\bar{b}}(\mu)}
 {[\mu^2-M_H^2]^2+[\mu\Gamma_H^{tot}(\mu)]^2}\,,
\end{eqnarray}
and then integrate it out over the 4-particle $p(b\bar{b})p$
invariant phase space. In Eq.~(\ref{BW}) the total Higgs decay
width into fermions and gluons is (see e.g. Ref.~\cite{Passarino})
\begin{eqnarray}\label{Hdecay}
 \Gamma_H^{tot}(M_H)&=&\Gamma_H^{f\bar{f}}(M_H)+\Gamma_H^{gg}(M_H)\,,\\
 \Gamma_H^{f\bar{f}}(M_H)&=&\frac{g^2M_H}{32\,\pi
M_W^2}\,\Bigg\{3\left[m_b^2(M_H)+m_c^2(M_H)\right]
\left[1+5.67\frac{\alpha_s(M_H)}{\pi}+
42.74\left(\frac{\alpha_s(M_H)}{\pi}\right)^2\right]
\nonumber\\
&+&m_{\tau}^2\Bigg\}\,,\nonumber\\
 \Gamma_H^{gg}(M_H)&=&\frac{g^2M_H^3}{288\,\pi
 M_W^2}\,\frac{\alpha_s^2(M_H)}{\pi^2}\Bigg\{1+17.91667\frac{\alpha_s(M_H)}{\pi}\Bigg\}
 \nonumber
\end{eqnarray}
and the partial Higgs decay width into $b\bar{b}$ channel is given by
\begin{eqnarray}\label{Hdecay-bb}
\Gamma_H^{b\bar{b}}(M_H)&=&\frac{3g^2M_H}{32\,\pi
M_W^2}\,m_b^2(M_H)\left[1+5.67\frac{\alpha_s(M_H)}{\pi}+
42.74\left(\frac{\alpha_s(M_H)}{\pi}\right)^2\right]\;.
\end{eqnarray}
Above, $g^2=0.42502$ is the electromagnetic coupling constant, and
$m_{c,b}(\mu)$ and $\alpha_s(\mu)$ are the running quark masses and
the QCD coupling constant, respectively. For example, at $M_H=100$
GeV, they are $m_c(M_H)=0.542$ GeV, $m_b(M_H)=2.676$ GeV and
$\alpha_s(M_H)=0.12121$. Substituting these values into
Eq.~(\ref{Hdecay}) we get the total decay width
$\Gamma_H^{tot}(M_H)=2.268$ MeV which coincides with the number
given in Ref.~\cite{Passarino}.

\section{Numerical results}

In this section we will present differential distributions for
central exclusive heavy quark dijets and Higgs production. We show also
distributions of $b$ and $\bar b$ quarks from the decay of the
Higgs boson. We start from the presentation of the results for heavy
quark ($c \bar c$ and $b \bar b$) production. Very important part of
the analysis below concerns the $b \bar b$ background below the $b
\bar b$ Higgs signal.

\subsection{Differential distributions for exclusive $c{\bar c}$ and
$b{\bar b}$ pair production}

\begin{figure}[!h]
\begin{minipage}{0.47\textwidth}
 \centerline{\includegraphics[width=1.0\textwidth]{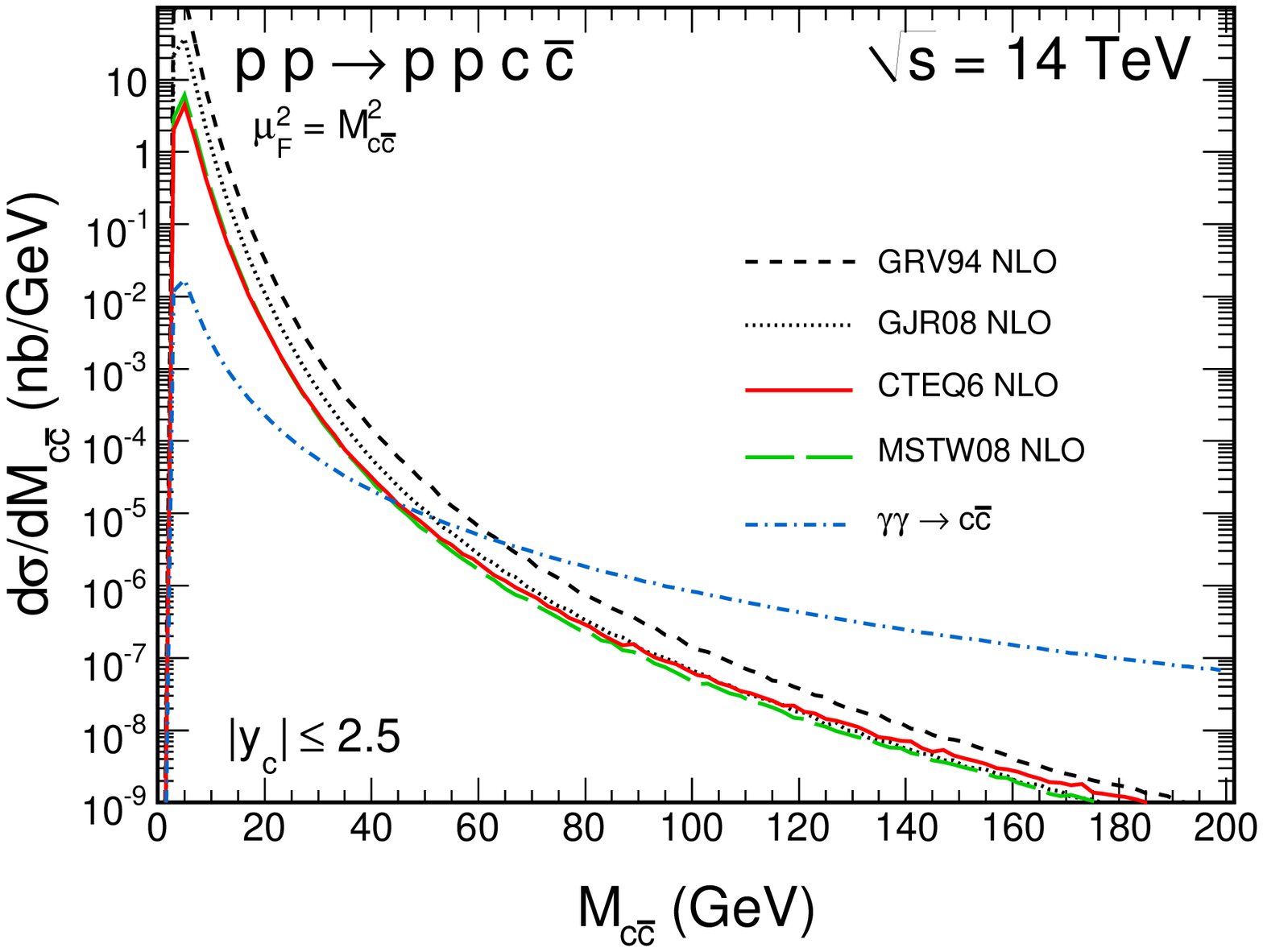}}
\end{minipage}
\hspace{0.5cm}
\begin{minipage}{0.47\textwidth}
 \centerline{\includegraphics[width=1.0\textwidth]{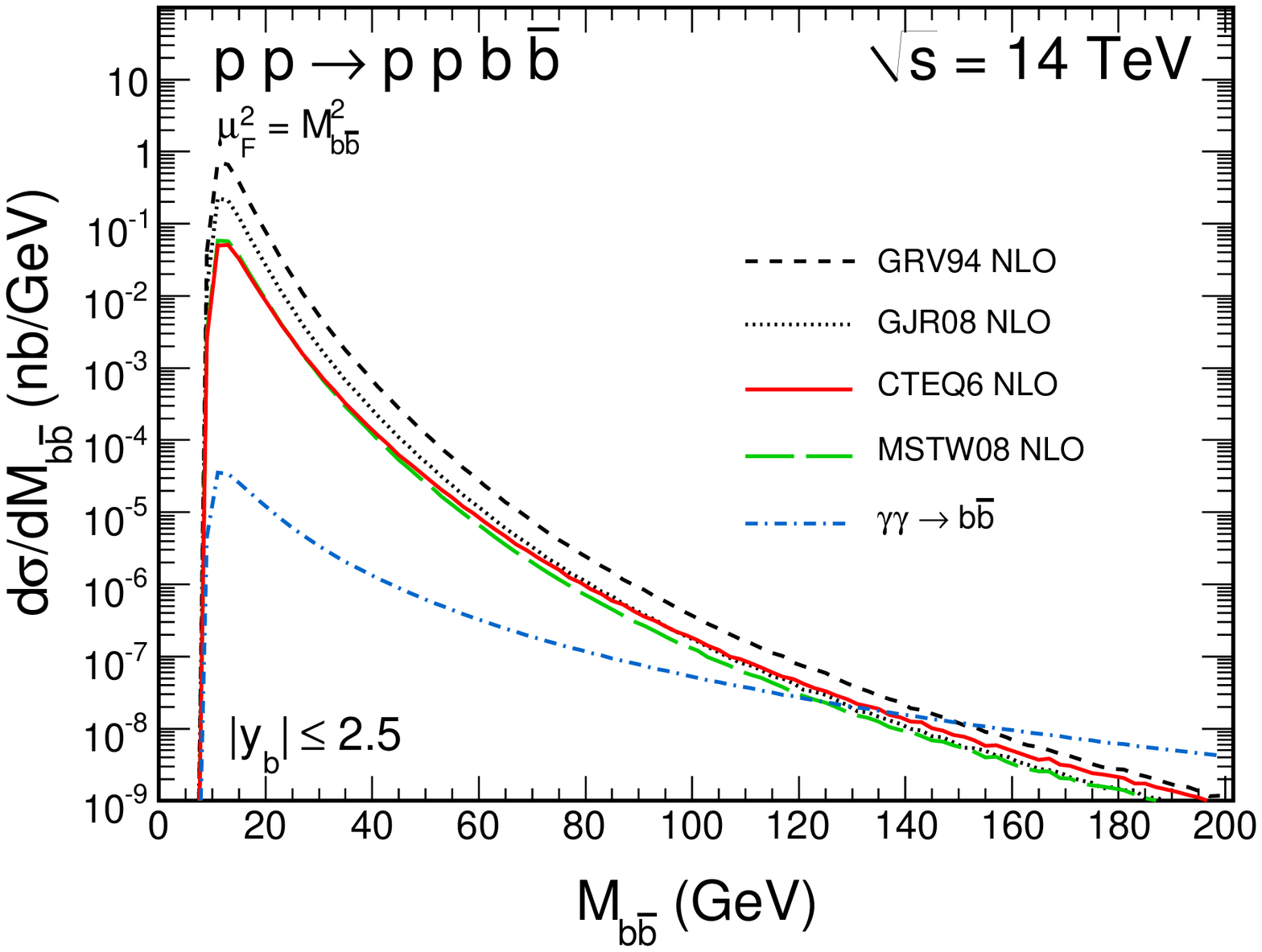}}
\end{minipage}
   \caption{
 \small Invariant mass $M_{q{\bar q}}$ distributions for the central
 exclusive production of $c \bar c$ (left panel) and $b \bar b$ (right
 panel) dijets at LHC. Results for diffractive component are shown for a few
 GDFs from the literature: GRV94 NLO \cite{GRV94} (dashed line), GJR08 NLO
 \cite{GJR} (dotted line),
 CTEQ6 NLO \cite{CTEQ} (solid line)
 and MSTW2008 NLO \cite{MSTW} (long-dashed line). QED
 $\gamma^*\gamma^*$ fusion contribution is illustrated by
 the dash-dotted line. Experimental cuts on quark (antiquark) rapidities
 $|y_q|\leq 2.5$ are included. The distribution is then integrated over whole
 invariant $q{\bar q}$ mass range $2m_q<M_{q{\bar q}}<200$ GeV in order to get
 the total cross section.
}
 \label{fig:dM-pdfs}
\end{figure}

We start our presentation from the distribution in heavy quark invariant
mass. In Fig.~\ref{fig:dM-pdfs} we show distributions for $c \bar c$
(left panel) and $b \bar b$ (right panel) for different gluon collinear
distributions used to generate UGDFs, for quark rapidities
$y_q, y_{\bar q} \in (-2.5,2.5)$.
We show distributions for QCD diffractive mechanism as well as for
the QED $\gamma^*\gamma^*$ fusion. Only at small invariant masses
is the distribution for $c{\bar c}$ higher than that for $b{\bar b}$.
The position of the peak of the distributions depends on the quark mass
and is placed slightly above $2m_q$, both for diffractive and QED
components. Relative contribution of QED
mechanism grows with increasing invariant $q{\bar q}$ mass,
and starts to dominate at $M_{q{\bar q}}\gtrsim 60$ GeV
for $c{\bar c}$ and $M_{q{\bar q}}\gtrsim 120$ GeV for $b \bar b$.

There is a strong sensitivity of the invariant mass distribution
on the gluon PDFs. In particular, the difference of the
results for the GRV94 NLO \cite{GRV94} and CTEQ6 NLO \cite{CTEQ} is up
to an order of magnitude in the peak. The QED contribution is found to
be smaller than that for the diffractive mechanism but it is very important as background
to the exclusive Higgs boson production. In this calculation
only the Dirac $F_1$ proton electromagnetic form factor is included.
The contribution of the Pauli $F_2$ proton electromagnetic form factor is expected to be negligible.

For the bulk of the $c \bar c$ or $b \bar b$ production
$M_{q \bar q}$ is only slightly larger than the quark masses.
Nevertheless what matters is that $M_{q \bar q}$ is in the
perturbative region. We can go down with transverse momenta
of gluons as low as $q_{\perp}^2 \sim$  0.4 GeV$^2$. Of course
it is not easy to predict what happens in the nonperturbative
region. In the present paper we concentrate on large quark-antiquark
invariant masses where the issue is not so important.

\begin{figure}[!h]
\begin{minipage}{0.47\textwidth}
 \centerline{\includegraphics[width=1.0\textwidth]{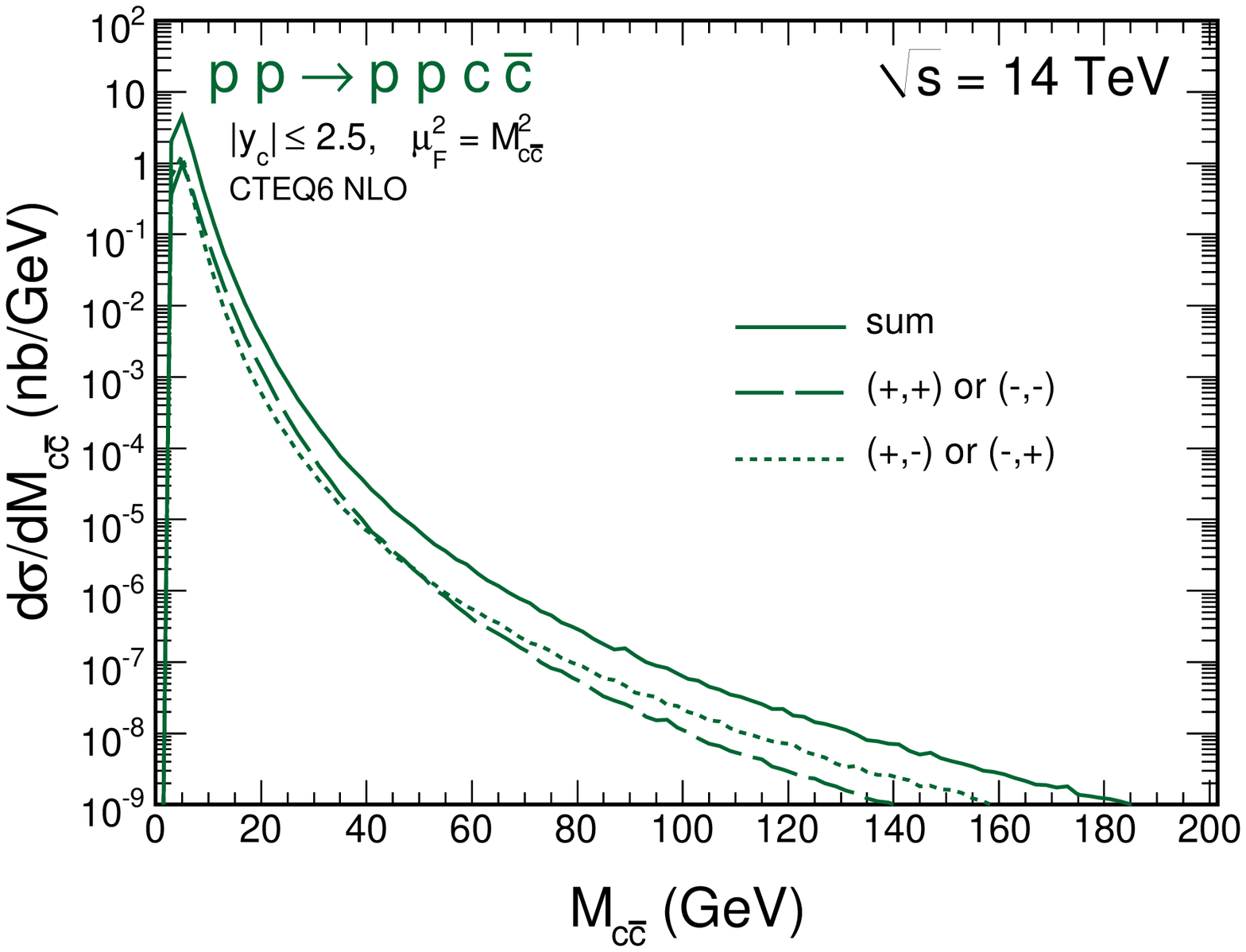}}
\end{minipage}
\hspace{0.5cm}
\begin{minipage}{0.47\textwidth}
 \centerline{\includegraphics[width=1.0\textwidth]{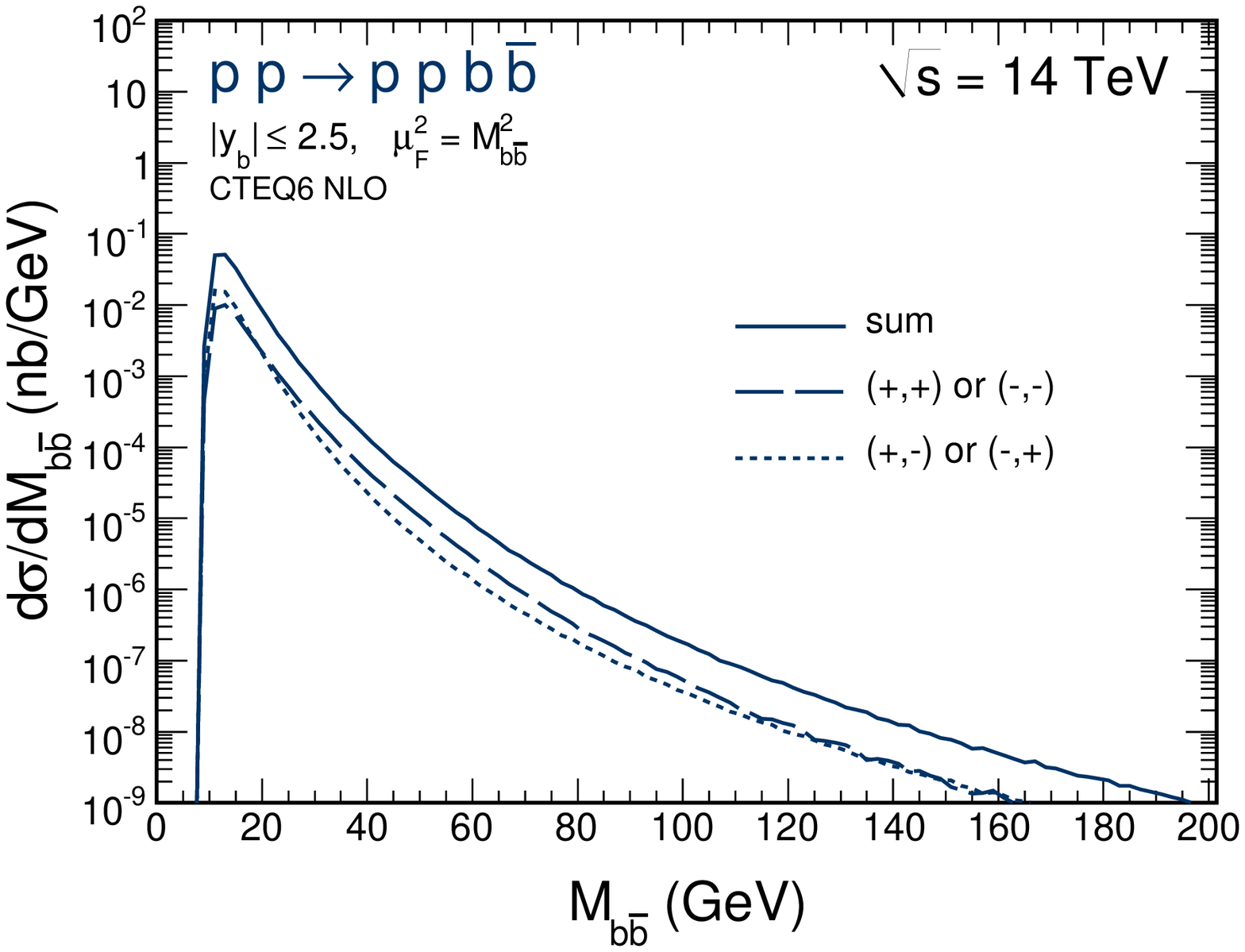}}
\end{minipage}
   \caption{
 \small Invariant mass distributions of centrally produced $c{\bar c}$ pair
 (left panel) and $b{\bar b}$ pair (right panel) for different (anti)quark
 helicities $\lambda_q\lambda_{\bar q}=++,\,-+$ and for the sum over
 all quark helicity states. CTEQ6 PDF was used here.
 Kinematical constraints are the same as in
 Fig.~\ref{fig:dM-pdfs}.}
 \label{fig:dsig_dM-helicities}
\end{figure}

There was recently \cite{KMR-bb} a discussion about the contribution
of different helicity states to the cross section. In
Fig.~\ref{fig:dsig_dM-helicities} we show individual contributions
for different quark helicities $\lambda_q\lambda_{\bar q}=++$ and
$-+$ (other helicity contributions are the same due to symmetry as
discussed above). The contributions of the same and opposite quark
helicities are rather similar in the broad range of the
quark-antiquark pair invariant masses. In particular, they are
almost identical in the region of typical light Higgs mass.

\begin{figure}[!h]
\begin{minipage}{0.47\textwidth}
 \centerline{\includegraphics[width=1.0\textwidth]{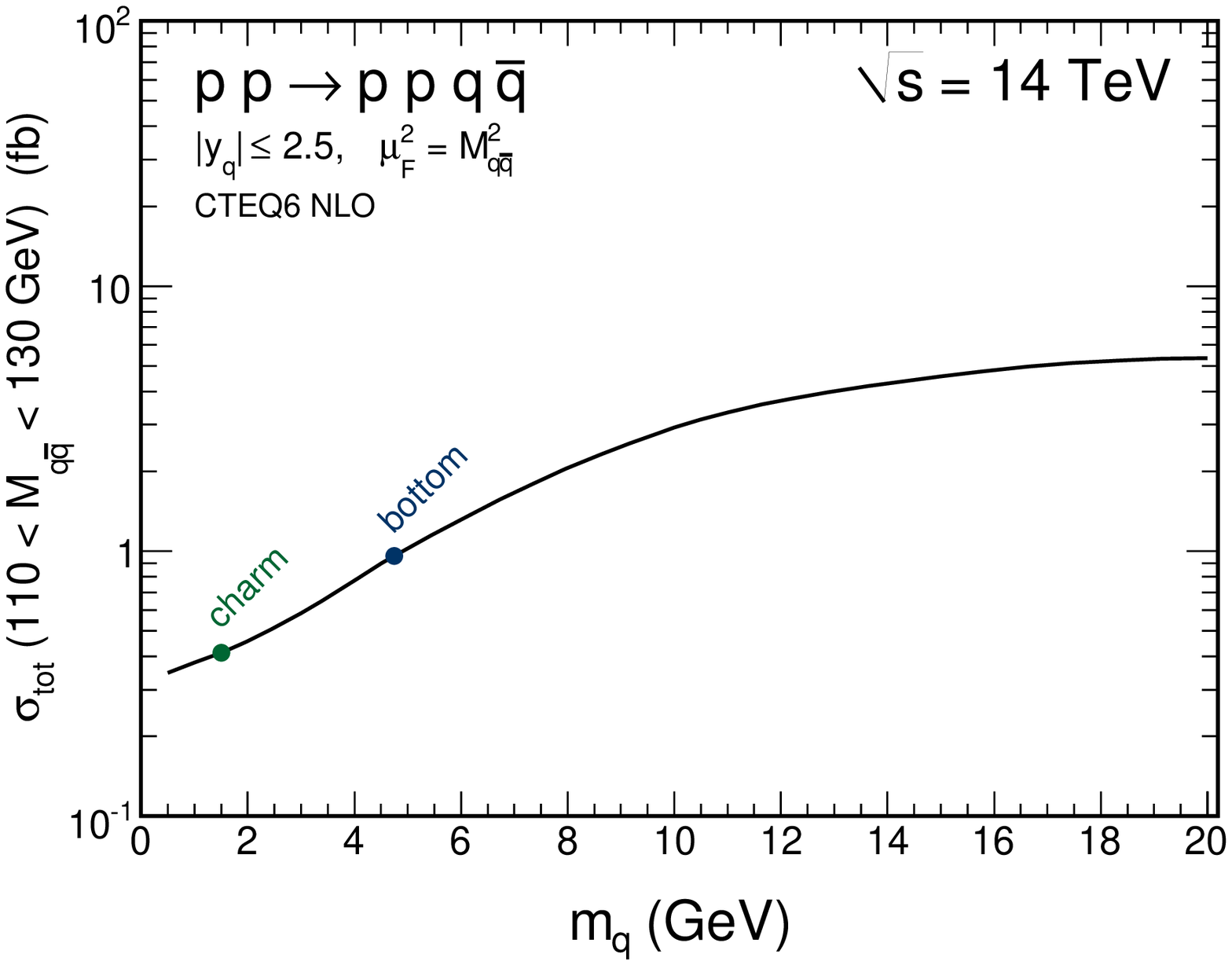}}
\end{minipage}
\hspace{0.5cm}
\begin{minipage}{0.47\textwidth}
 \centerline{\includegraphics[width=1.0\textwidth]{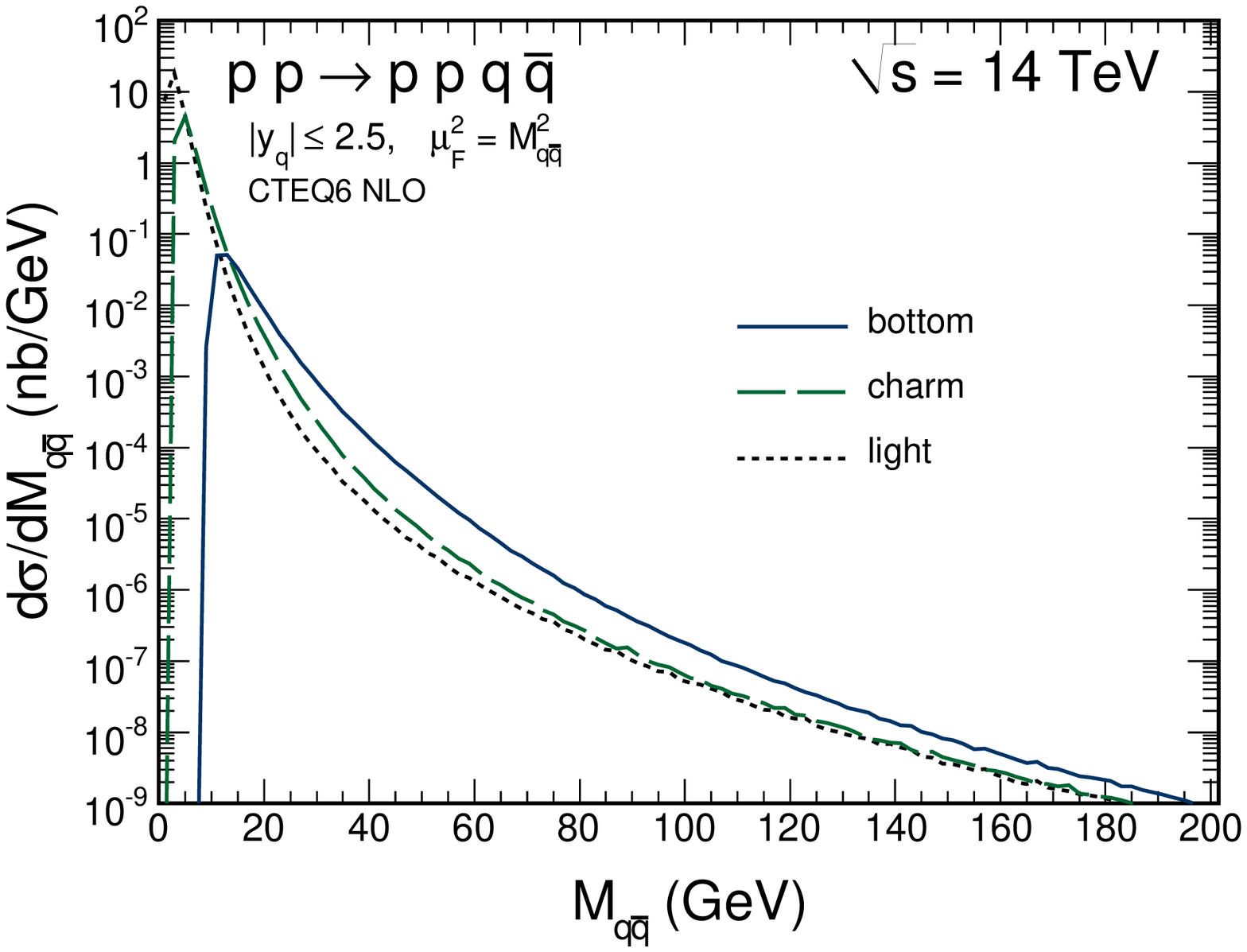}}
\end{minipage}
   \caption{
 \small The EDD $q{\bar q}$ cross section in the vicinity of the Higgs mass
  as a function of quark mass(left panel) and invariant mass distributions of the EDD $q{\bar q}$
  production for different quark masses. Kinematical constraints are the same as
 in Fig.~\ref{fig:dM-pdfs}. }
 \label{fig:dsig_dM-mq}
\end{figure}

The cross section for the diquark production strongly depends on the
quark masses thus underlying the importance on the finite $b$-quark
masses for the CEP Higgs background evaluation. This is encoded in
the matrix elements discussed in the theory section. In
Fig.~\ref{fig:dsig_dM-mq} we have collected the results for the
total (left panel) and differential in quark-antiquark invariant
mass (right panel) cross sections. The total cross section slightly
grows with the quark mass. The growth is, however, slower than in
Ref.~\cite{KMR-bb} where the matrix element is proportional to the
quark mass. Taken a typical misidentification probability and the
fact that light quark cross sections are smaller than that for $b
\bar b$ in the Higgs region, the latter is in practice the only
troublesome background. Our result is very interesting in the
context of diffractive dijet production. Recently the CDF
collaboration has measured the corresponding cross section at the
Tevatron \cite{CDF-dijets}. The quark jets are usually neglected and
only gluonic jets are included in theoretical calculations (for the
gluonic jets analysis, see e.g. Ref.~\cite{Cudell:2008gv}). The
quark jets contribution to the CDF data will be discussed elsewhere
\cite{MPS2011}.

\begin{figure}[!h]
\begin{minipage}{0.47\textwidth}
 \centerline{\includegraphics[width=1.0\textwidth]{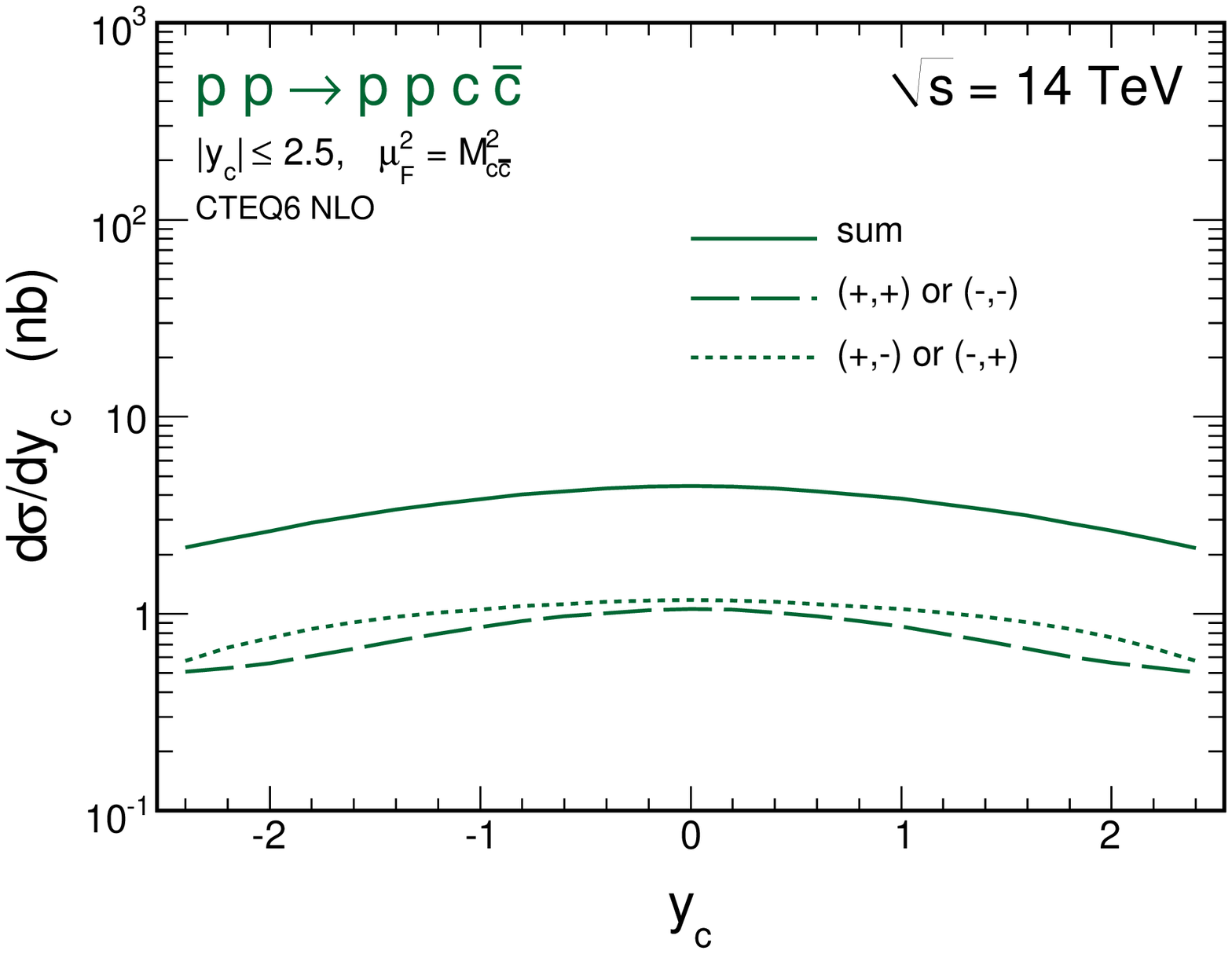}}
\end{minipage}
\hspace{0.5cm}
\begin{minipage}{0.47\textwidth}
 \centerline{\includegraphics[width=1.0\textwidth]{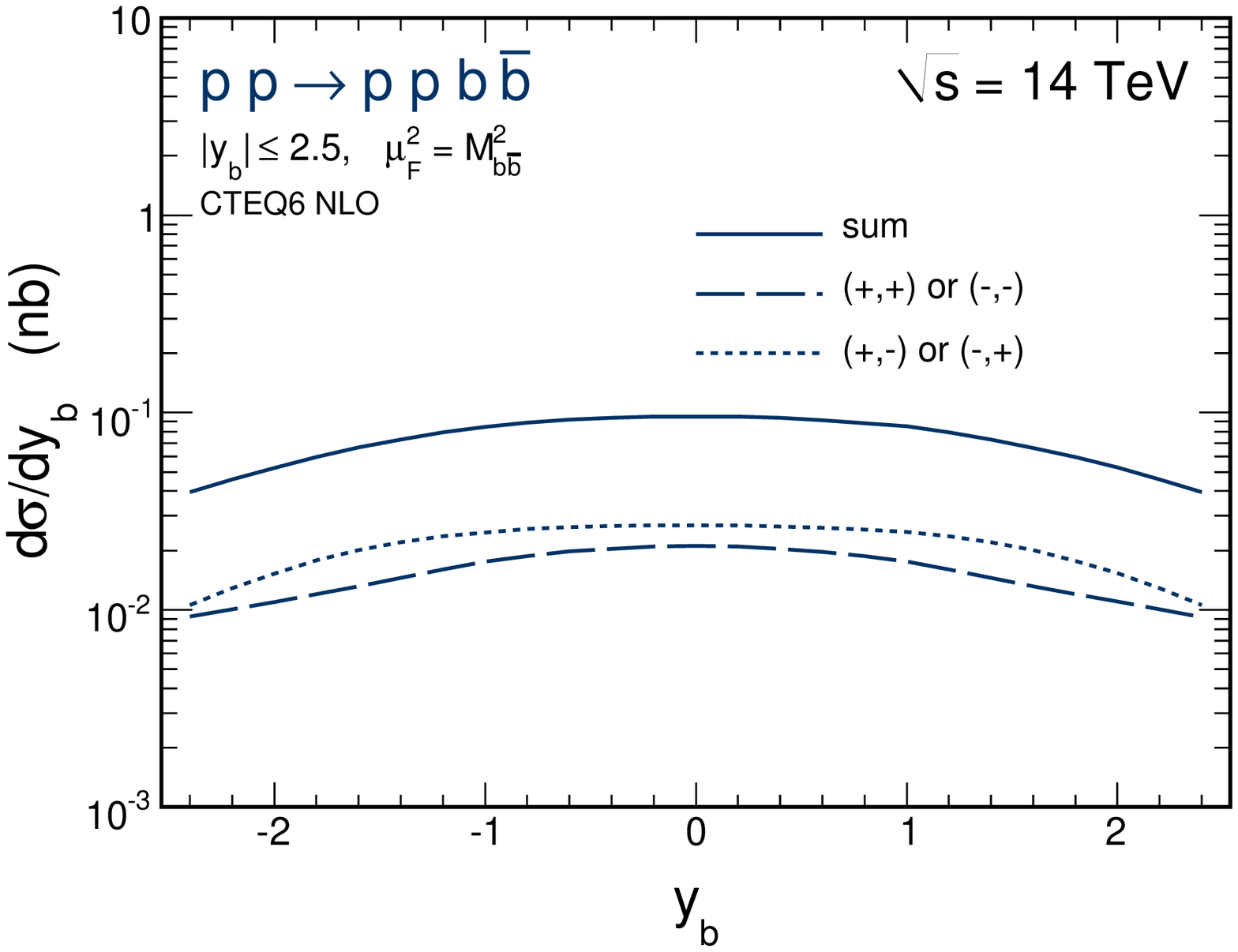}}
\end{minipage}
   \caption{
 \small Differential distributions in quark (antiquark) rapidity
 of EDD $c$ or $\bar c$ (left panel) and $b$ or $\bar b$ (right panel).
 Kinematical constraints are the same as
in Fig.~\ref{fig:dM-pdfs}.}
 \label{fig:dsig_dy}
\end{figure}

Let us come now to the rapidity distributions. The distribution in
quark (antiquark) rapidity in the detector interval, integrated over
whole invariant mass range of $M_{q{\bar q}}\in (2m_q,200)$ GeV, is
shown in Fig.~\ref{fig:dsig_dy}. The distribution for charm quarks
is flatter than that for the bottom quarks. For comparison, we also
show the corresponding distributions for different combinations of
quark helicities $\lambda_q\lambda_{\bar q}=++$ and $-+$. As for the
quark-antiquark invariant mass distributions, the same and opposite
helicity contributions are similar. The
rapidity distributions shown are dominated by the low
quark-antiquark invariant masses.

\begin{figure}[!h]
\begin{minipage}{0.47\textwidth}
 \centerline{\includegraphics[width=1.0\textwidth]{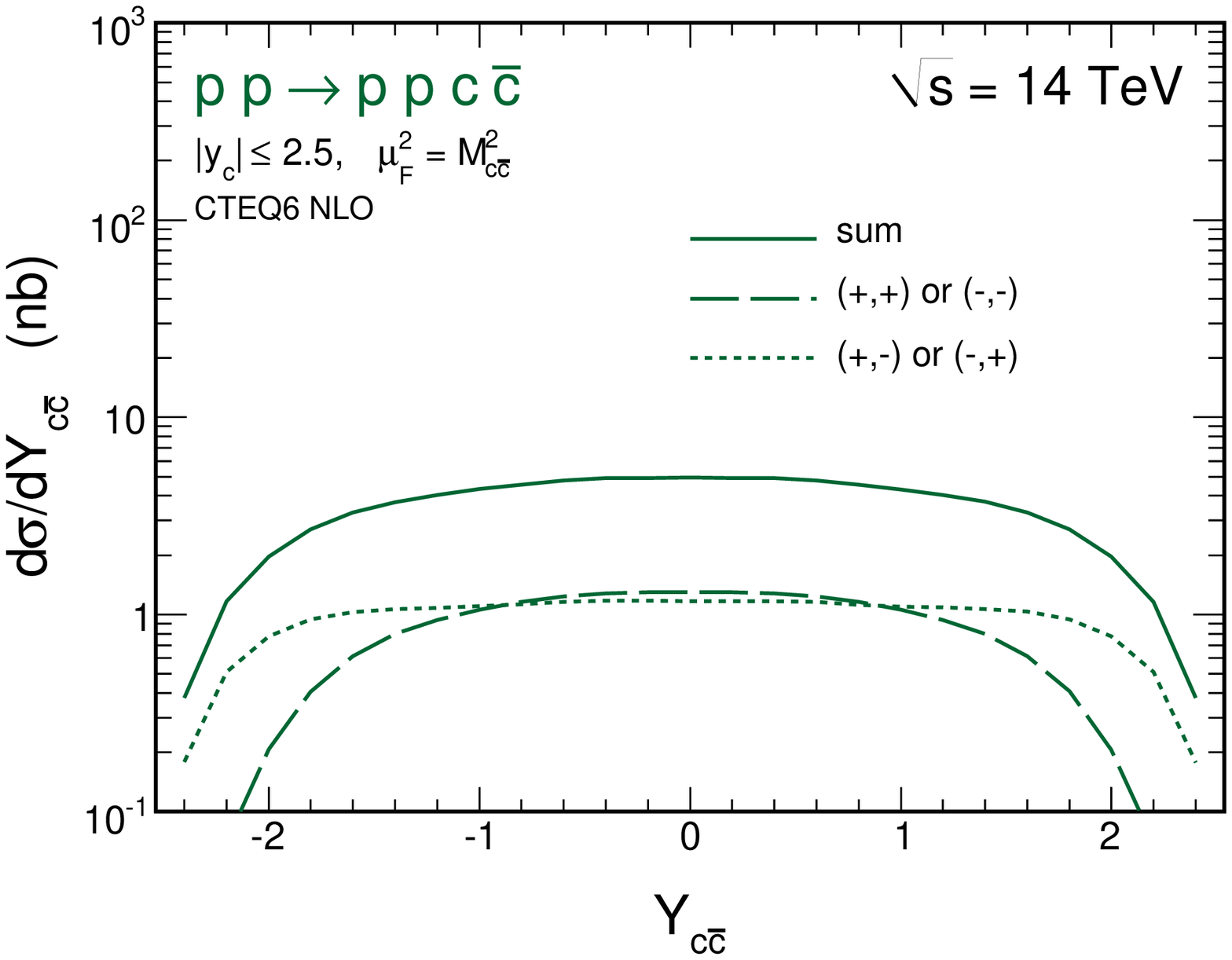}}
\end{minipage}
\hspace{0.5cm}
\begin{minipage}{0.47\textwidth}
 \centerline{\includegraphics[width=1.0\textwidth]{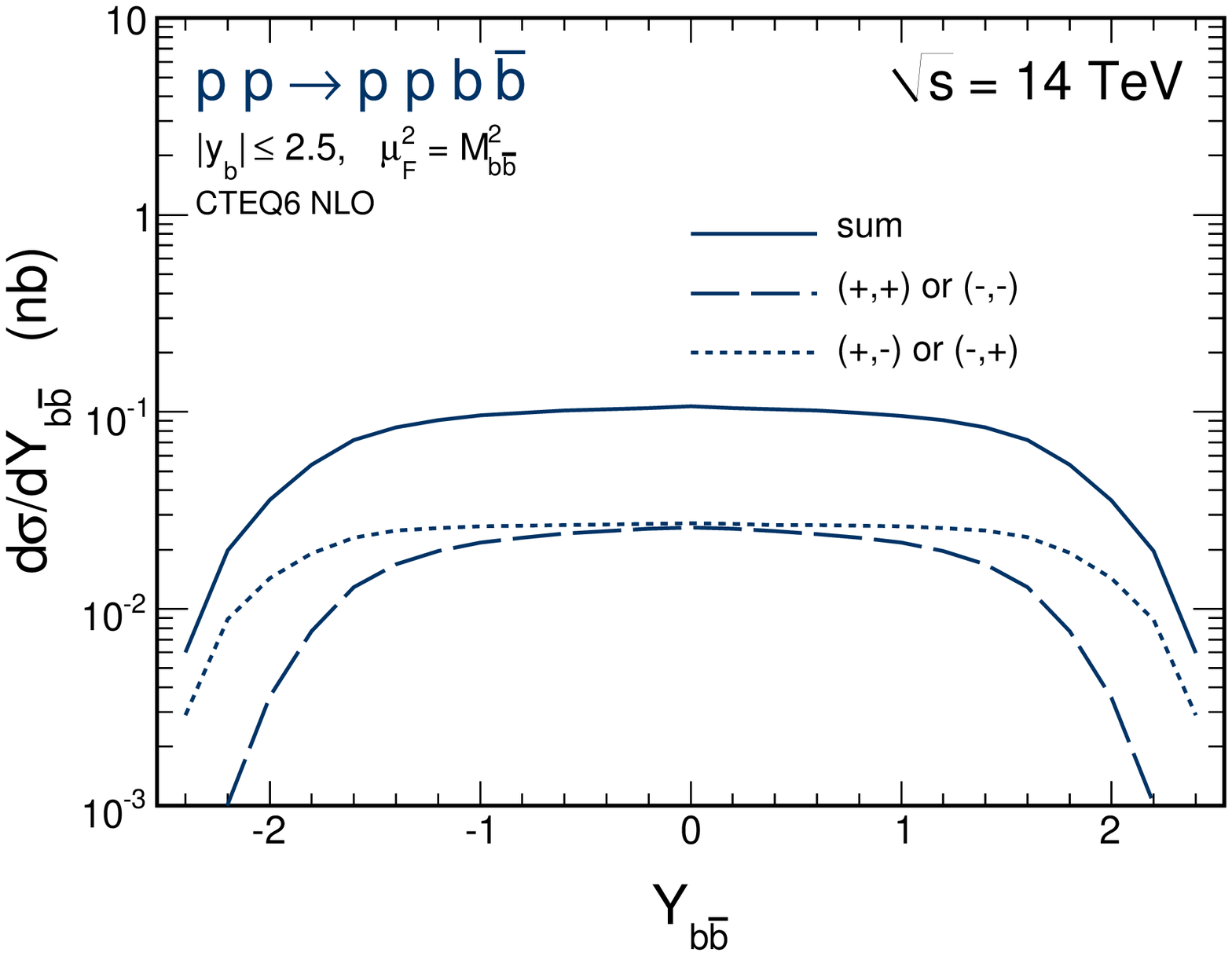}}
\end{minipage}
   \caption{
 \small Differential distributions in rapidity
 of the $c{\bar c}$ pair $Y_{c{\bar c}}$ (left panel) and $b{\bar b}$ pair $Y_{b{\bar b}}$ (right
 panel). Kinematical constraints are the same as in
Fig.~\ref{fig:dM-pdfs}.}
 \label{fig:dsig_dysum}
\end{figure}

In Fig.~\ref{fig:dsig_dysum} we show distribution in the rapidity
of the $q \bar q$ pair. Here the distribution of the opposite
quark helicity is much flatter than the distribution of the same quark
helicity. This distribution may, however, be slightly biased by
the limitation of the individual rapidities of the quark and antiquark.

\begin{figure}[!h]
\begin{minipage}{0.47\textwidth}
 \centerline{\includegraphics[width=1.0\textwidth]{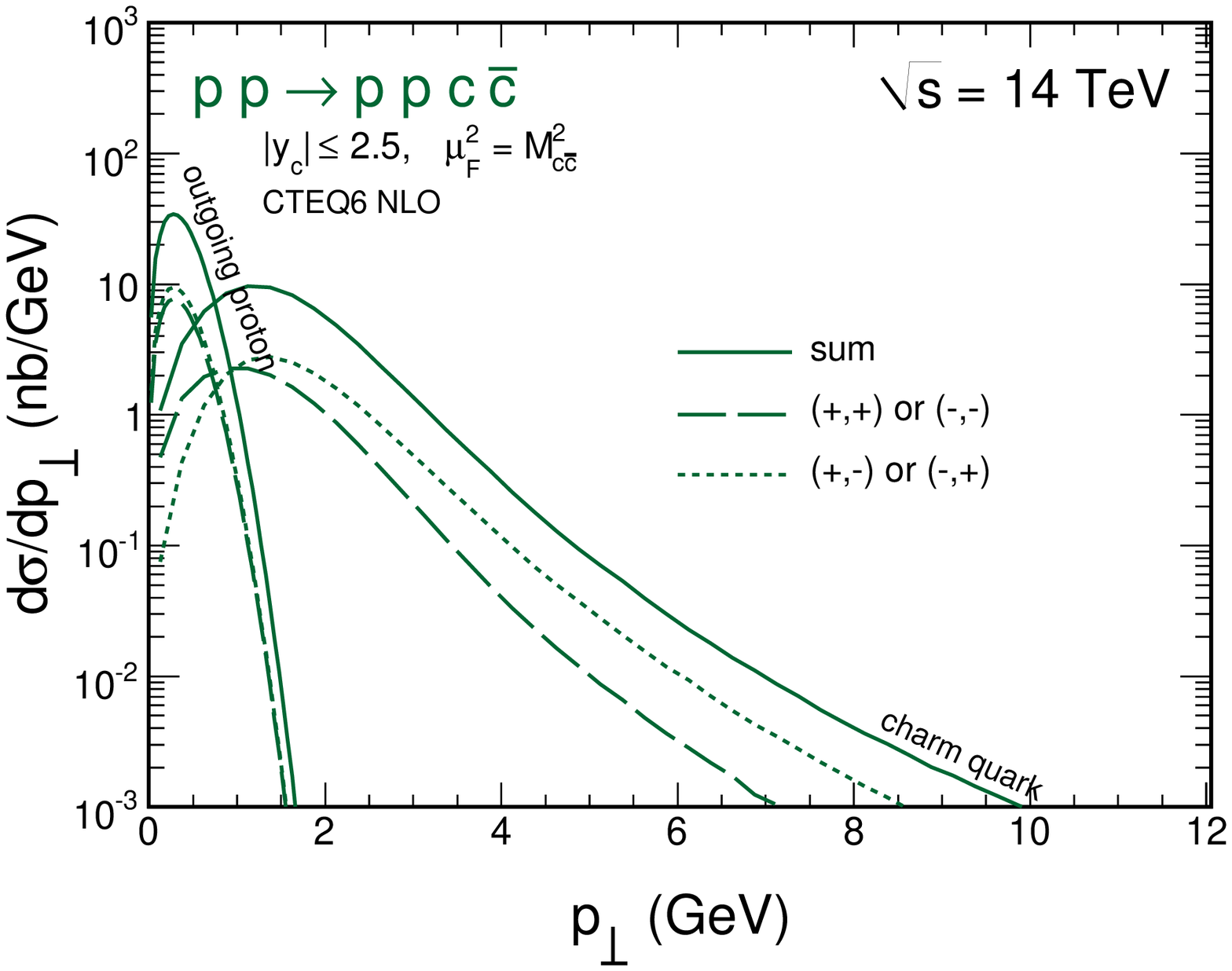}}
\end{minipage}
\hspace{0.5cm}
\begin{minipage}{0.47\textwidth}
 \centerline{\includegraphics[width=1.0\textwidth]{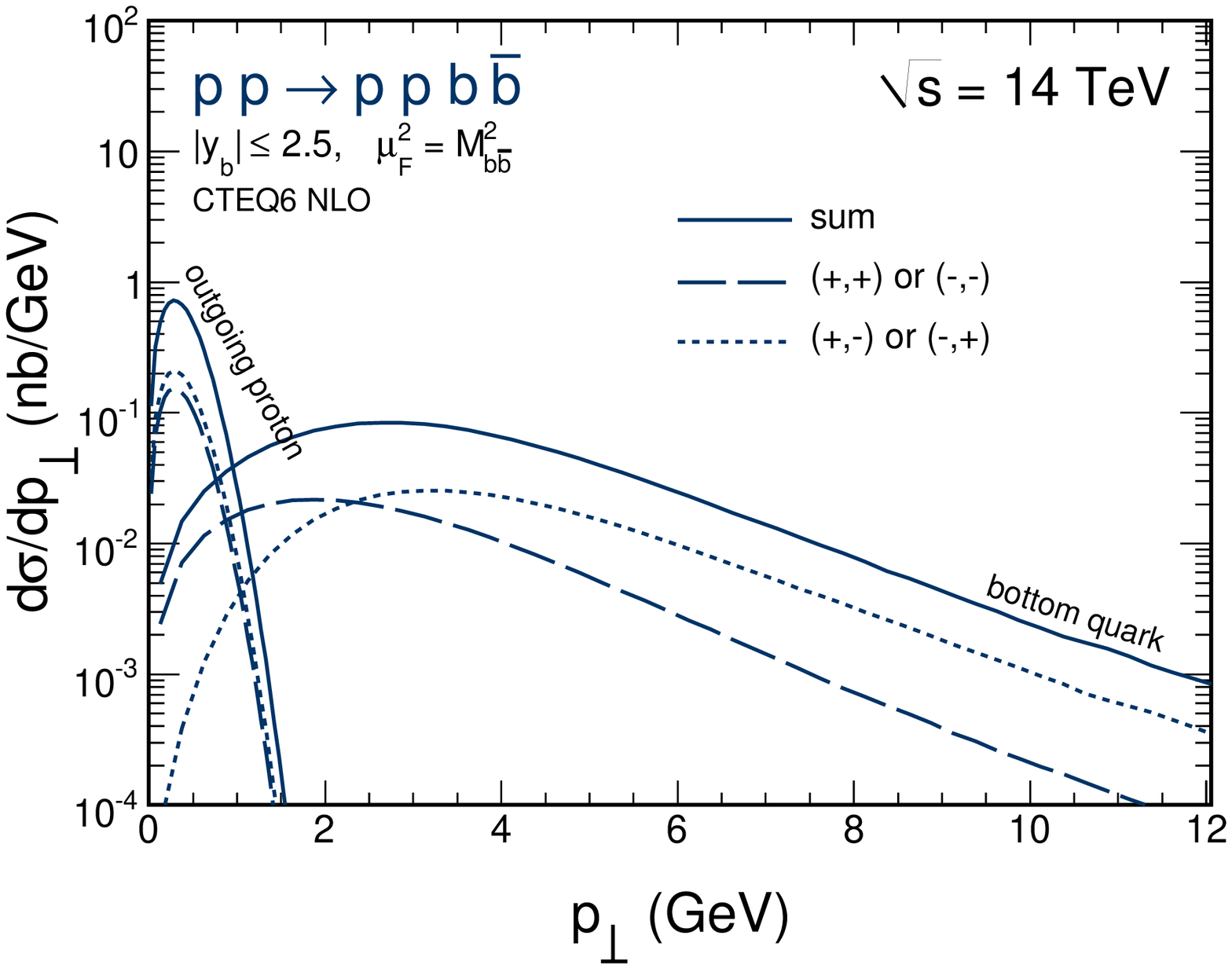}}
\end{minipage}
   \caption{
 \small  Differential distributions in transverse momenta of quark (antiquark)
 (solid line) and outgoing protons (dashed line) of EDD $c{\bar c}$ (left panel)
 and $b{\bar b}$ (right panel) cross section.
 Kinematical constraints are the same as in Fig.~\ref{fig:dM-pdfs}.}
 \label{fig:dsig_dpt}
\end{figure}

Let us come now to transverse momentum distributions. In
Fig.~\ref{fig:dsig_dpt} we show distribution in quark
(antiquark) transverse momenta. These distributions are extended to
large transverse momenta with the peak at about 2 GeV for $c/\bar{c}$ and 3 GeV for $b/\bar{b}$. This is fully
perturbative effect and is encoded in the $g^* g^* \to q {\bar q}$
matrix elements discussed in Sections III and IV. One can clearly
see the dominance of the opposite sign helicities contribution at large transverse
momenta. For comparison, we show also distribution in proton
transverse momenta (dashed line). In contrast to quarks (antiquarks)
$p_{\perp}$-distributions, they are order of magnitude narrower and
concentrated below 1 GeV with maximum at about 0.3 GeV. These
distributions are controlled by a nonperturbative proton form factor
and are thus sensitive to internal structure of the proton.

\begin{figure}[!h]
\begin{minipage}{0.47\textwidth}
 \centerline{\includegraphics[width=1.0\textwidth]{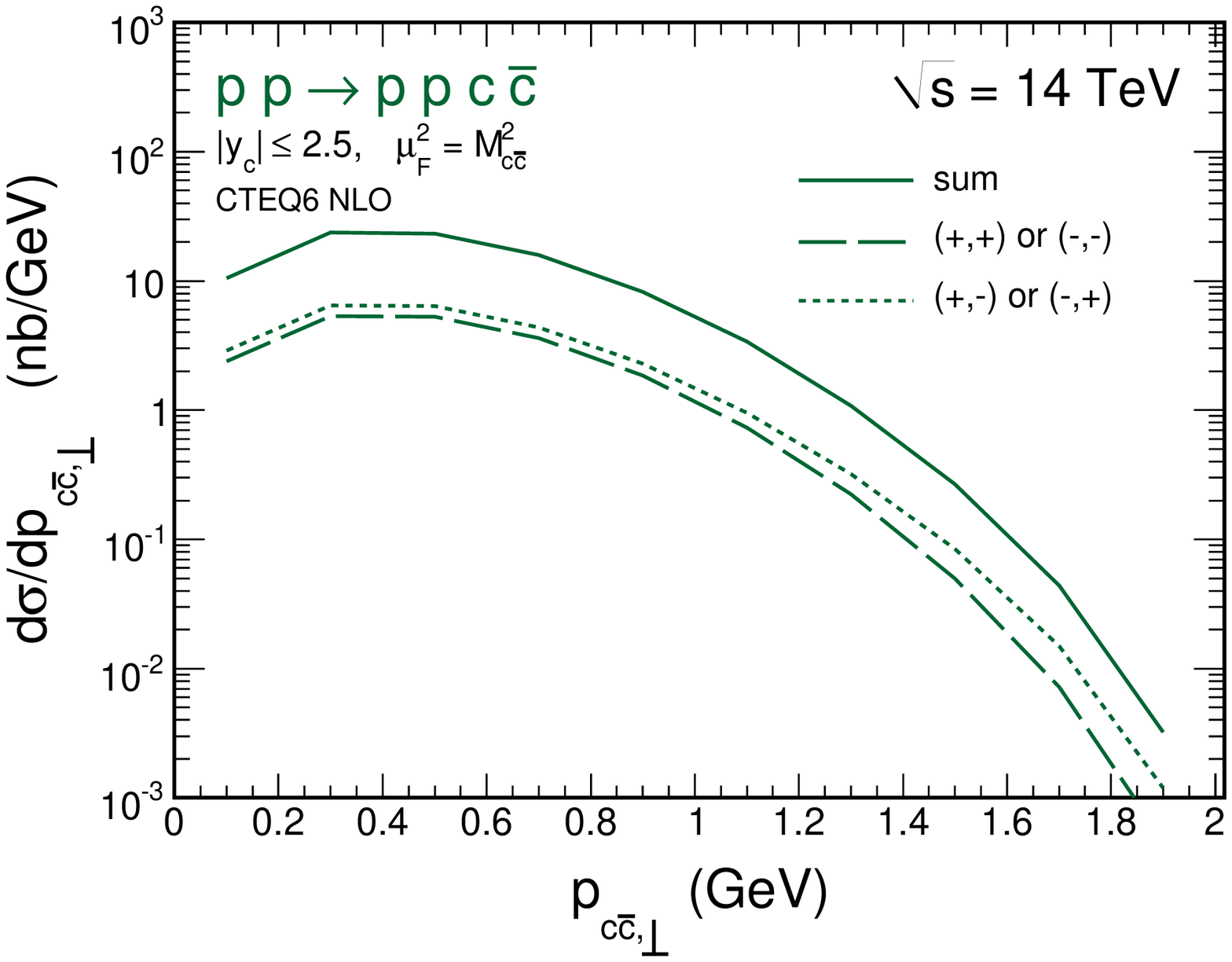}}
\end{minipage}
\hspace{0.5cm}
\begin{minipage}{0.47\textwidth}
 \centerline{\includegraphics[width=1.0\textwidth]{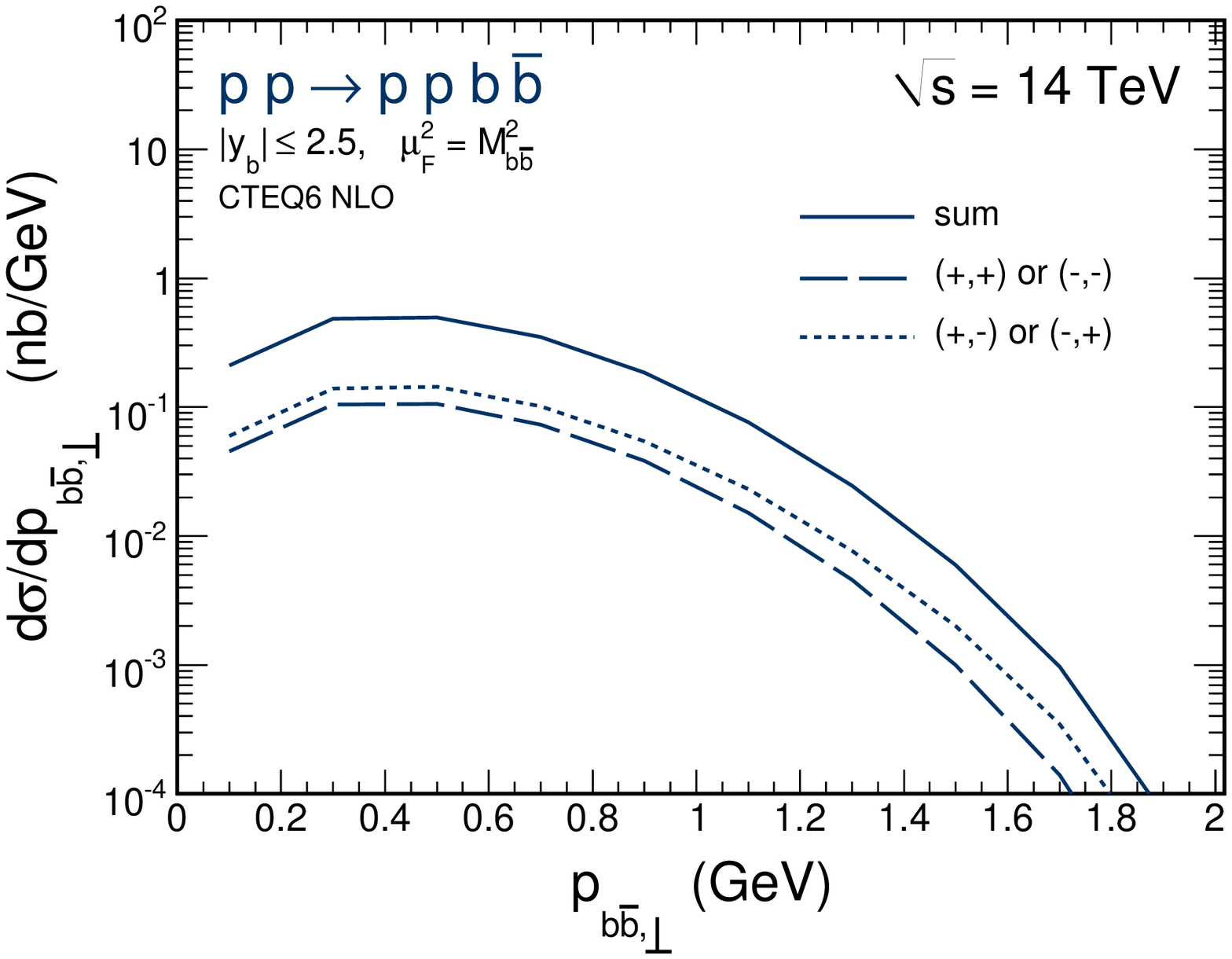}}
\end{minipage}
   \caption{
 \small Differential distributions in transverse momenta of
 quark-antiquark pair $|{\bf p}_{q{\bar q}\perp}|$.
 Kinematical constraints are the same as in Fig.~\ref{fig:dM-pdfs}.}
 \label{fig:dsig_dptsum}
\end{figure}

The distribution in the total transverse momentum of the $q{\bar q}$
pair $|{\bf p}_{q{\bar q}\perp}|$ (by definition ${\bf p}_{q{\bar
q}\perp} = {\bf p}_{q\perp} + {\bf p}_{{\bar q}\perp}$) is shown in
Fig.~\ref{fig:dsig_dptsum}. It is much narrower than that for the
individual quark (antiquark). The maximum of the cross section is at
about 0.5 GeV. Similarly as the distributions in the proton transverse momentum
this distribution is fully nonperturbative and
related to the slope of the nucleon form factors.

Finally, let us turn to azimuthal angle correlations. We will consider
correlations between outgoing quark jets, as well as between
outgoing protons.

\begin{figure}[!h]
\begin{minipage}{0.47\textwidth}
 \centerline{\includegraphics[width=1.0\textwidth]{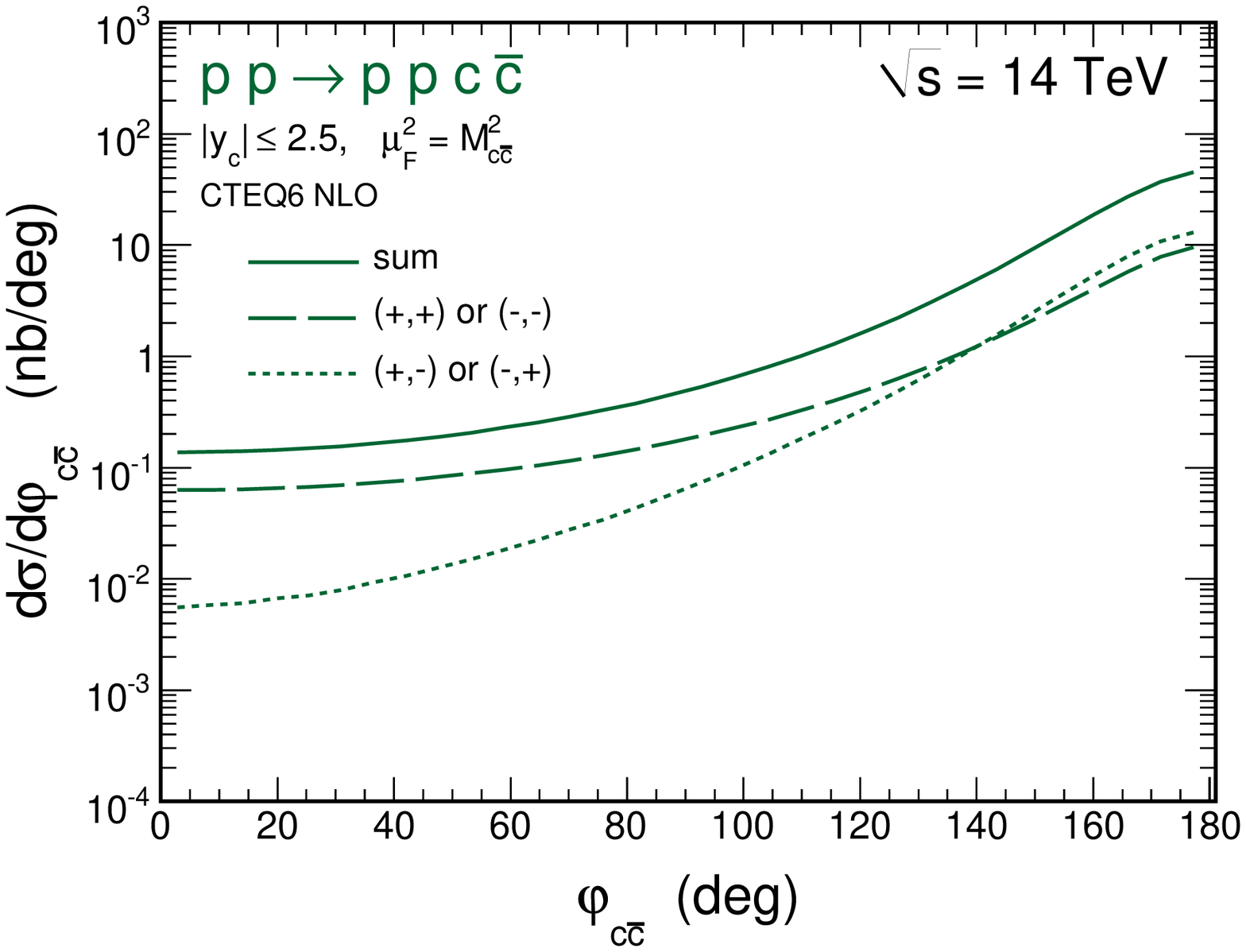}}
\end{minipage}
\hspace{0.5cm}
\begin{minipage}{0.47\textwidth}
 \centerline{\includegraphics[width=1.0\textwidth]{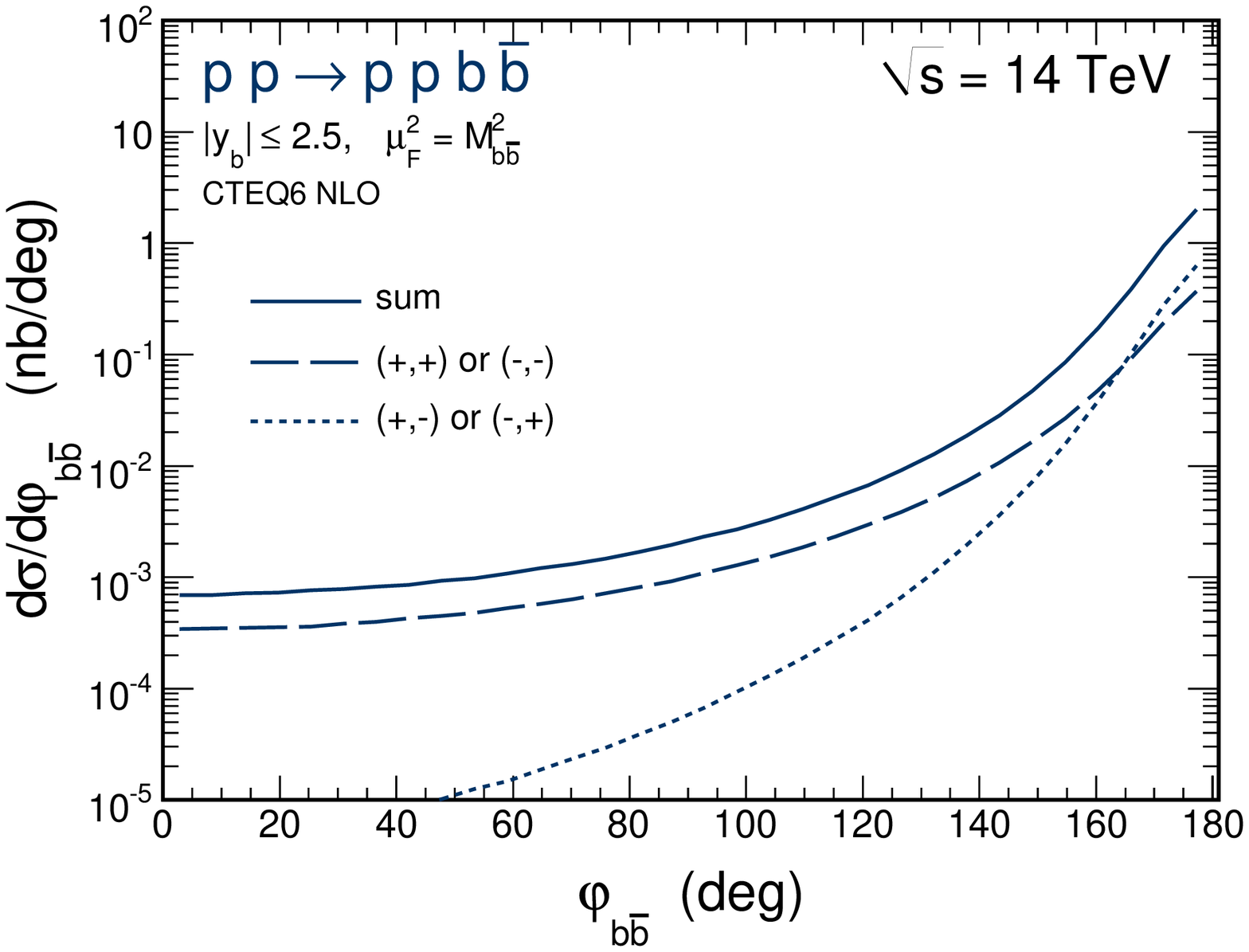}}
\end{minipage}
   \caption{
 \small Differential distributions in azimuthal angle between $c,\,{\bar c}$
 jets (left panel) and $b,\,{\bar b}$ jets (right panel).
 Kinematical constraints are the same as in Fig.~\ref{fig:dM-pdfs}.}
 \label{fig:dsig_dphiqq}
\end{figure}
\begin{figure}[!h]
\begin{minipage}{0.47\textwidth}
 \centerline{\includegraphics[width=1.0\textwidth]{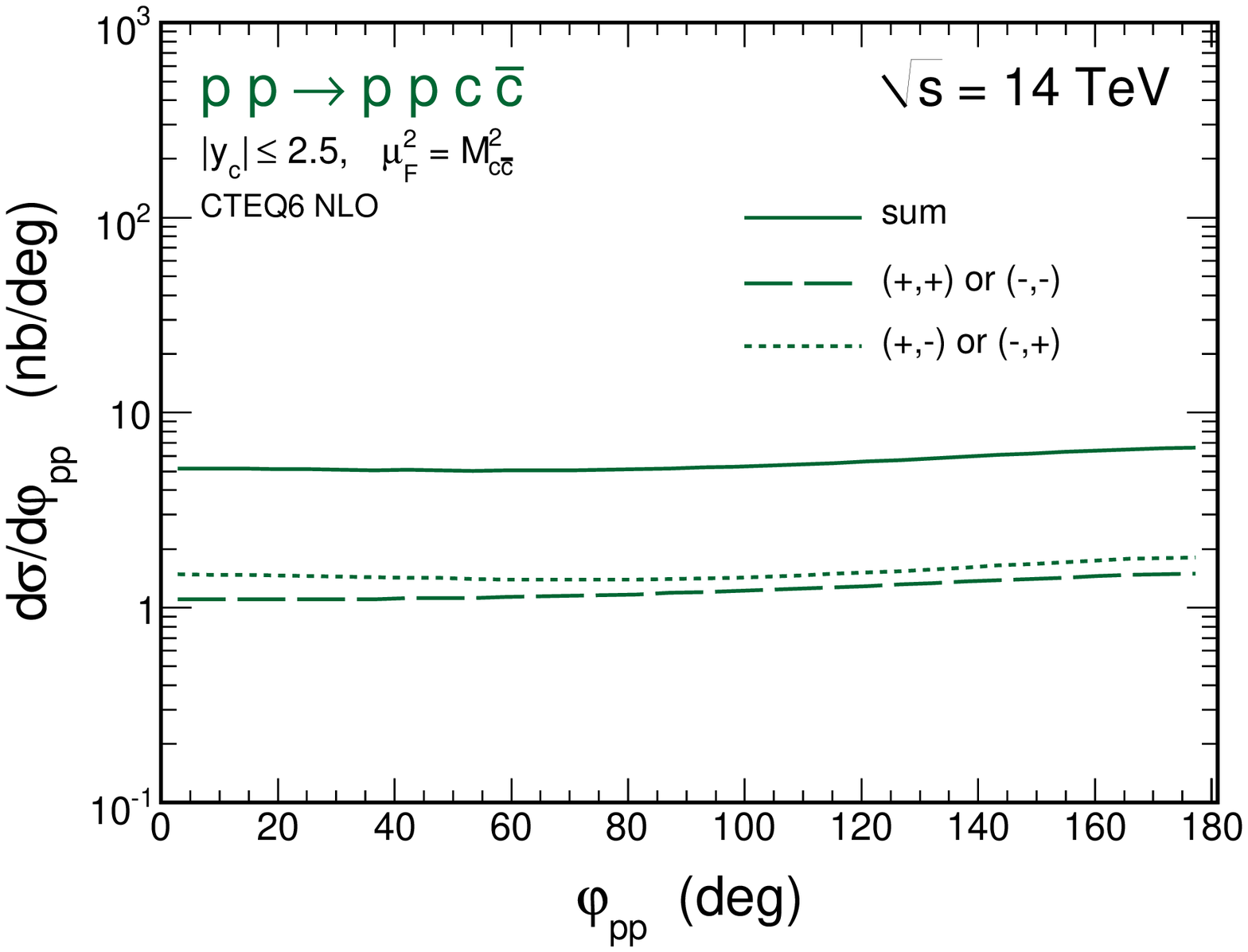}}
\end{minipage}
\hspace{0.5cm}
\begin{minipage}{0.47\textwidth}
 \centerline{\includegraphics[width=1.0\textwidth]{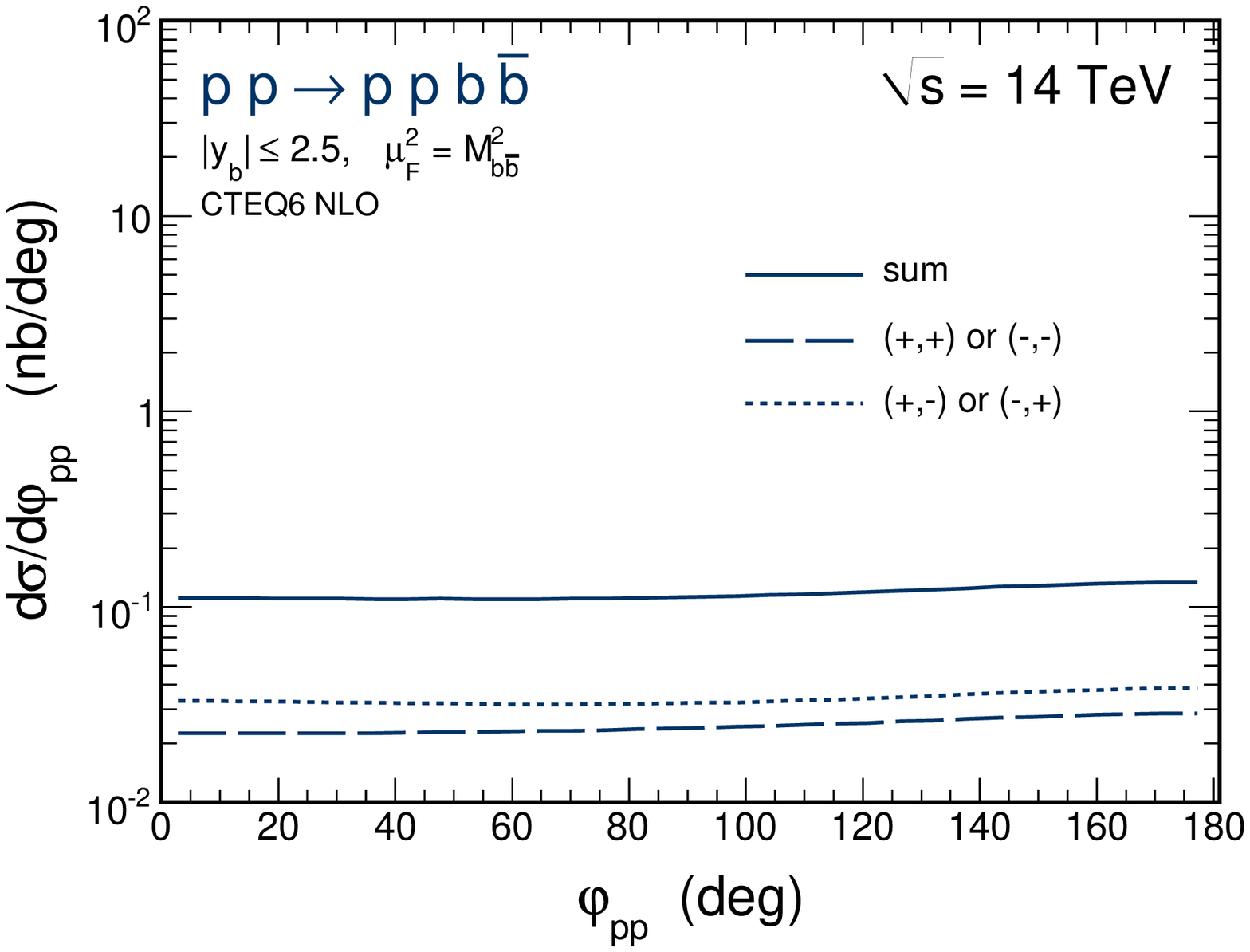}}
\end{minipage}
   \caption{
 \small Differential distributions in angle between outgoing protons corresponding to $c{\bar c}$
 EDD (left panel) and $b{\bar b}$ EDD (right panel).
 Kinematical constraints are the same as in Fig.~\ref{fig:dM-pdfs}.}
 \label{fig:dsig_dphipp}
\end{figure}

In Fig.~\ref{fig:dsig_dphiqq} we show correlations between outgoing
jets without extra cuts on jets transverse momenta. Even without
such cuts the quark and antiquark are strongly correlated with a
preference for the back-to-back configuration. The deviation from the
back-to-back configuration is caused by the transverse momenta of
gluons in the ladder. If there were no transverse momenta of initial gluons,
final jets would be back-to-back which follows from the kinematics of the process.
There is a stronger helicity correlation
for the opposite quark helicities than that for the same quark
helicities. The correlation would even increase when imposing extra
cuts on quark (antiquark) transverse momenta.

In Fig.~\ref{fig:dsig_dphipp} we show correlations between outgoing
protons. In contrast to quarks (antiquarks) protons are almost
decorrelated. This can be understood post factum taken
complicated gluonic ladders spanned between protons and quarks.
The soft rescattering effects could further modify the distribution
(see e.g. \cite{KMR00,KRP05}).

\begin{figure}[!h]
\begin{minipage}{0.47\textwidth}
 \centerline{\includegraphics[width=1.0\textwidth]{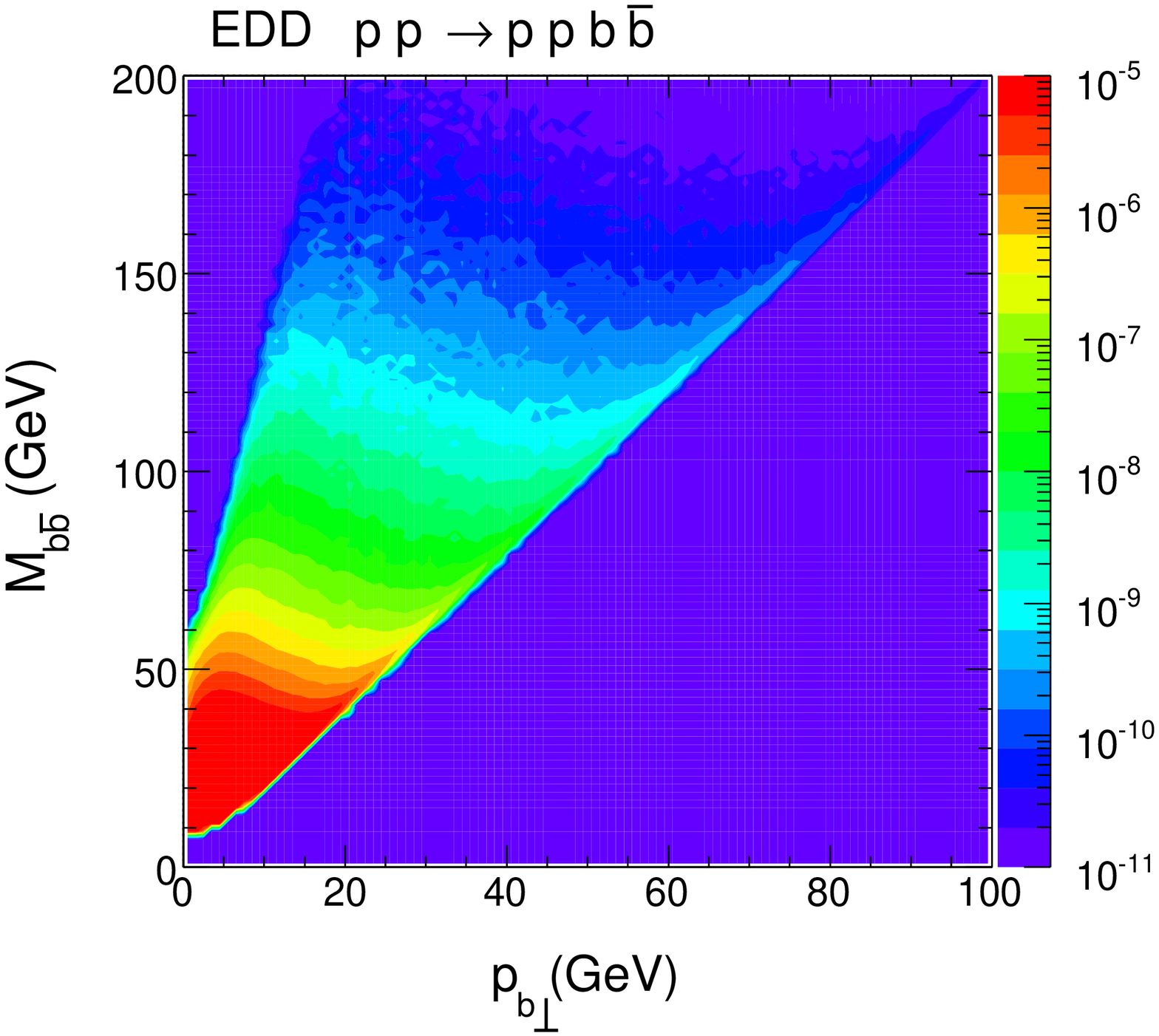}}
\end{minipage}
\begin{minipage}{0.47\textwidth}
 \centerline{\includegraphics[width=1.0\textwidth]{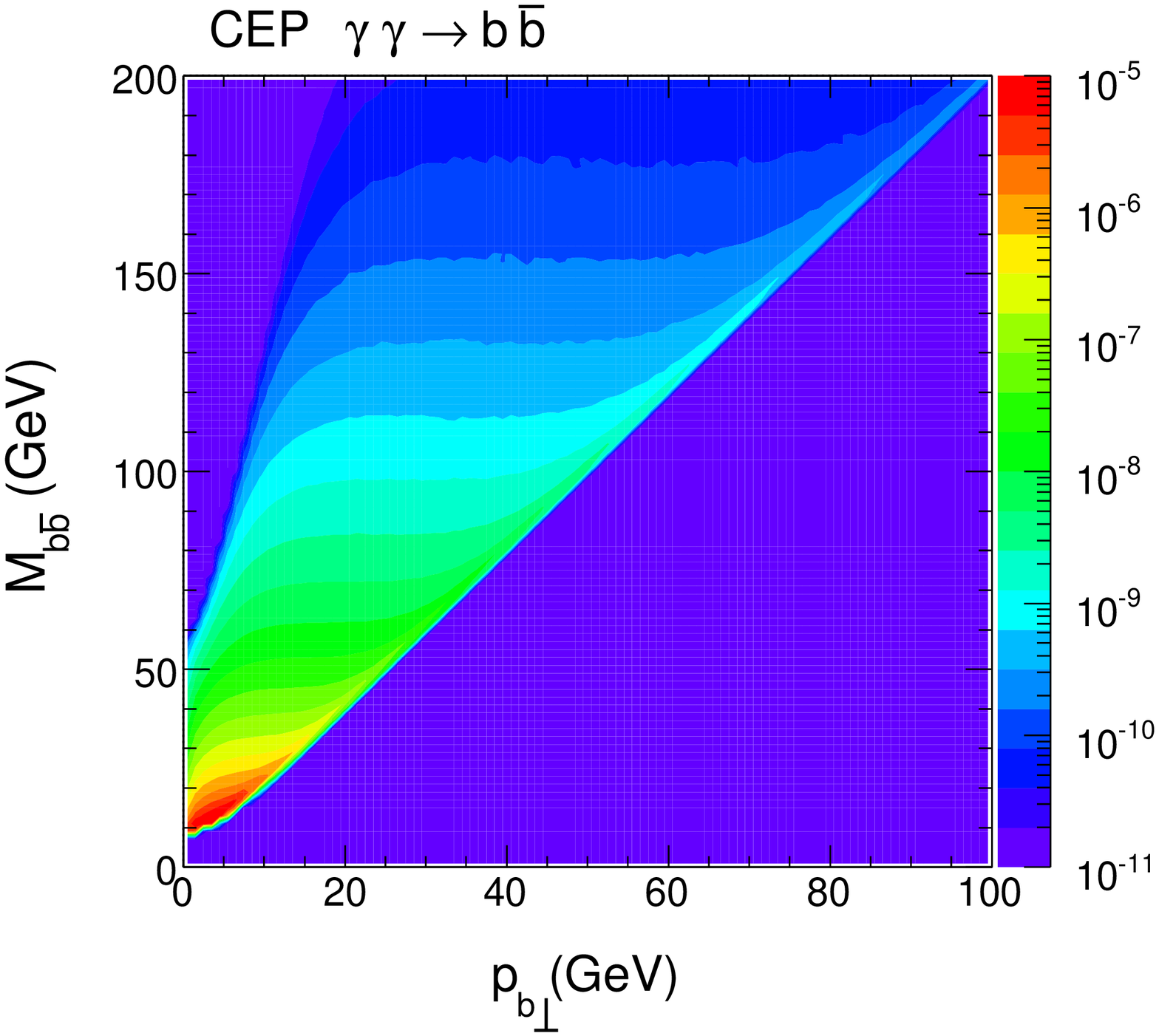}}
\end{minipage}
   \caption{
 \small Two-dimensional distribution in invariant $b{\bar b}$ mass
 $M_{b{\bar b}}$ and (anti)quark transverse momentum $p_{b\perp}$.
 Kinematical constraints are the same as in Fig.~\ref{fig:dM-pdfs}.}
 \label{fig:M34p3t_2D}
\end{figure}

Finally, we would like to show a two-dimensional distribution which
is very useful when discussing background to the
exclusive Higgs boson production in the $b \bar b$ channel. In
Fig.~\ref{fig:M34p3t_2D} we show the distribution in $M_{b \bar b}$
and transverse momentum of the quark ($p_{b\perp}$) for EDD (left panel)
and QED (right panel) mechanisms. One can clearly see that a fixed
mass (e.g. mass of the Higgs) can be obtained both for high and low
transverse momenta of the quark jets. The latter case can be
realized when the quark rapidities are large. For EDD contribution
one can remove such cases by imposing cuts on jet transverse
momenta. One could equivalently limit quark (antiquark) rapidities.
We will return to these correlations when discussing the Higgs
background.

\subsection{Central exclusive production of Higgs boson}

Due to a very large hard scale of the process $\sim M_H$, the
influence of typically small gluon virtualities in the amplitude of
hard subprocess amplitude (\ref{proj-excl}), as well as the role of
form factor $G_2$, in the exclusive diffractive Higgs production
turned out to be quite small, in analogy to the inclusive case
\cite{incl-Higgs}.

The integrated cross section of diffractive Higgs
production at the LHC energy $\sqrt{s}$ = 14 TeV, taking into account the
``effective'' gap survival factor $\langle S^2\rangle\simeq0.03$
\cite{KMR_Higgs}, calculated for typical Higgs mass $M_H=120$ GeV is
$\sigma_{tot}\lesssim 1$ fb. Our result is smaller than that found by the Durham group.
As discussed above this is mostly due to different choice
of the scale of the Sudakov form factor.
The result of Cudell, Hernandez, Ivanov and Dechambre \cite{Cudell:2008gv}
is closer to our result but still slightly bigger.
This is probably due to different unintegrated gluon distribution.
In particular, the Ivanov-Nikolaev UGDF used in their analysis
includes also a nonperturbative piece fitted to the data.

\begin{figure}[!h]
\begin{minipage}{0.45\textwidth}
 \centerline{\includegraphics[width=1.0\textwidth]{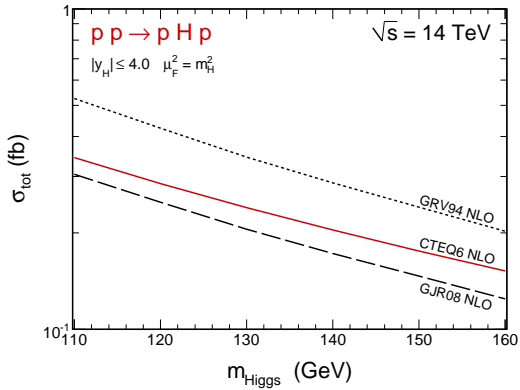}}
\end{minipage}
   \caption{
\small Total cross section for exclusive Higgs production for
different gluon PDFs from the literature. The calculation was done
including off-shell effects.
}
\label{fig:total_mH}
\end{figure}

In Fig.~\ref{fig:total_mH} we show the total cross section for
exclusive production of Higgs boson as a function of the Higgs mass
for different gluon distributions for $\sqrt{s}$ = 14 TeV. The
difference between different gluon PDFs comes mainly from a
different lower cut-off parameter for gluon transverse momenta in
different gluon distributions. This is necessary and is dictated by
the construction of different UGDFs. In particular, different groups
choose different initial scale for QCD evolution and going below it
often leads to unphysical solutions (negative glue for instance).
This forces one to put lower cut-off at the value of the initial
scale. The cross section for exclusive Higgs production obtained
here is rather small\footnote{Similarly small cross sections have
been obtained very recently \cite{Cudell:2010} when this paper was
already finished.}.

We have made the calculation of the cross section in
the limit of real gluons in the hard part (\ref{Vhiggsre})
($\sigma_H^{\mathrm{on}}$), as well as with an account of gluon
virtualities (\ref{proj-excl}) ($\sigma_H^{\mathrm{off}}$).
Contribution of non-zeroth $q_1^2,\,q_2^2$ in form factors $G_{1,2}$
turns out to be negligibly small; difference between
$\sigma_H^{\mathrm{on}}$ and $\sigma_H^{\mathrm{off}}$ is formed
mainly by the second form factor $G_2$, and gives about 6 $\%$, so
it is much smaller than other theoretical uncertainties of the approach.
The overall uncertainty of $0^+$ Higgs CEP cross section was estimated in
Ref.~\cite{KMR_Higgs} to be up to a factor of 2.5.

\begin{figure}[!h]
\begin{minipage}{0.45\textwidth}
\includegraphics[width=1.2\textwidth]{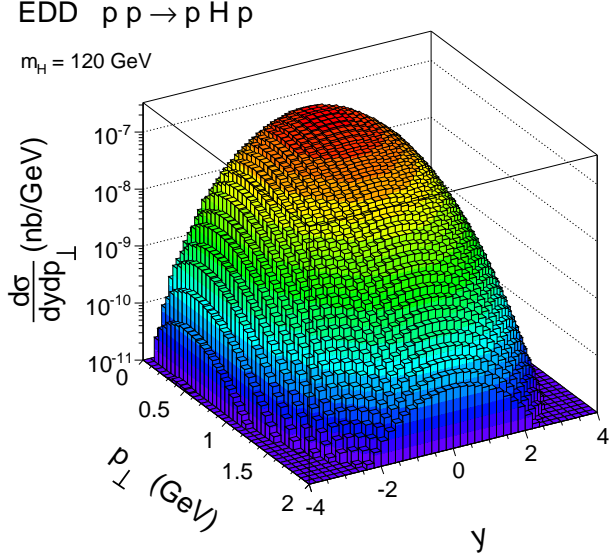}
\end{minipage}
\caption{
A two-dimensional distribution in the Higgs rapidity ($y$) and
Higgs transverse momentum ($p_\perp$) for CTEQ6 gluon PDF.
}
\label{fig:ypt_higgs}
\end{figure}

\begin{figure}[!h]
\begin{minipage}{0.47\textwidth}
 \centerline{\includegraphics[width=1.0\textwidth]{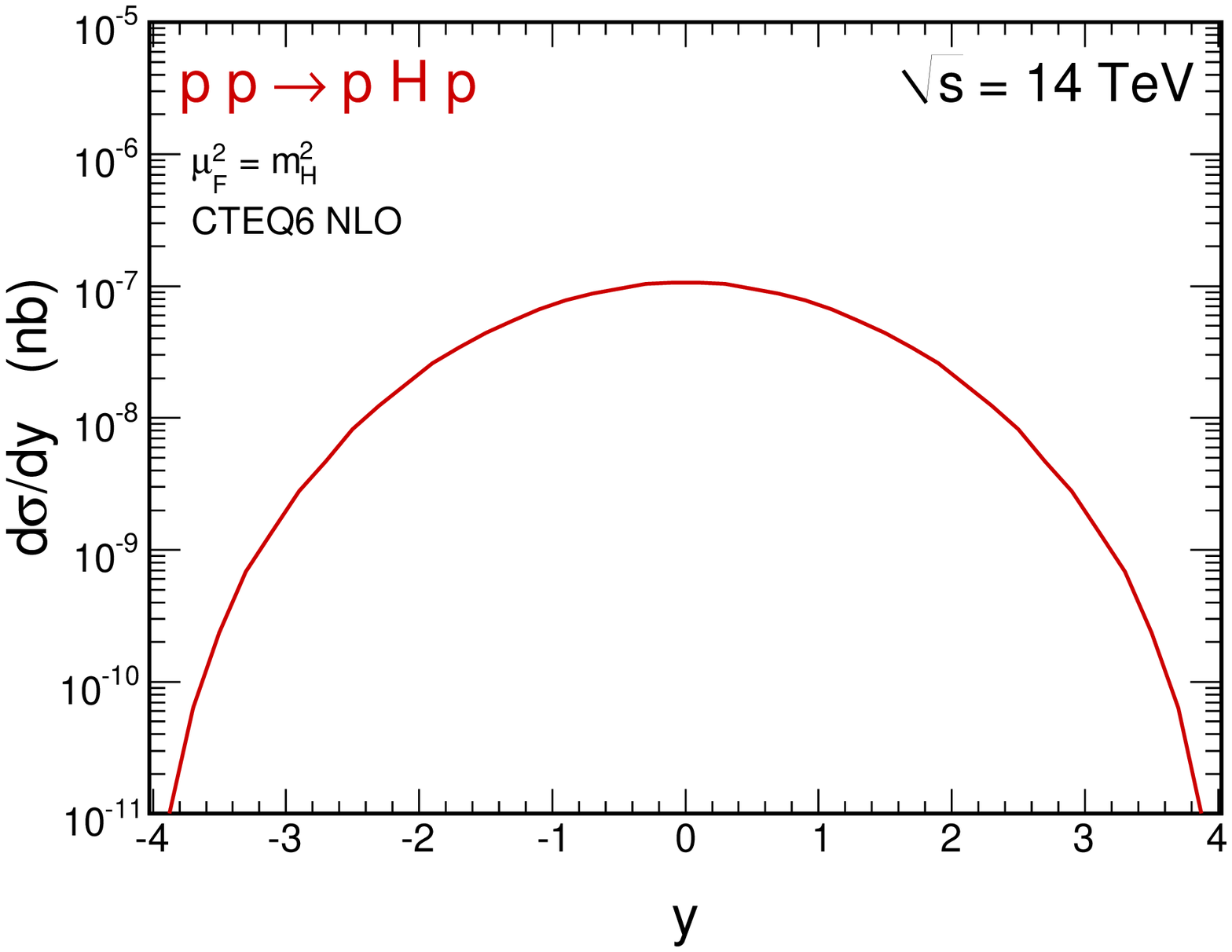}}
\end{minipage}
\hspace{0.3cm}
\begin{minipage}{0.47\textwidth}
 \centerline{\includegraphics[width=1.0\textwidth]{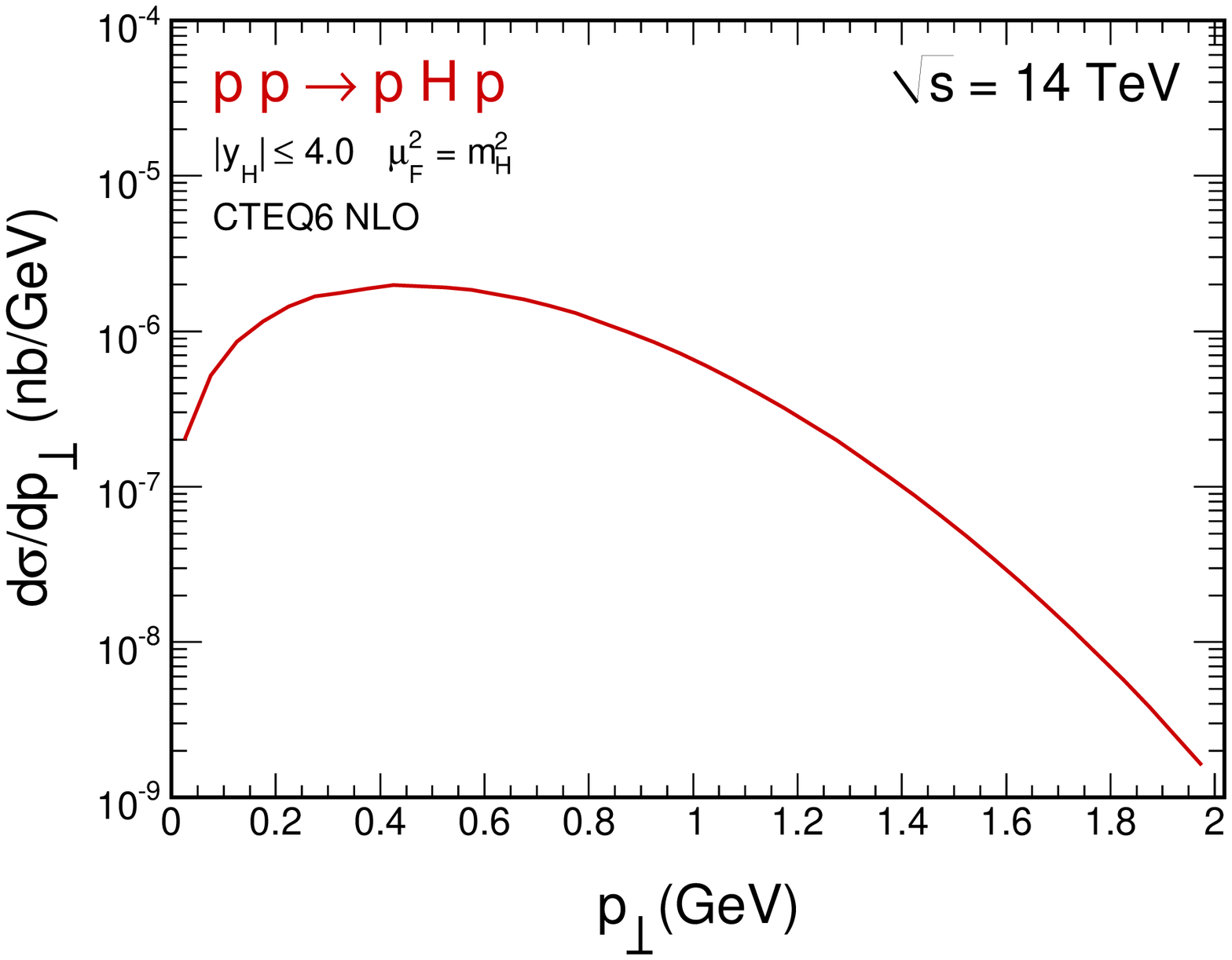}}
\end{minipage}
\caption{
Rapidity (left) and transverse momentum (right) distributions
of the Higgs boson. CTEQ6 PDF was used in this calculation.
}

\label{fig:yorpt}
\end{figure}

In Fig.~\ref{fig:ypt_higgs} we show a two-dimensional distribution
of the Higgs in its rapidity and transverse momentum.
The Higgs production is concentrated around rapidity
$y =$ 0 and the cross section quickly drops with Higgs transverse
momentum. In Fig.~\ref{fig:yorpt} we show respective projections on
rapidity (left panel) and transverse momentum (right panel).
The maximum of the transverse momentum dependence occurs at about 0.4 GeV.
The distribution reflects a convolution of the nucleon form factors, i.e.
is of purely nonperturbative nature.

\begin{figure}
\begin{minipage}{0.47\textwidth}
 \centerline{\includegraphics[width=1.0\textwidth]{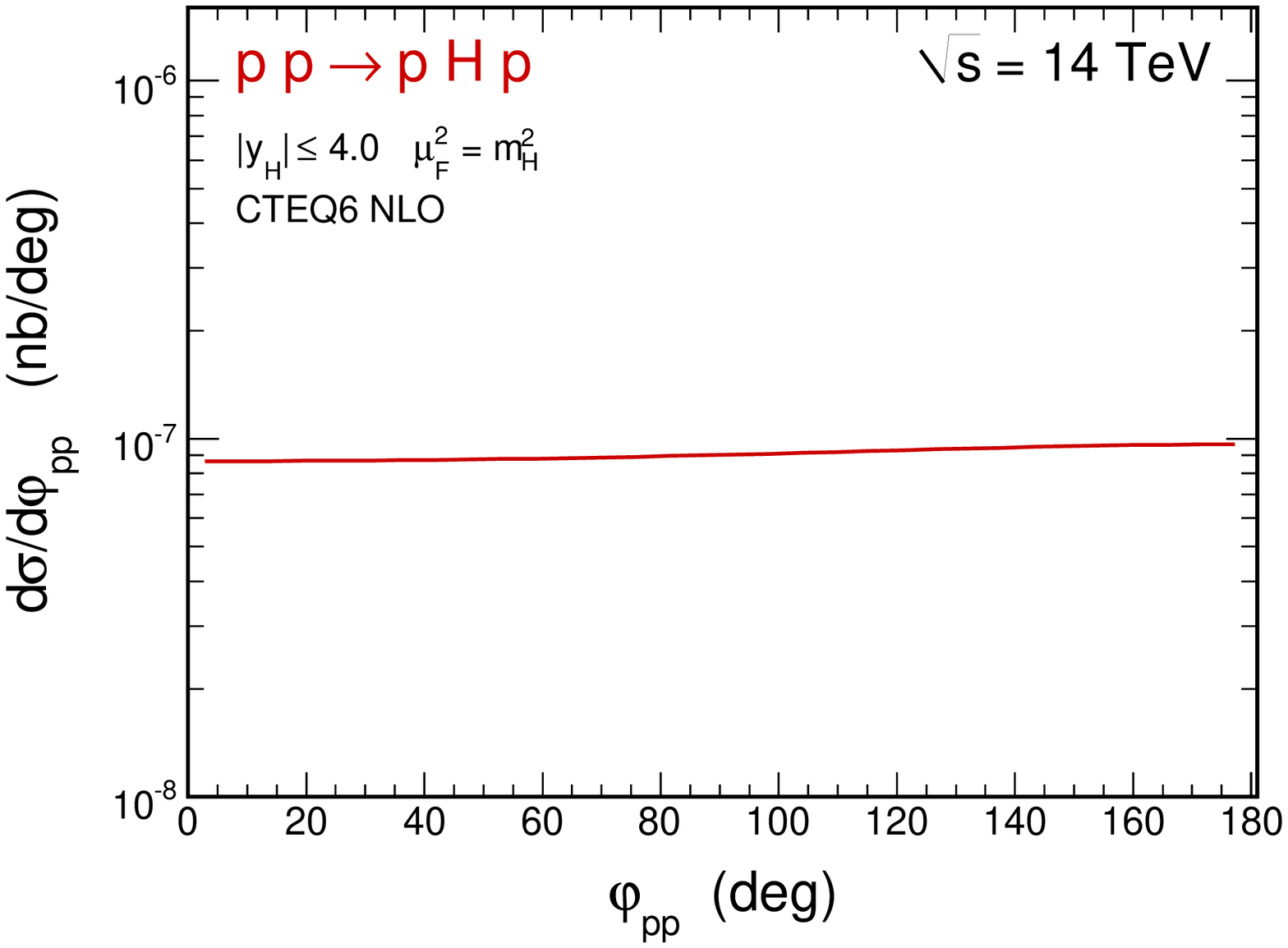}}
\end{minipage}

\caption{
Differential distribution in angle between outgoing protons for
central exclusive Higgs boson production. CTEQ6 PDF was used in this calculation.}
\label{fig:higgs_azimuthal}
\end{figure}

Finally, we focus on angular correlations (see
Fig.~\ref{fig:higgs_azimuthal}). In the figure we show distribution
in azimuthal angle between outgoing protons. As for the exclusive
production of heavy quarks there is a very small correlation between
outgoing protons.

Note that the distribution in relative azimuthal angle between protons $\phi_{pp}$ strongly differs
from the distributions in azimuthal angle $\phi_{{\bf q}_{1}{\bf q}_{2}}$ between
interacting gluons $\sim\cos^2\phi_{{\bf q}_{1}{\bf q}_{2}}$ due to the convolution with
momentum transfer $q_{0\perp}$ of the screening gluon.
Only in the case when Higgs boson production is
governed by e.g. pomeron-pomeron (or $\gamma^*\gamma^*$) fusion, the
angle between pomerons (photons) coincides with the angle between outgoing
protons, and corresponding distribution has $\cos^2\phi_{pp}$-dependence.
This fundamental difference of the two mechanisms was observed also in
other processes \cite{our-eta,PST_chic0,PST_chic12}.

We have not discussed yet the influence of the off-shell effects
in the matrix element on
differential distributions. In the limit of real gluons the
differential distributions are practically unchanged, so the
corresponding off-shell effects are hard to observe taken the other
theoretical uncertainties. Moreover,
unlike for inclusive production case \cite{incl-Higgs}, shapes of
differential distributions of Higgs CEP are not sensitive to
the off-shell effects since they get averaged out effectively when the
off-shell matrix element (\ref{proj-excl}) is integrated over ${\bf
q}_{0\perp}$ in the diffractive amplitude (\ref{ampl}).

\subsection{Irreducible $b{\bar b}$ background for exclusive Higgs production}

\begin{figure}[!h]
\begin{minipage}{0.47\textwidth}
 \centerline{\includegraphics[width=1.0\textwidth]{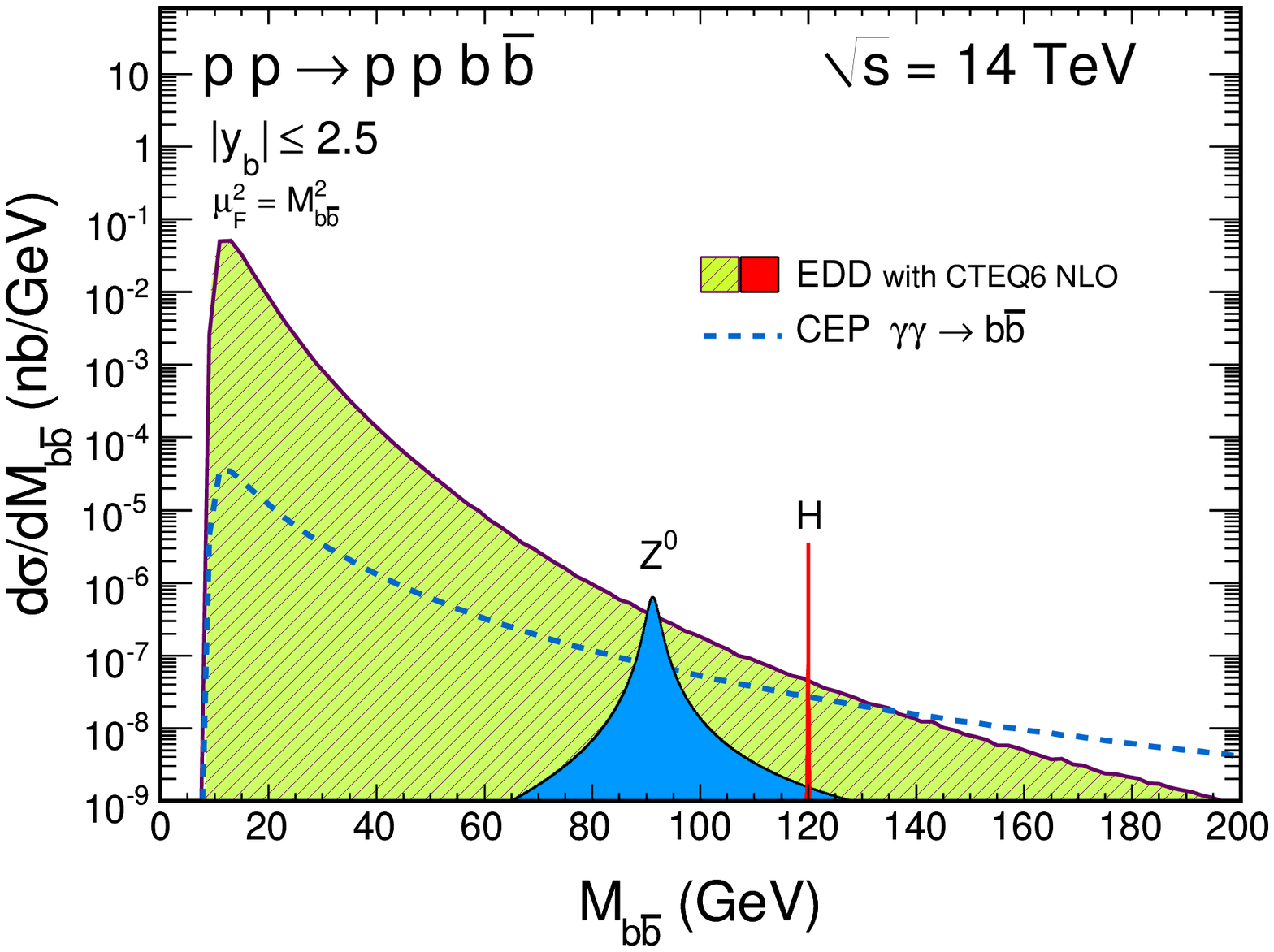}}
\end{minipage}
\hspace{0.5cm}
\begin{minipage}{0.47\textwidth}
 \centerline{\includegraphics[width=1.0\textwidth]{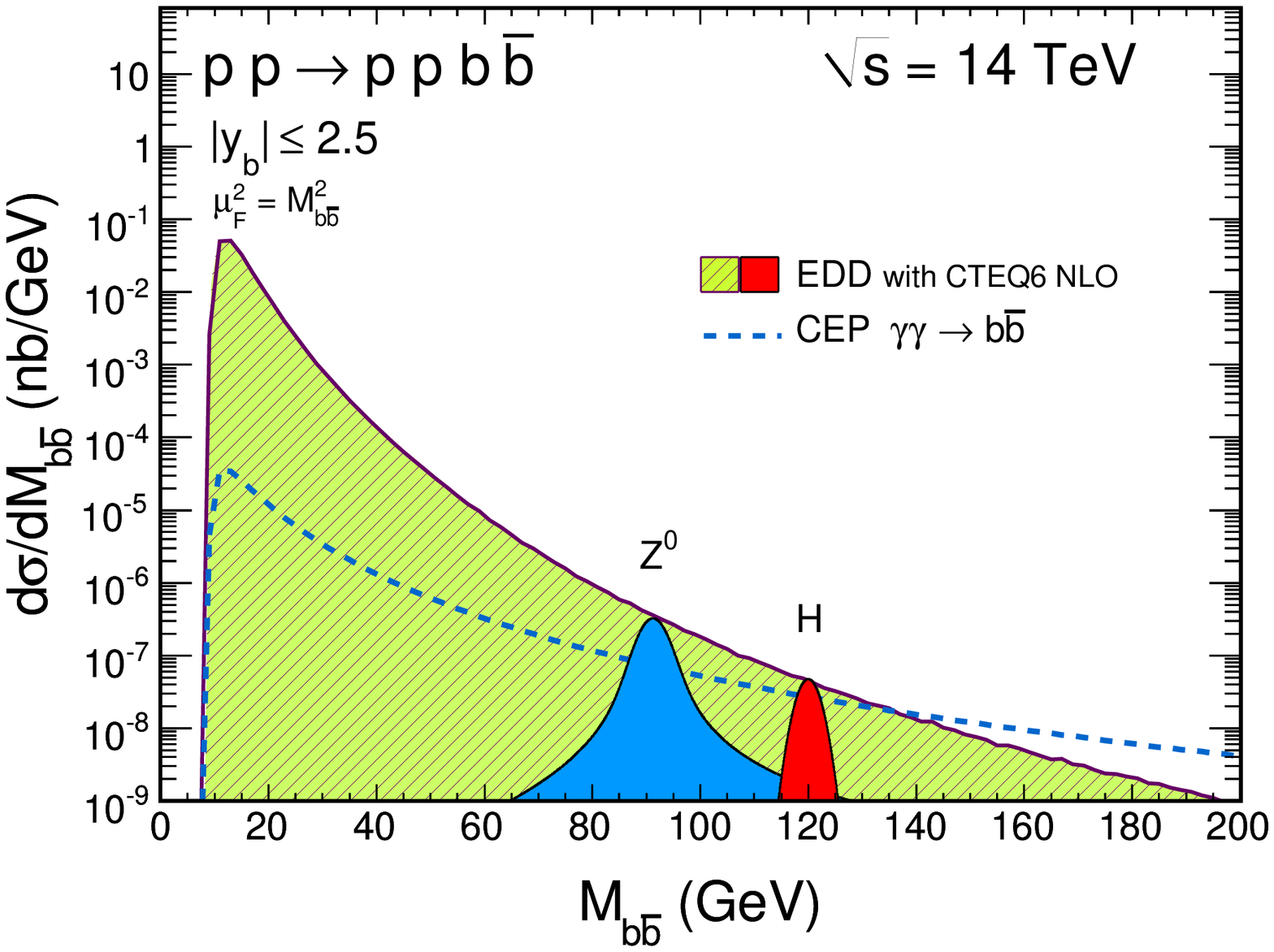}}
\end{minipage}
   \caption{
 \small The $b \bar b$ invariant mass distribution for $\sqrt{s}$ =
14 TeV and for $b$ and $\bar b$ jets from Higgs decay in the
rapidity interval $-2.5 < y_b < 2.5$ corresponding to the ATLAS
detector. The absorption effects for the Higgs boson and the
background were taken into account by multiplying cross section by the gap
survival factor $\langle S^2\rangle$ = 0.03. The left panel shows
purely theoretical predictions, while the right panel includes
experimental effects due to experimental uncertainty in invariant
mass measurement. The left peaks (bumps) correspond to the $Z^0$
contribution and the right ones to the Higgs contribution.}
 \label{fig:sigbkg-ratio}
\end{figure}

Now we turn to the analysis of the $b\bar{b}$ continuum as a background for the
$b\bar{b}$ Higgs signal.
In the left panel of Fig.~\ref{fig:sigbkg-ratio} we show contributions
of several CEP mechanisms to the $b \bar b$ quark invariant mass distribution.
The diffractive $b \bar b$ and Higgs contributions were calculated for
a selected (CTEQ6~\cite{CTEQ}) collinear gluon distribution.
The QED mechanism is also shown by the short-dashed line.
Natural decay width, calculated as in
Ref.~\cite{Passarino}, was assumed in this calculation,
see the sharp peak at $M_{b {\bar b}}$ = 120
GeV (assumed arbitrarily for illustration) which is not excluded at
present by the Higgs searches at LEP \cite{LEP_Higgs_searches} and
Tevatron \cite{Tevatron_Higgs_searches}.

As was already mentioned above, the phase space integrated cross
section for the Higgs production, including absorption effects with
$\langle S^2\rangle=0.03$ is less than 1 fb which is somewhat smaller than that
predicted by the Durham group \cite{KMR_Higgs}. The main reason is different choice
of the scale in the Sudakov form factor $\mu^{2}=M_{H}^{2}$ instead of
$\mu^{2}=M_{H}^{2}/4$ as used by the Durham group. The first choice was advocated
recently by theoretical studies in Ref.~\cite{CF09}. The result shown in
Fig.~\ref{fig:sigbkg-ratio} includes also the branching fraction
BR($H \to b {\bar b}) \approx$ 0.8 and the rapidity restrictions.
The second much broader Breit-Wigner type peak corresponds to the
exclusive production of the $Z^0$ boson with the cross section
calculated as in Ref.~\cite{CSS09}. The exclusive cross section for
$\sqrt{s}$ = 14 TeV is 16.61 fb including absorption (28.71 fb
without absorption effects). The branching fraction BR($Z^0 \to b
{\bar b}) \approx$ 0.15 has been included in addition. In contrast
to the Higgs case the absorption effects for the $Z^0$ production
are much smaller \cite{CSS09}. The sharp peak corresponding to the
Higgs boson clearly sticks above the background. In the above
calculations we have assumed an ideal no-error measurement.

In reality the situation is, however, much worse as both protons and
the $b$ and $\bar b$ jets are measured with a certain
precision which automatically leads to a smearing of experimental
distribution in $M_{b \bar b}$.
Much better resolution can be obtained by measuring missing mass
than from the direct measurement of heavy quark (antiquark) jet momenta.
In the following in spite we will present distribution in $M_{b \bar b}$
(it will mean experimentally distribution in missing mass ($M_{pp}$)).
The two are identical when there are no errors on kinematical variables.
While the smearing is negligible for the background, it leads to
a significant modification of the Breit-Wigner peaks, especially of
the sharp one for the Higgs boson. In the present paper the
experimental effects are included in the simplest way by a
convolution of the theoretical distributions and experimental
resolution function
\begin{equation}
\frac{d\sigma^{exp}}{dM_{q{\bar q}}}(M_{q{\bar q}})=\int d\mu
\frac{d\sigma^{th}}{dM_{q{\bar q}}}(\mu)G(\mu-M_{q{\bar q}}) \;,
\end{equation}
where the experimental resolution function is taken as the Gaussian
function
\begin{equation}
G(x) = \frac{1}{\sqrt{2 \pi} \sigma} \exp\left( \frac{x^2}{2\sigma^2
} \right) \;,
\end{equation}
with $\sigma$ = 2 GeV, which realistically represents the
experimental situation \cite{Pilkington_private,Royon_private} and
is determined mainly by the precision of measuring forward protons.
In the right panel we show the invariant mass distribution when the
invariant mass smearing is included. Now the bump corresponding to
the Higgs boson is below the $b \bar b$ background. With the
experimental resolution assumed above the identification of the
Standard Model Higgs looks rather difficult. The situation for
some scenarios beyond the Standard Model may be better
\cite{MPS_Higgs_beyondSM}.

Below we wish to discuss how to improve the situation by imposing extra cuts.
Before we establish how to impose cuts let us consider first a few
two-dimensional distributions which may help to come to the final solution.

\begin{figure}[!h]
\begin{minipage}{0.328\textwidth}
 \centerline{\includegraphics[width=1.0\textwidth]{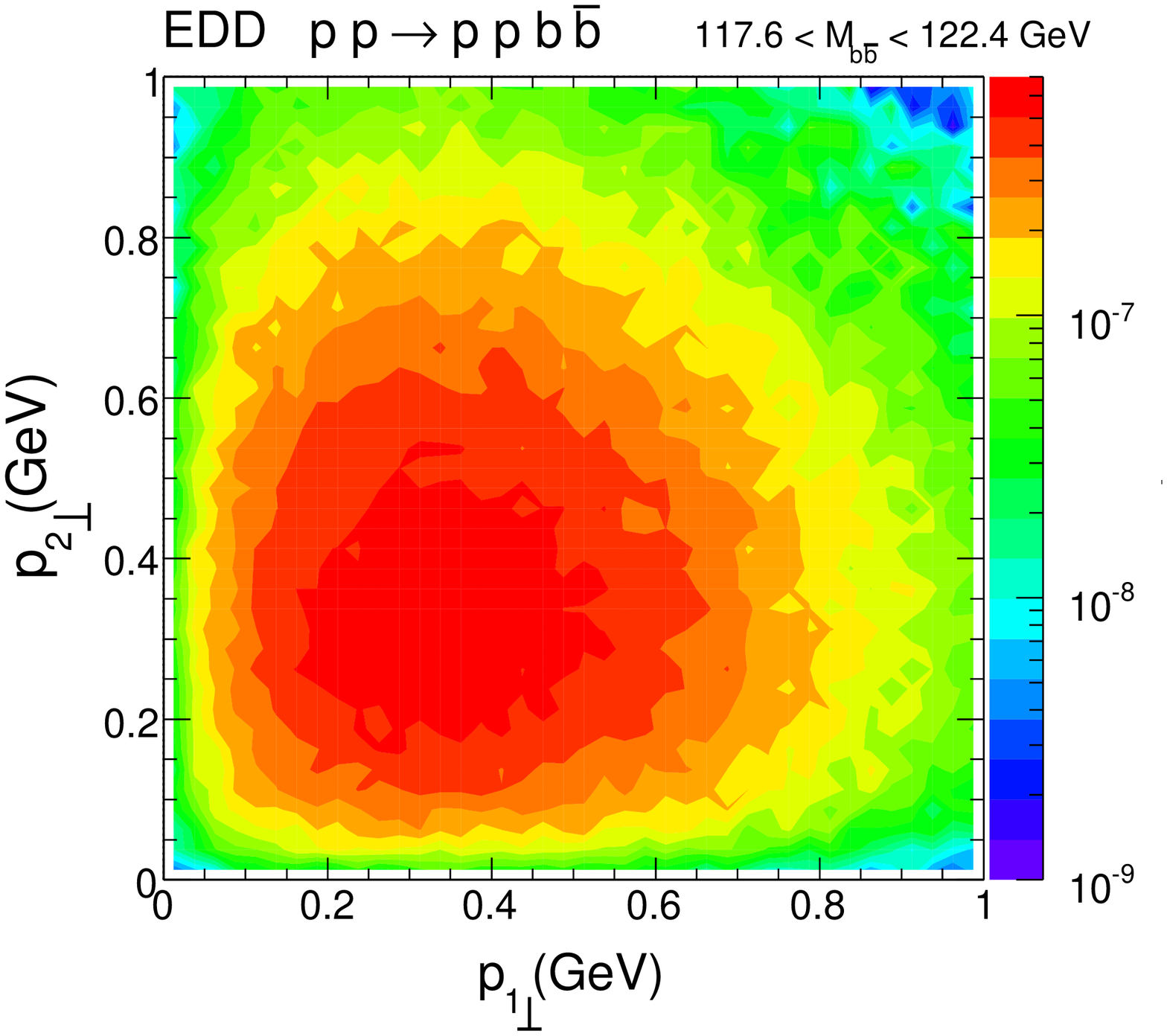}}
\end{minipage}
\begin{minipage}{0.328\textwidth}
 \centerline{\includegraphics[width=1.0\textwidth]{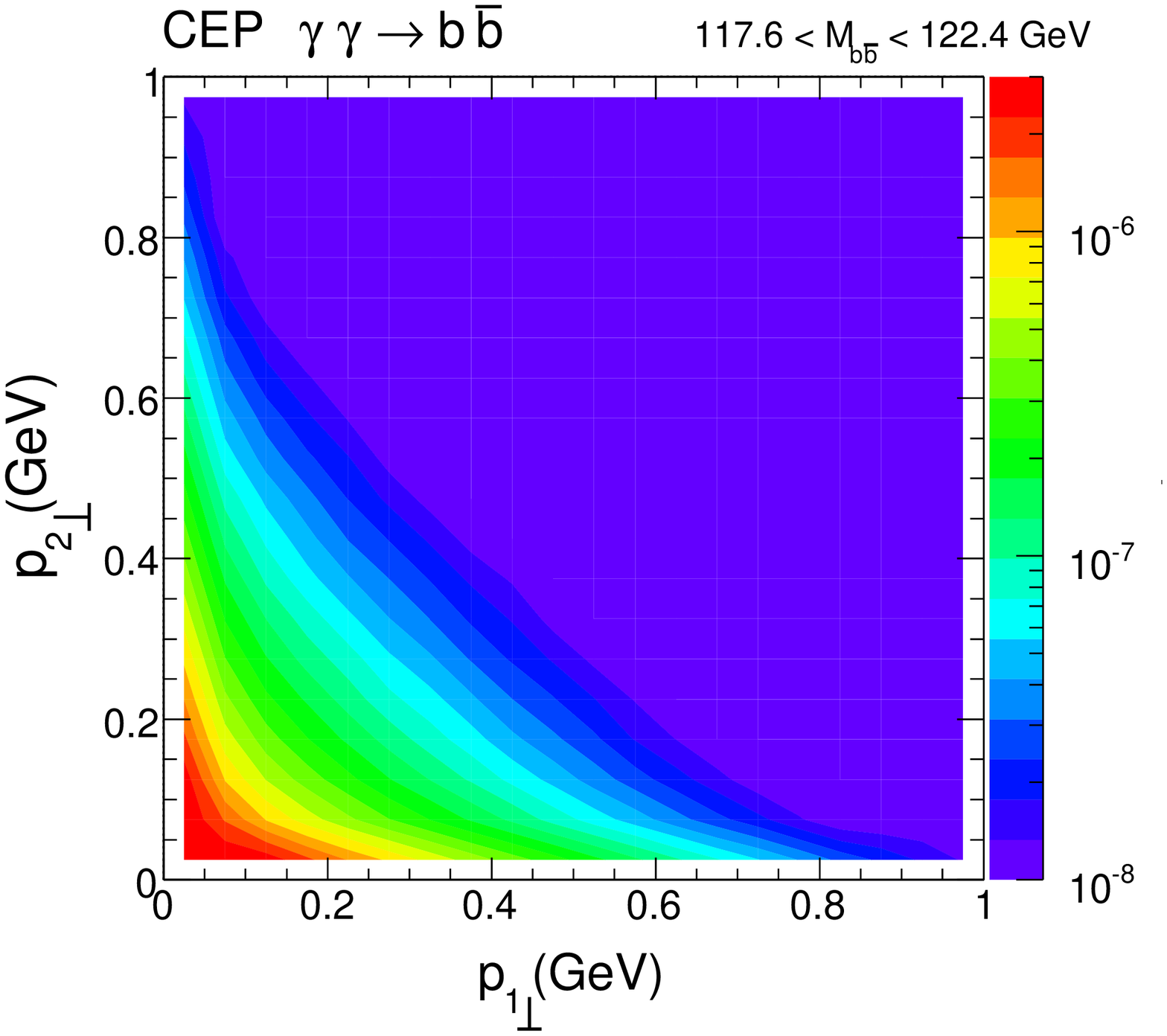}}
\end{minipage}
\begin{minipage}{0.328\textwidth}
 \centerline{\includegraphics[width=1.0\textwidth]{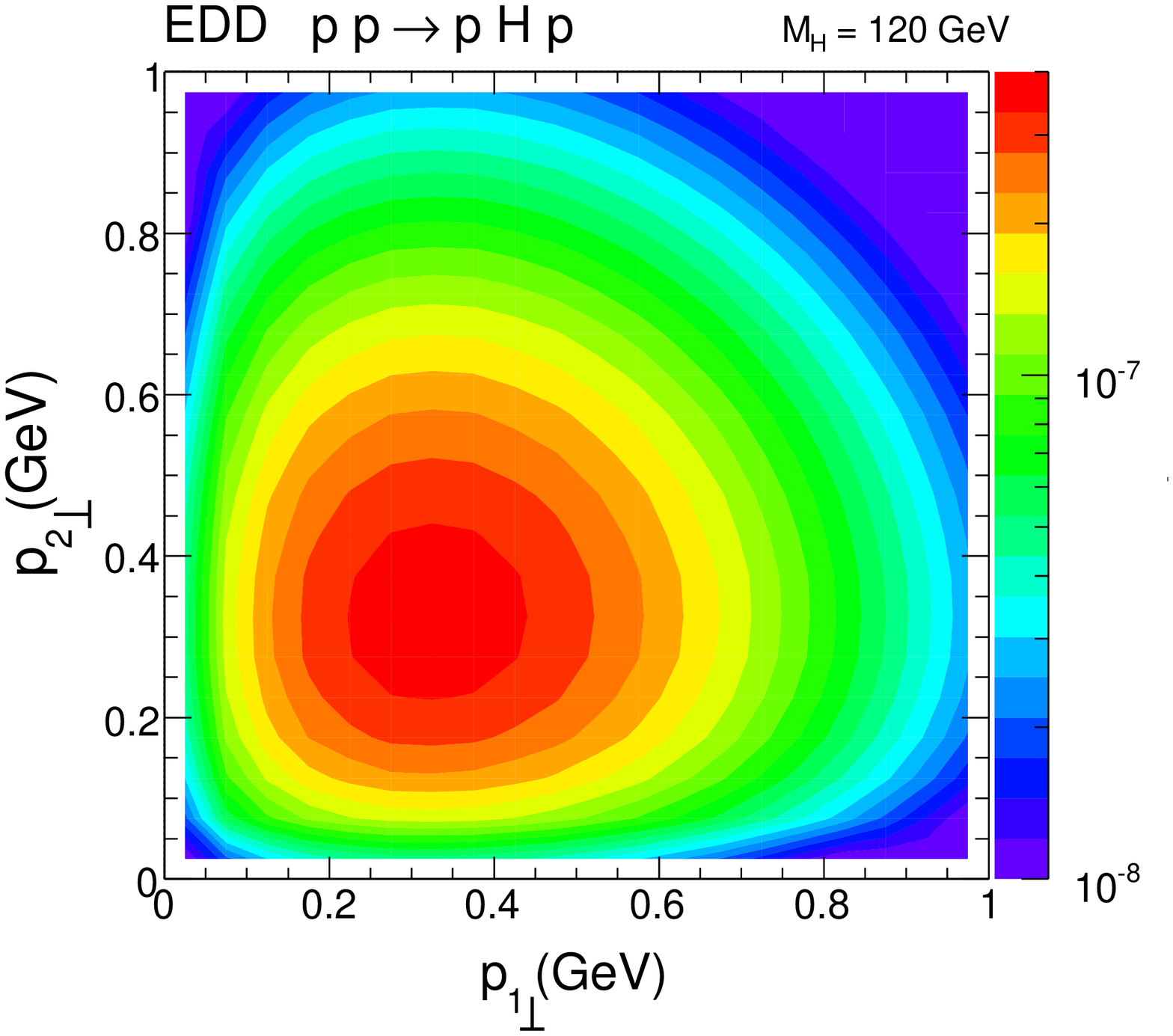}}
\end{minipage}
   \caption{
 \small Two-dimensional distributions in outgoing proton momenta $p_{1,2 \perp}$
 of the $b{\bar b}$ EDD (left) and QED (middle) continua integrated
 in the window 117.6 GeV $< M_{b \bar b} <$ 122.4 GeV
 and Higgs CEP signal (right panel). Kinematical constraints are the same as
in Fig.~\ref{fig:sigbkg-ratio}. } \label{fig:p1tp2t}
\end{figure}

Let us start from two-dimensional distributions in proton transverse
momenta. In Fig.~\ref{fig:p1tp2t} we show distributions for the
diffractive (left panel), photon-photon (middle panel) and for
$b$ and $\bar b$ from the Higgs boson decay (right panel) contributions. In the
case of the EDD and QED continua we are limited to a very
restrictive range of invariant masses (117.6 GeV $< M_{b \bar b} <$
122.4 GeV) around the chosen Higgs mass in order to facilitate a
comparison of the signal and background. While the distributions for
the diffractive $b \bar b$ continuum and Higgs are rather similar
the distribution for the photon-photon production differs considerably.
While the first two ones are peaked at sizeable transverse momenta
of about 0.3 GeV, the photon-photon contribution is peaked at
extremely small proton transverse momenta due to photon propagators.
Cutting off extremely small proton transverse momenta would allow to
get rid of the photon-photon contribution to a large extent. It is
not completely clear if this can be done easily experimentally. A
Monte Carlo study including the experimental apparatus seems to be
required.

\begin{figure}[!h]
\begin{minipage}{0.328\textwidth}
 \centerline{\includegraphics[width=1.0\textwidth]{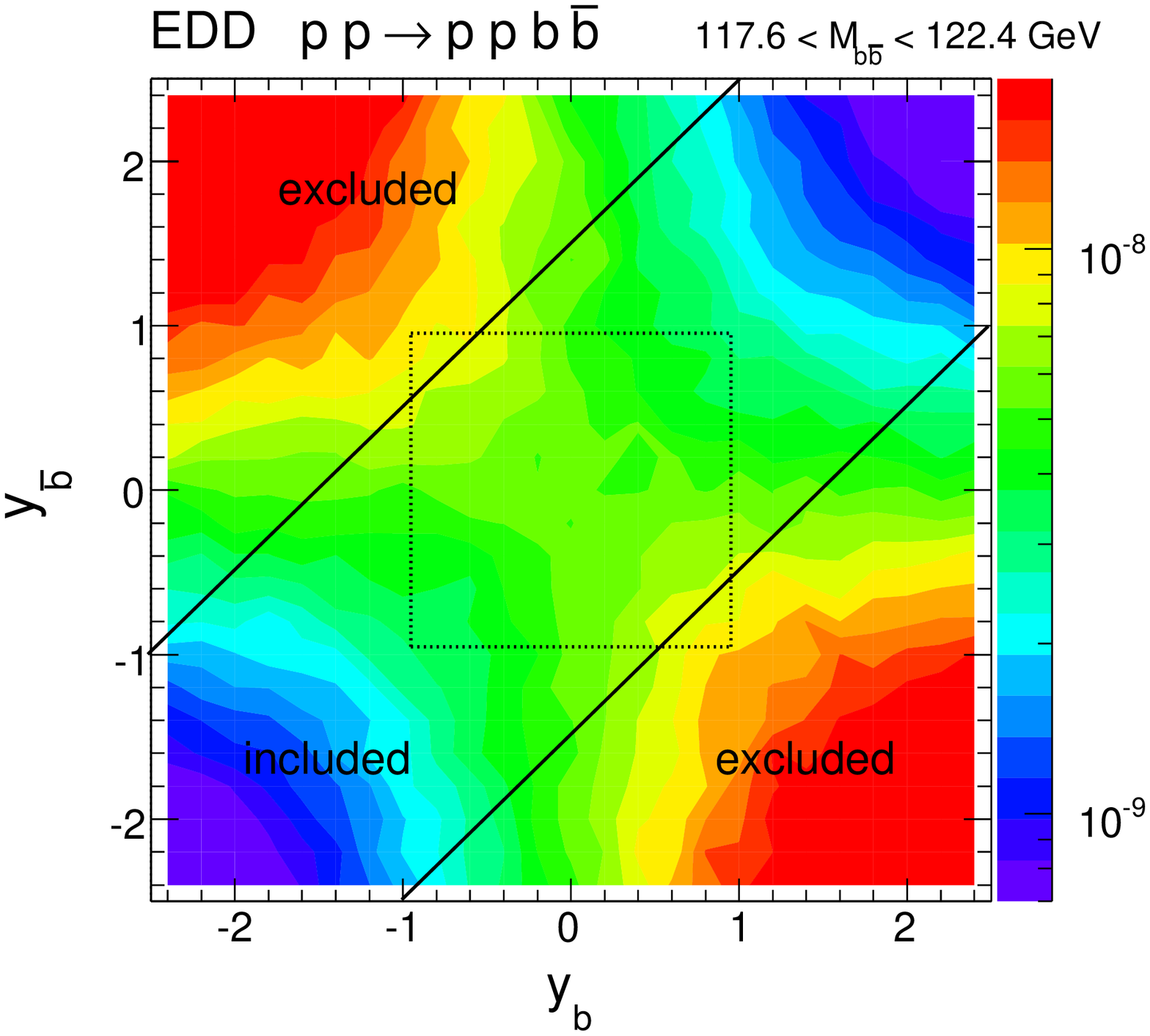}}
\end{minipage}
\begin{minipage}{0.328\textwidth}
 \centerline{\includegraphics[width=1.0\textwidth]{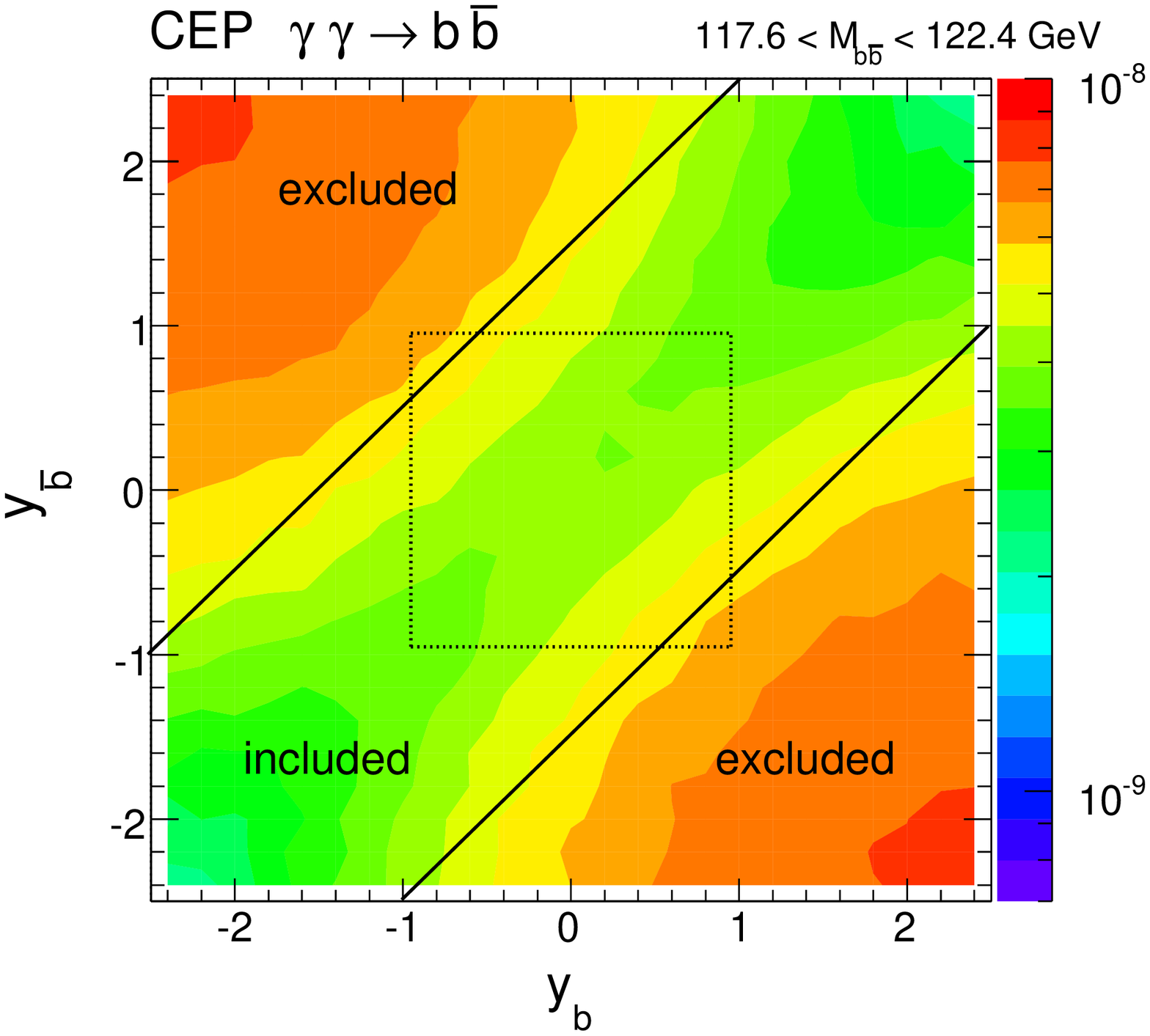}}
\end{minipage}
\begin{minipage}{0.328\textwidth}
 \centerline{\includegraphics[width=1.0\textwidth]{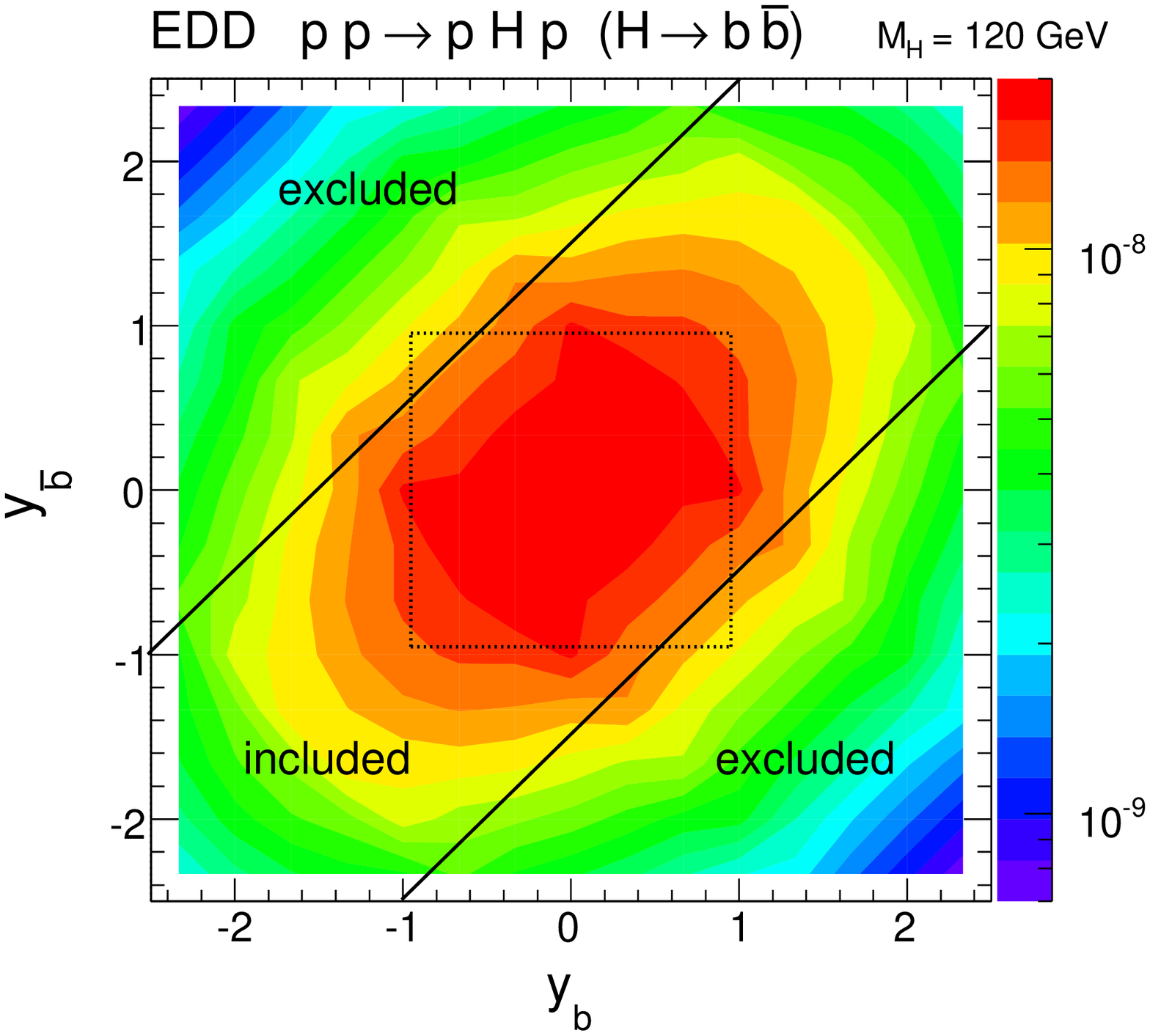}}
\end{minipage}
   \caption{
 \small
 Two-dimensional distributions in $b$ and ${\bar b}$ rapidities
 integrated in the window 117.6 GeV $< M_{b \bar b} <$ 122.4 GeV
 for the diffractive QCD background (left panel), $\gamma^*\gamma^*$ contribution (middle panel)
 and Higgs CEP (right panel) .
 Kinematical constraints are the same as in Fig.~\ref{fig:sigbkg-ratio}.
}
 \label{fig:y3y4}
\end{figure}

Next let us consider two-dimensional distributions in rapidities of
the quark ($y_b$) and antiquark ($y_{\bar b}$). In
Fig.~\ref{fig:y3y4}, as in the previous case, we show separately
distributions for diffractive continuum (left panel), photon-photon
continuum (middle panel) and Higgs (right panel) contributions. The
problem of the background subtraction looks here fairly favorable. In
the case of the $b\bar{b}$ continuum production the cross section is maximal
when quark and antiquark have opposite rapidities at the edges of
the main detector (ATLAS, CMS). This is completely different for the
Higgs contribution where the maximum occurs when $y_b, y_{\bar b}
\sim$ 0. Two windows suggesting how to get rid of the major part of
the background are shown in Fig.~\ref{fig:y3y4}: the square marked
by the dashed line and the area between two parallel lines at
45$^0$. The consequences of such cuts will be discussed in the
following.

\begin{figure}[!h]
\begin{minipage}{0.55\textwidth}
\includegraphics[width=1.0\textwidth]{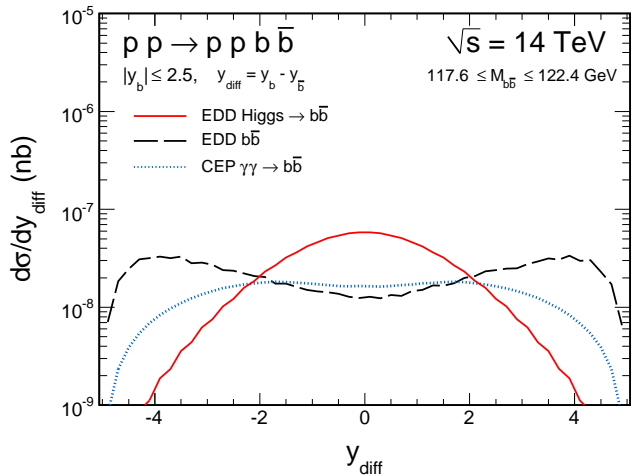}
\end{minipage}
   \caption{
\small Distribution in the difference of the quark and antiquark
rapidities.
Please note an extra cut on the $b \bar b$ invariant mass.
Kinematical constraints are the same as
in Fig.~\ref{fig:sigbkg-ratio}.
}
 \label{fig:ydiff_comparison}
\end{figure}

The situation can be also quantified in a one-dimensional
plot in a function of the difference of the quark and antiquark
rapidities (see Fig.~\ref{fig:ydiff_comparison}). The distributions for the signal and background are very
different. Imposing a cut on $y_{diff}$ can significantly improve
the signal-to-background ratio.

\begin{figure}[!h]
\begin{minipage}{0.55\textwidth}
\includegraphics[width=1.0\textwidth]{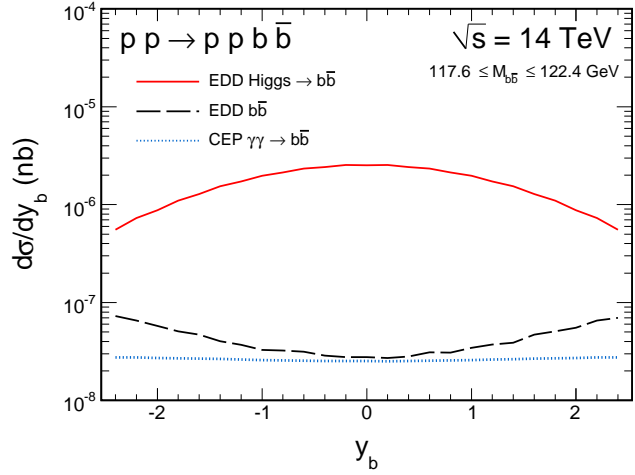}
\end{minipage}
   \caption{
\small Distribution in the quark/antiquark rapidity. Please note an
extra cut on the $b \bar b$ invariant mass. Kinematical constraints
are the same as in Fig.~\ref{fig:sigbkg-ratio}. }
 \label{fig:y3_comparison}
\end{figure}

In Fig.~\ref{fig:y3_comparison} we show the distribution in the
$b$-quark rapidity from Higgs decay and from a narrow region of $b
\bar b$ invariant mass (given in the figure) for the diffractive $b
\bar b$ and photon-photon components. While the Higgs contribution
is concentrated at $y_b \sim$ 0 the diffractive component has maxima
at the edges of the central detector. The $\gamma \gamma$
contribution is rather flat across the range of the central
detector. The different distributions in the $b$-quark rapidity of
the different components suggest that limiting to midrapidities
(i.e. not using the whole range of the detector) may help in
improving the signal-to-background ratio.

\begin{figure}[!h]
\begin{minipage}{0.55\textwidth}
\includegraphics[width=1.0\textwidth]{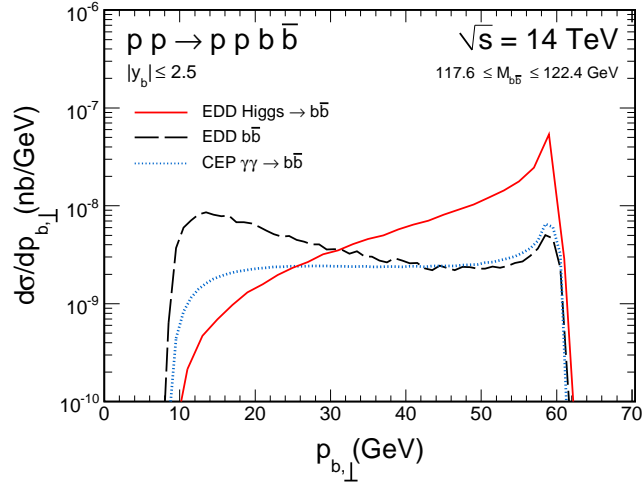}
\end{minipage}
   \caption{
 \small Distribution in the jet transverse momentum for different components.
Please note an extra cut on the $b \bar b$ invariant mass.
Kinematical constraints are the same as
in Fig.~\ref{fig:sigbkg-ratio}.}
 \label{fig:p3t_comparison}
\end{figure}

Further useful handles to improve the situation are the jet transverse momenta which can be
measured in the central detector. The importance of the cuts on the
jet transverse momenta is illustrated in
Fig.~\ref{fig:p3t_comparison}. Again we show the three components.
While the signal (Higgs) contribution is peaked at the transverse
momenta being half of the Higgs mass, the background contributions
are flat or even have local maxima at low transverse momenta.
Imposing therefore a lower cut on jet transverse momenta can again
significantly improve the signal-to-background ratio without losing
too much of the signal itself. Also, from experimental point of view
the $b$ ($\bar b$) jets can be well identified only above a certain
cut on their transverse momenta.

\begin{figure}[!h]
\begin{minipage}{0.47\textwidth}
 \centerline{\includegraphics[width=1.0\textwidth]{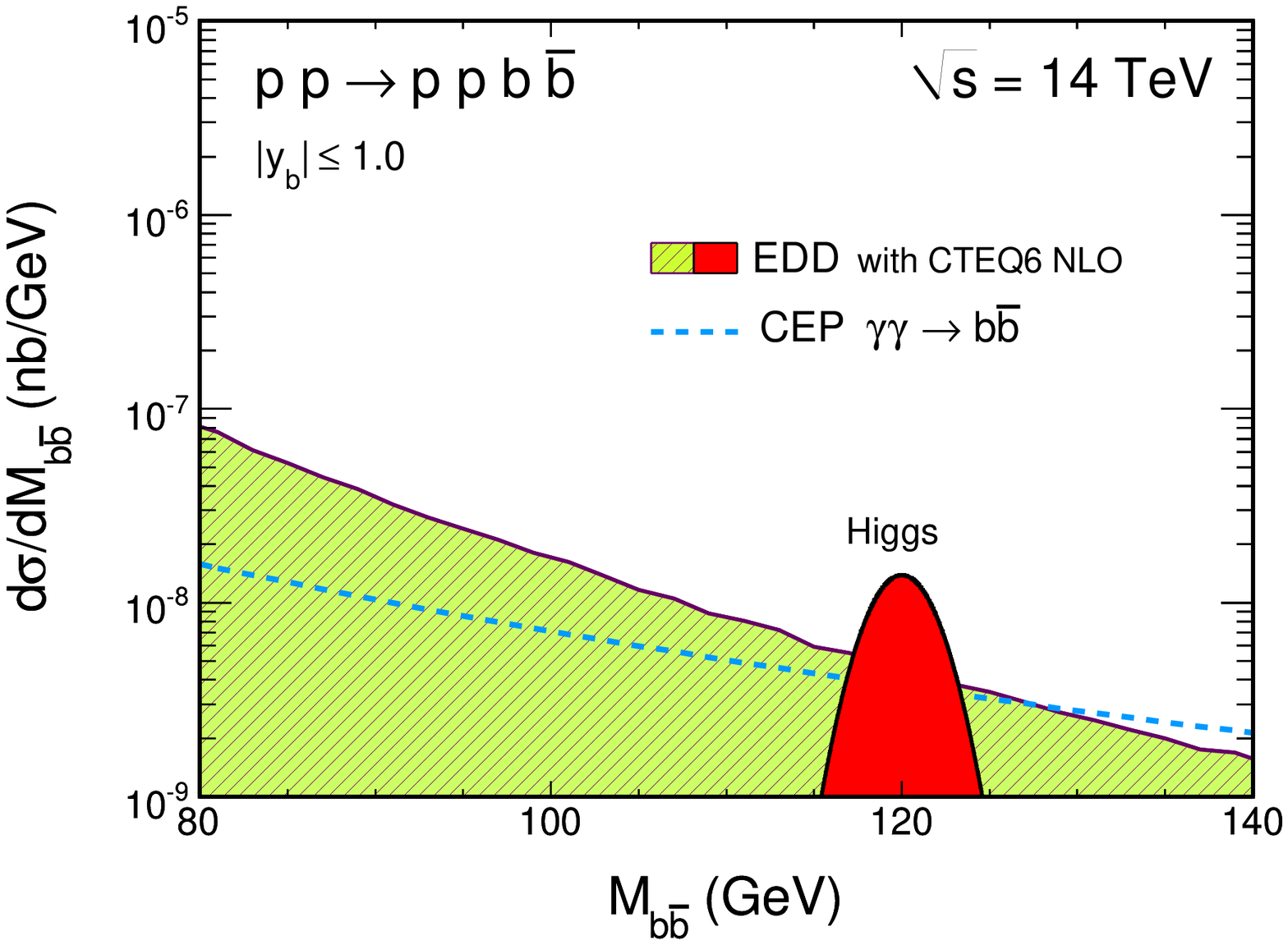}}
\end{minipage}
\hspace{0.5cm}
\begin{minipage}{0.47\textwidth}
 \centerline{\includegraphics[width=1.0\textwidth]{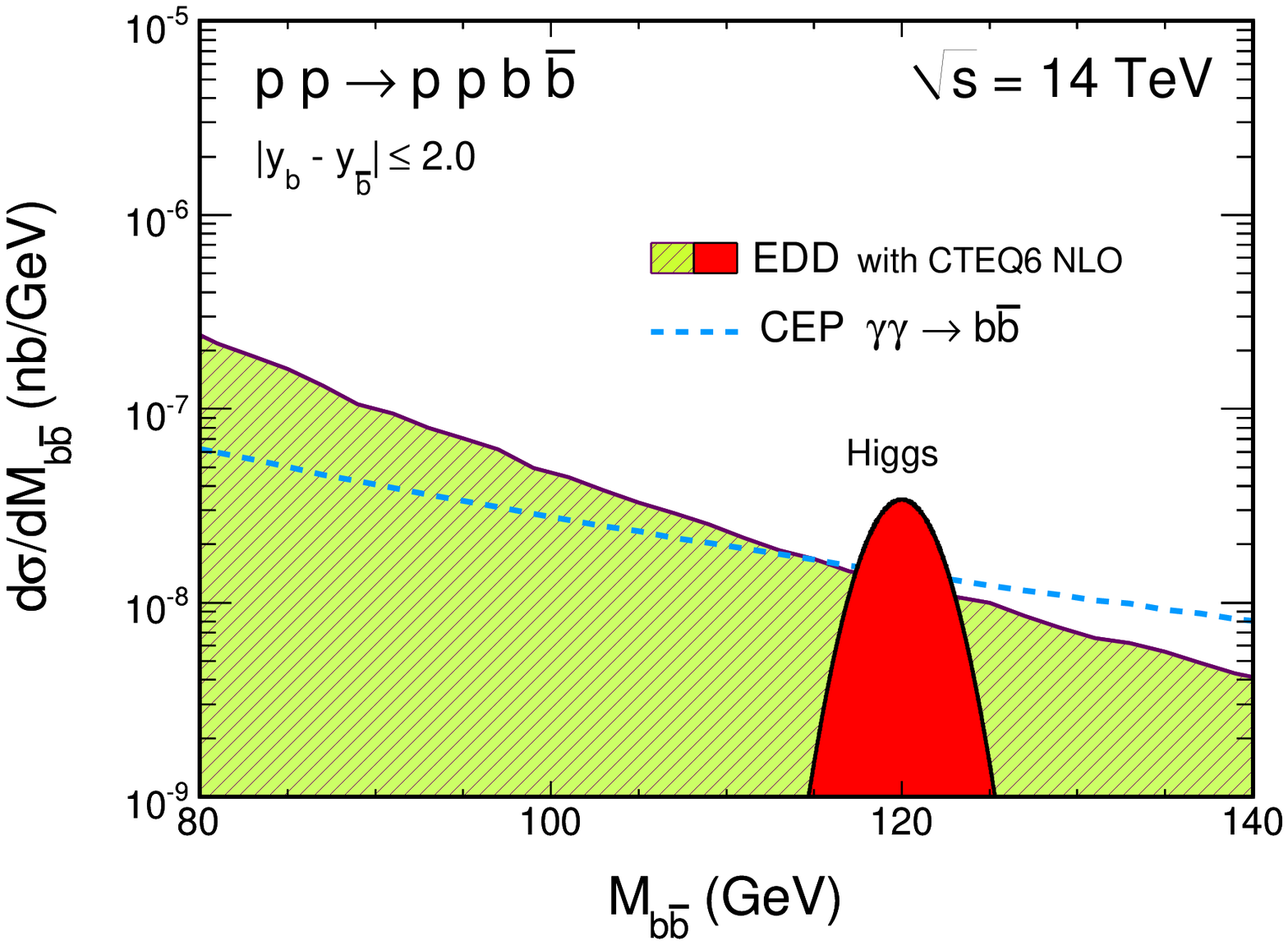}}
\end{minipage}
\hspace{0.5cm}
\begin{minipage}{0.47\textwidth}
 \centerline{\includegraphics[width=1.0\textwidth]{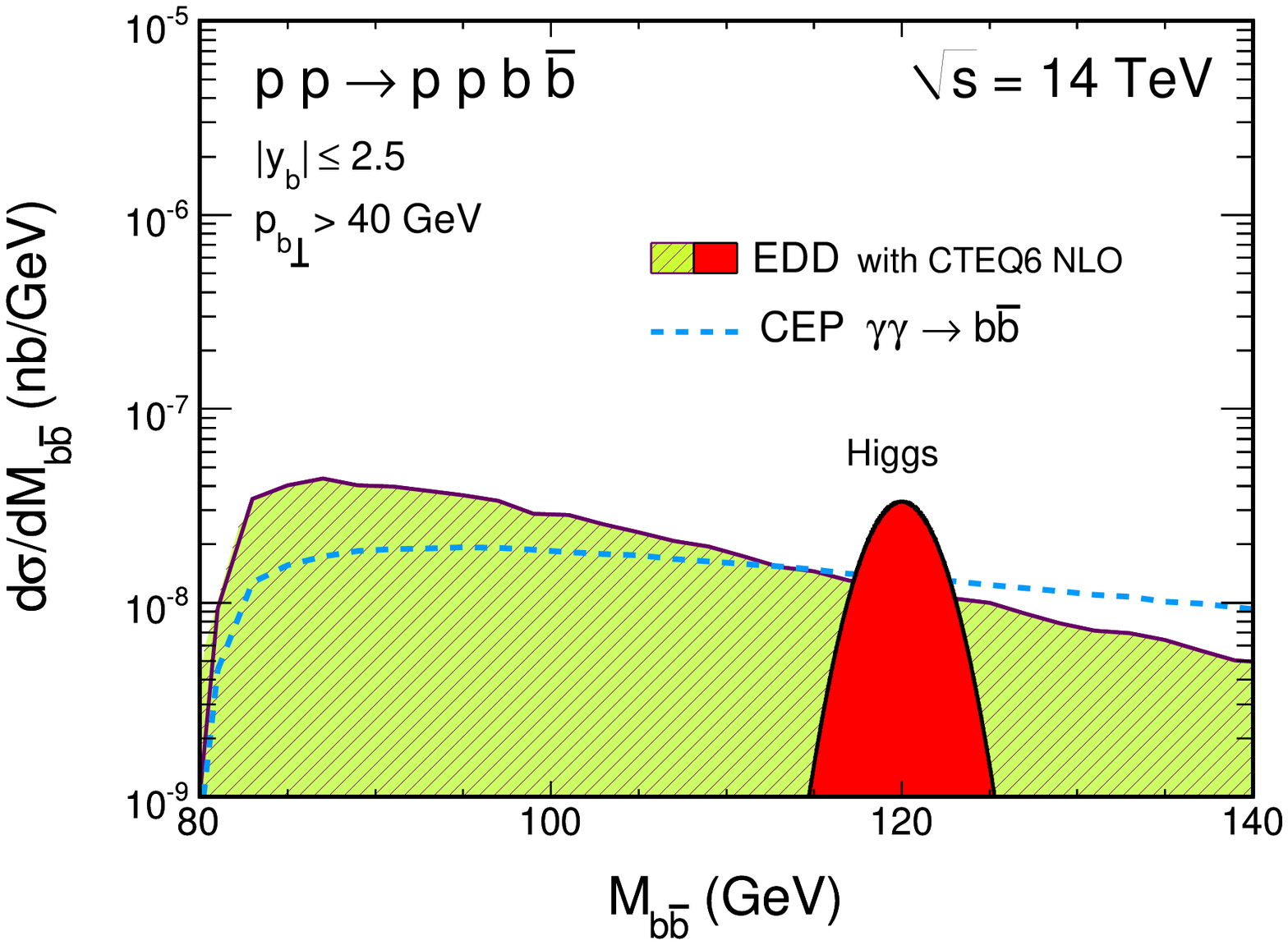}}
\end{minipage}
\hspace{0.5cm}
\begin{minipage}{0.47\textwidth}
 \centerline{\includegraphics[width=1.0\textwidth]{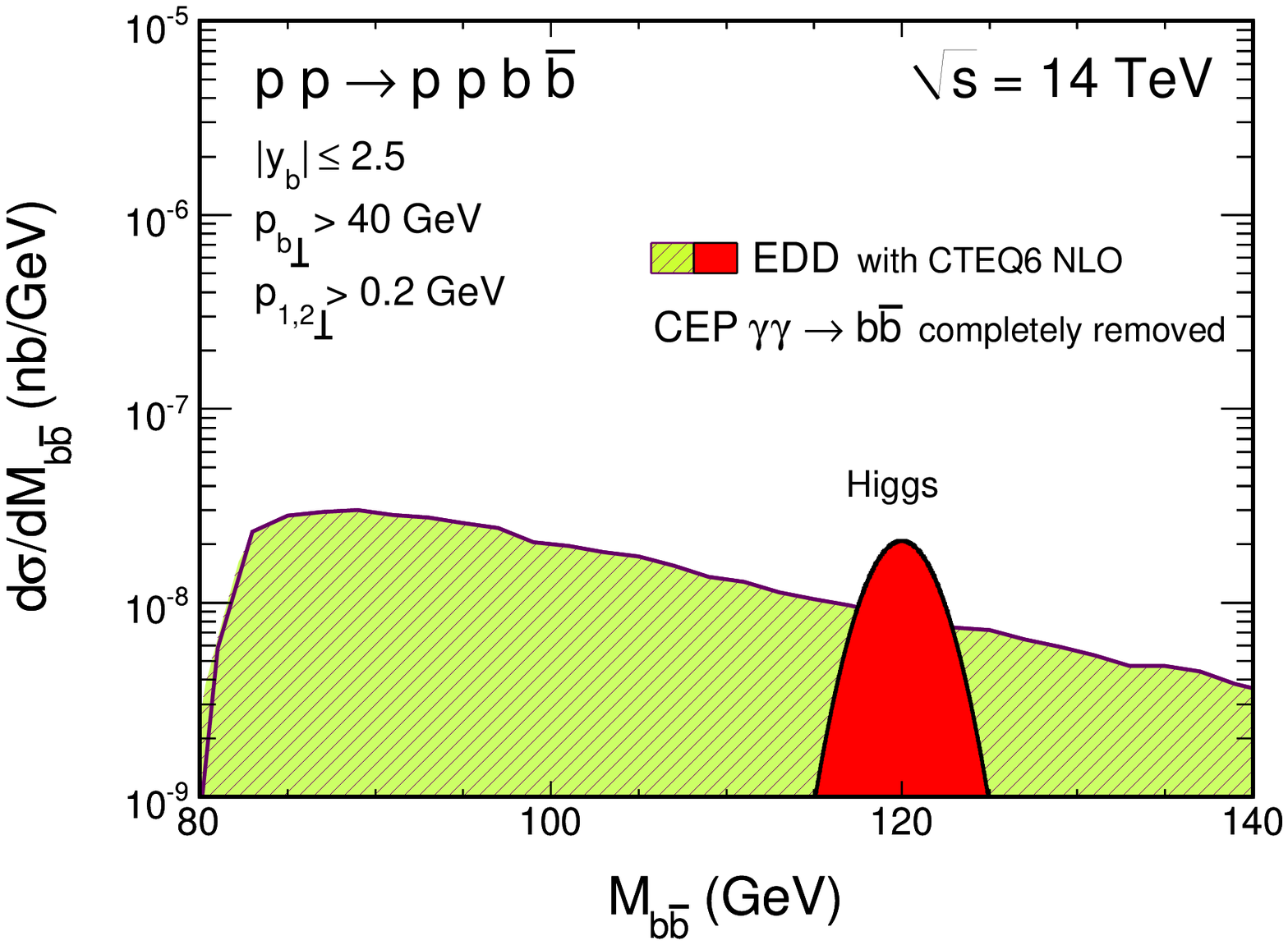}}
\end{minipage}
   \caption{
 \small The $b \bar b$ invariant mass distribution for $\sqrt{s}$ =
14 TeV and for $b$ and $\bar b$ jets from Higgs decay with different
limitations in the (anti)quark rapidity $y_q$, (anti)quark transverse momenta and outgoing proton
momenta $p_{1,2\perp}$.}
 \label{fig:dsig_dM_cuts}
\end{figure}

Now we wish to quantify the effect of cuts on the $b \bar b$
invariant mass (missing mass experimentally) distribution. We shall
impose cuts in order not to loose too much Higgs signal.
In Fig.~\ref{fig:dsig_dM_cuts} we show the results for several
scenarios (cuts). Here we omit the $Z^0$ contribution and
concentrate solely on the Higgs signal. In the left upper corner we
show result with the cut only on quark and antiquark rapidities (the
square in Fig.~\ref{fig:y3y4}) i.e. not making use of the whole
coverage of the main LHC detectors. The signal is now above the
diffractive background. We also show, by the thin dashed line, the
photon-photon background which is only slightly smaller than the
diffractive one. In the upper right corner we show the result for
the cut on the quark and antiquark rapidity difference (see parallel
thick solid lines in Fig.~\ref{fig:y3y4}). The signal-to-background
ratio is here similar, except that the cross sections are larger. In
the lower left corner we show the situation with the lower cut on
both quark and antiquark jets. The situation is similar as for the
rapidity cuts. In order to eliminate the photon-photon contribution
in the lower right corner we impose in addition a lower cut on proton
transverse momenta. Now the signal clearly sticks above the
background and the contribution of the photon-photon continuum is
negligible. The cross section for the Higgs boson with the cuts is only 2-3
times smaller than that without the cuts.

\begin{figure}[!h]
\begin{minipage}{0.328\textwidth}
\includegraphics[width=1.0\textwidth]{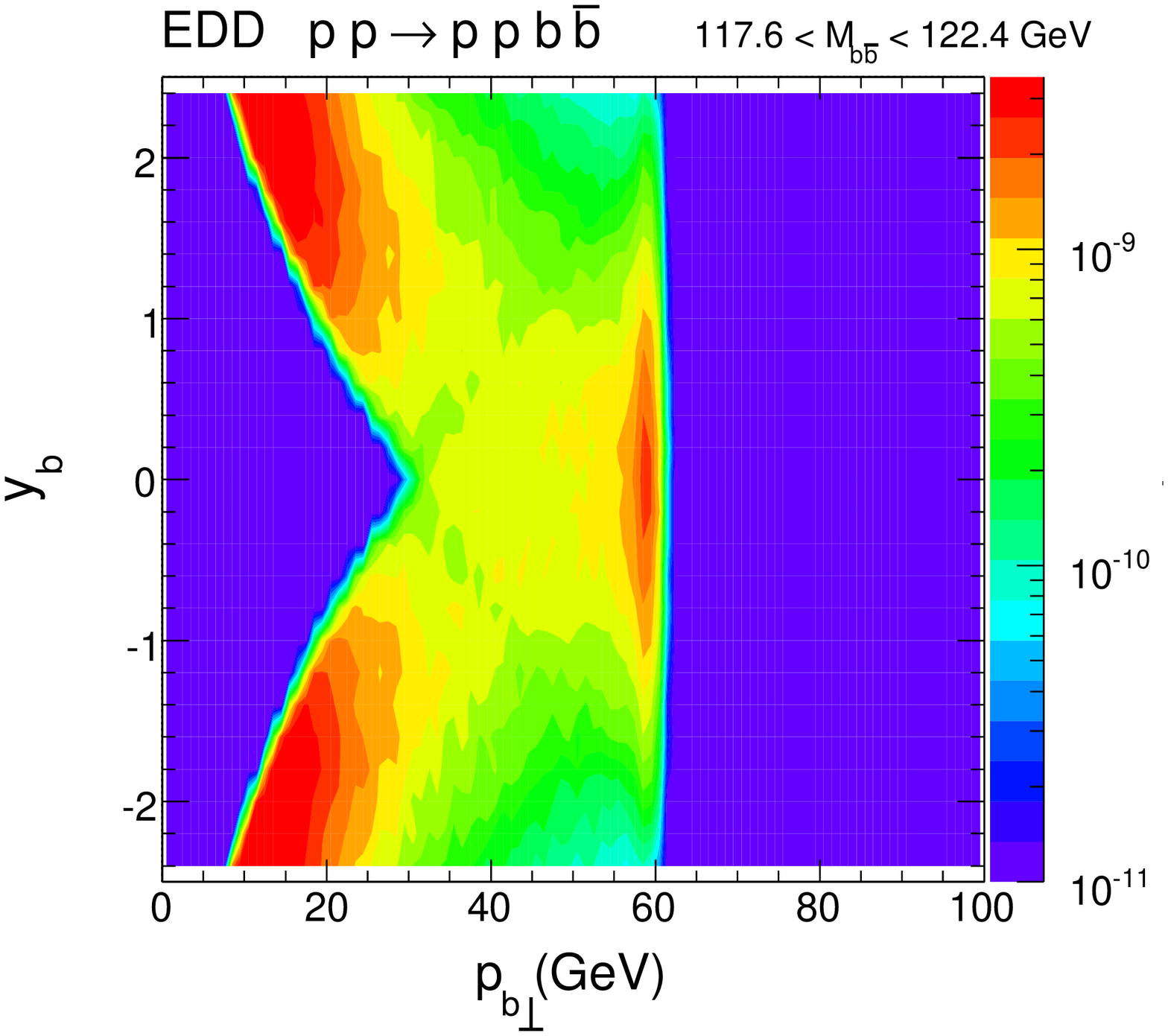}
\end{minipage}
\begin{minipage}{0.328\textwidth}
\includegraphics[width=1.0\textwidth]{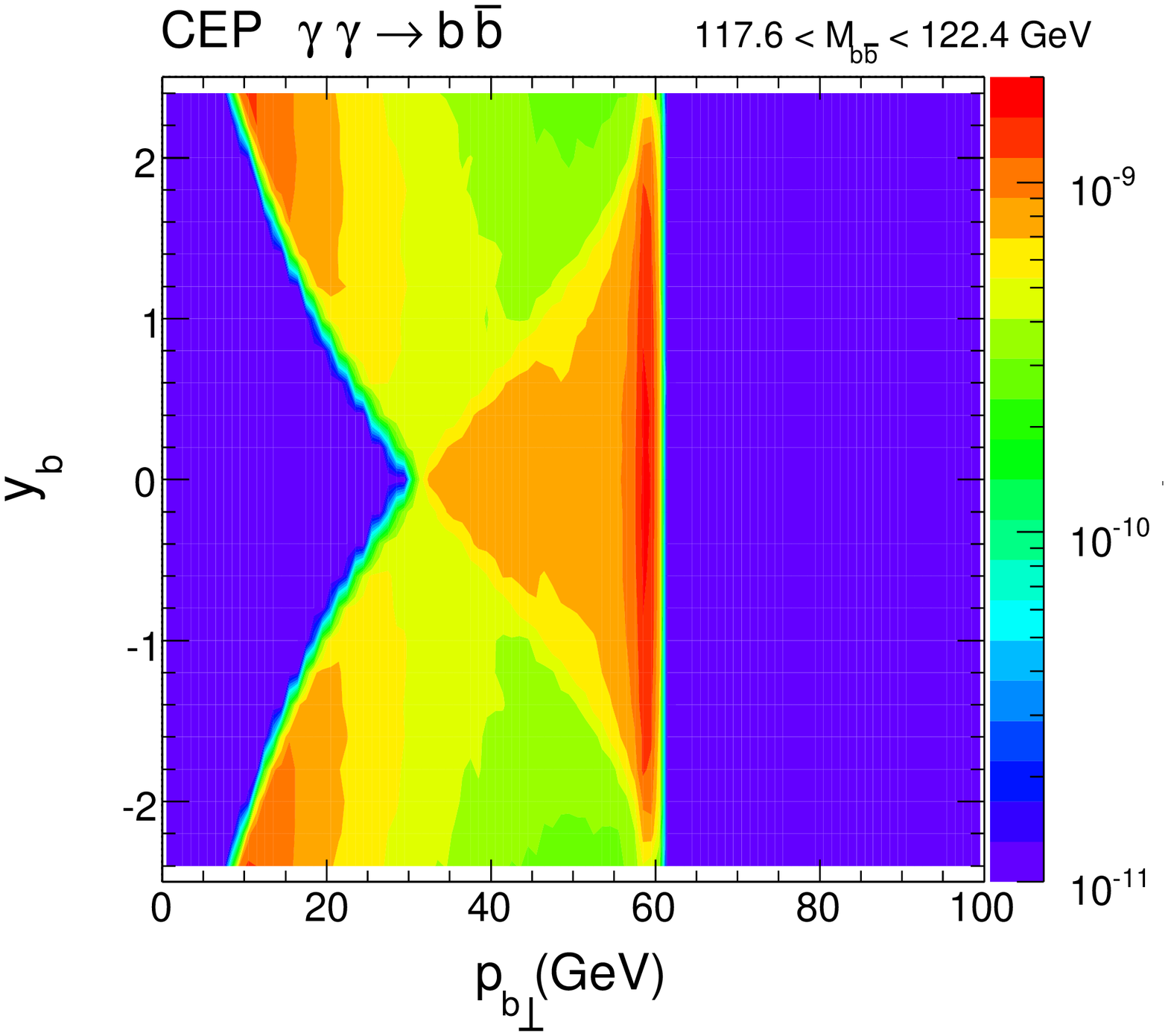}
\end{minipage}
\begin{minipage}{0.328\textwidth}
\includegraphics[width=1.0\textwidth]{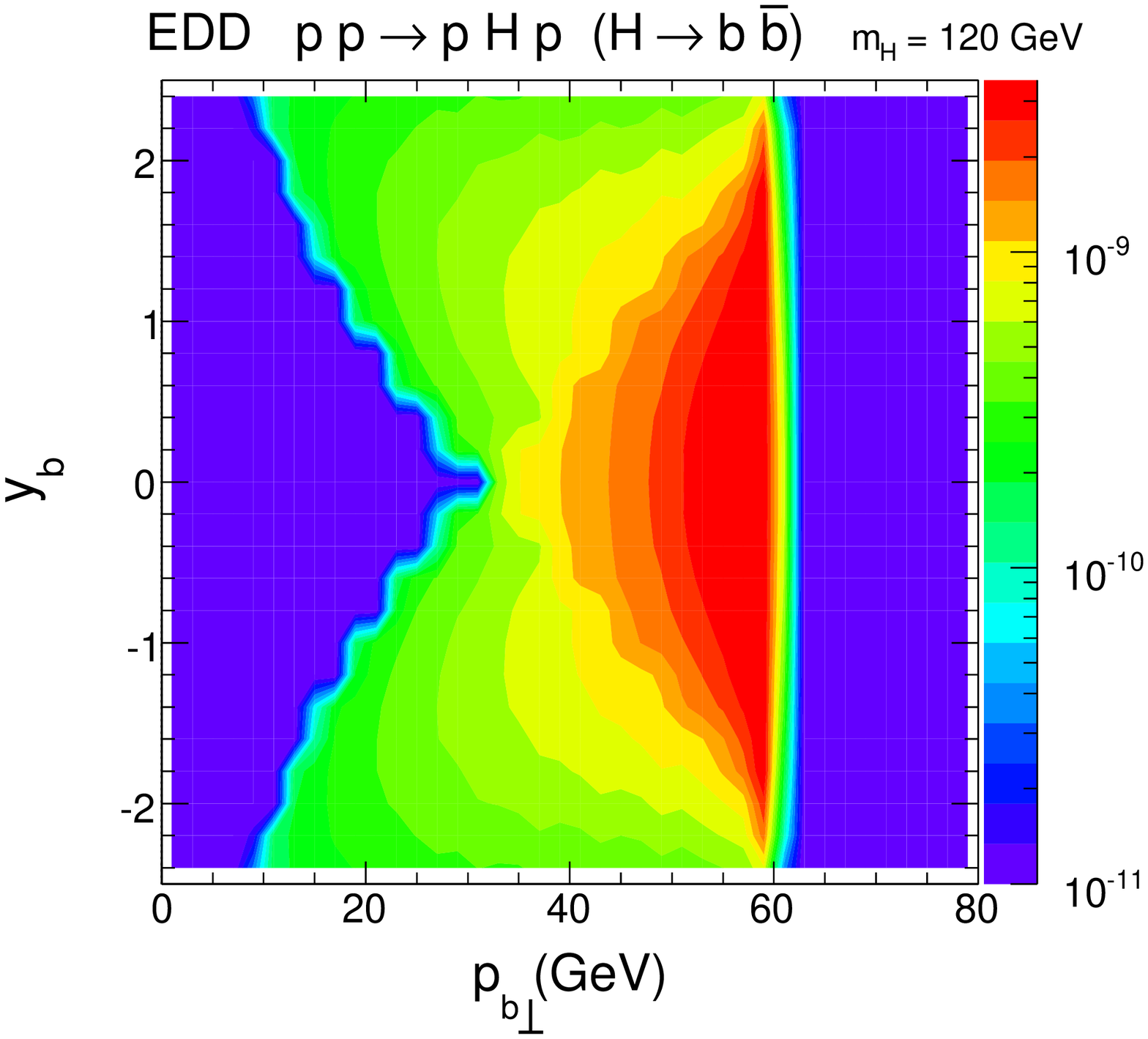}
\end{minipage}
   \caption{Two-dimensional distribution in quark rapidity
and quark transverse momentum for different mechanisms.
}
\label{fig:y3p3t}
\end{figure}

The simultaneous inclusion of cuts on quark (antiquark) rapidities
and transverse momenta does not improve further the
signal-to-background ratio. It is enough in practice to include only
one of them depending on experimental convenience. Why it is so is
discussed in Fig.~\ref{fig:y3p3t} where we show two-dimensional distributions in
$b$-quark rapidity and transverse momentum for the $b \bar b$
invariant mass window around the Higgs mass. Here three well separated maxima
are seen. One can extract them either by cuts on quark/antiquark
rapidities or by cuts on quark/antiquark transverse momenta. This
explains the equivalence of the cuts. Unlike the corresponding distribution not restricted to $M_{b\bar b}$ around
Higgs, the distributions with the restriction have more complicated
structure. While for the Higgs decay $b$-quarks are produced
predominantly at midrapidities with transverse momenta $p_{\perp} \sim
M_H/2$. The two-dimensional distributions for the diffractive
continuum and the $\gamma^*\gamma^*$ fusion subprocess have three
maxima: one as for the Higgs decay (smaller) and two other with much
smaller transverse momenta and larger rapidities (bigger). For
diffractive $b\bar b$ continuum the maxima at small $p_{\perp}$ and large
rapidities are much bigger. Identification of the maxima
experimentally would be a confirmation of the present predictions.
Imposing appropriate cuts in either rapidity or transverse momentum
would allow to get rid of a large fraction of the diffractive background.

\begin{figure}[!h]
\begin{minipage}{0.328\textwidth}
\includegraphics[width=1.0\textwidth]{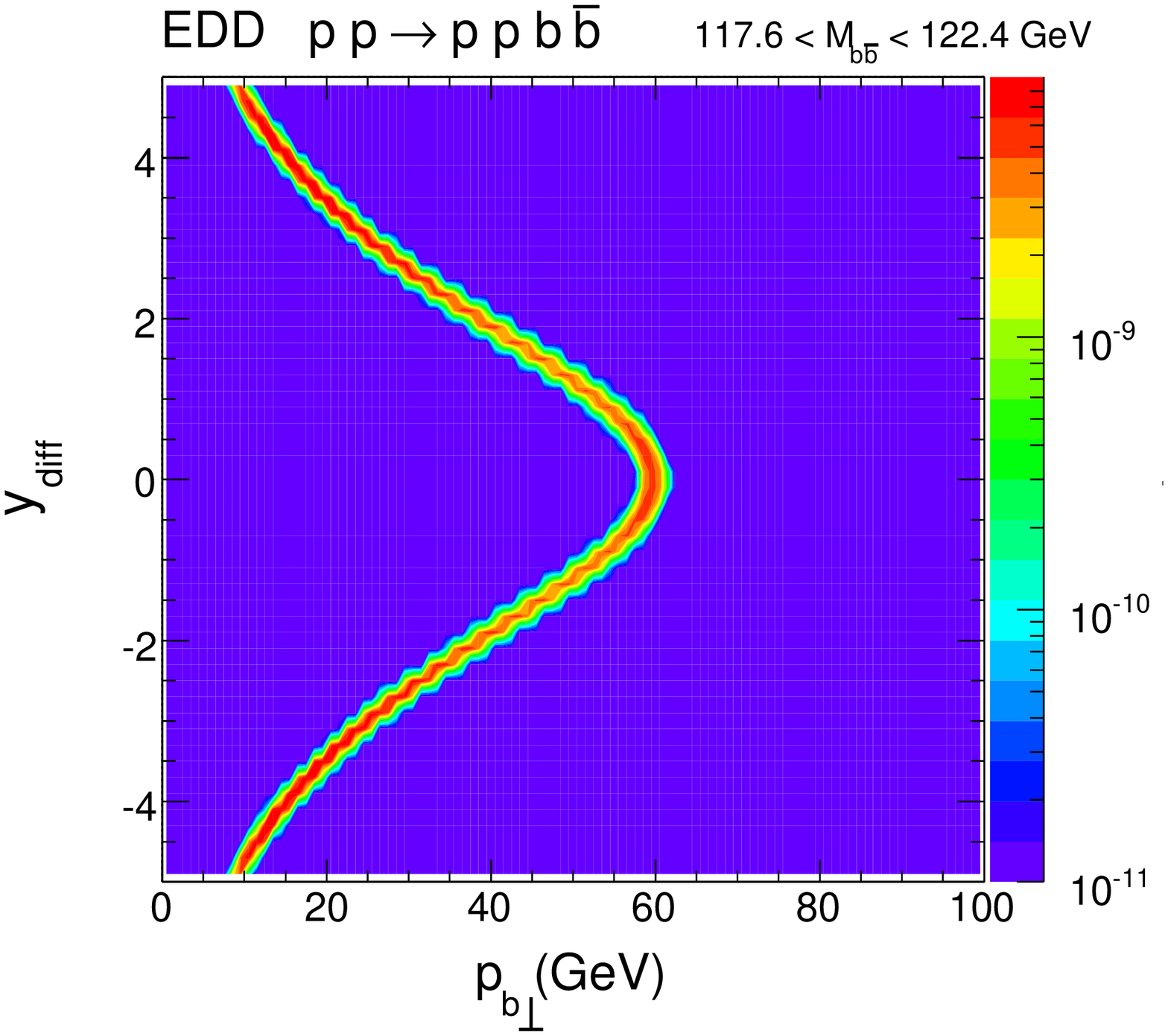}
\end{minipage}
\begin{minipage}{0.328\textwidth}
\includegraphics[width=1.0\textwidth]{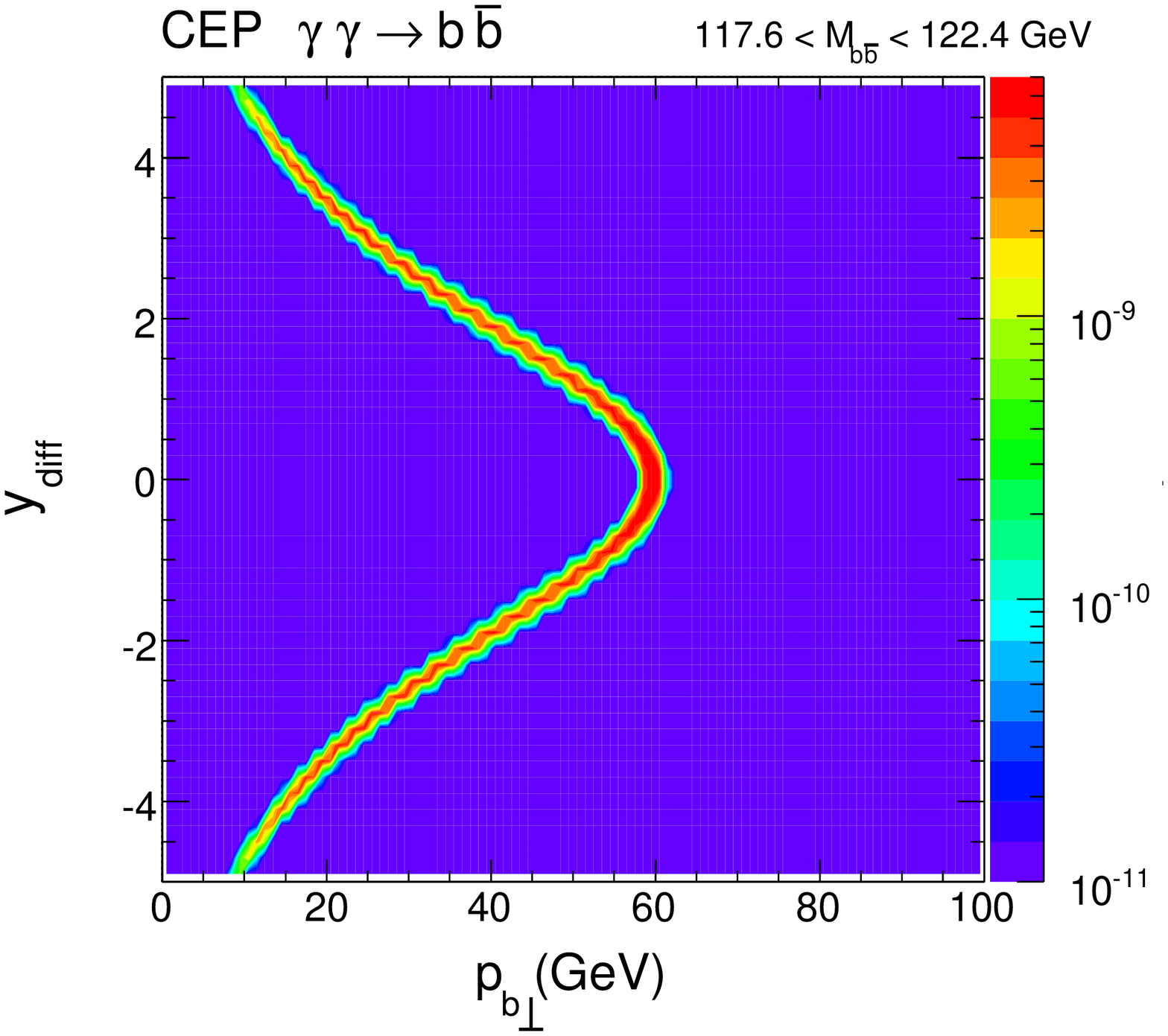}
\end{minipage}
\begin{minipage}{0.328\textwidth}
\includegraphics[width=1.0\textwidth]{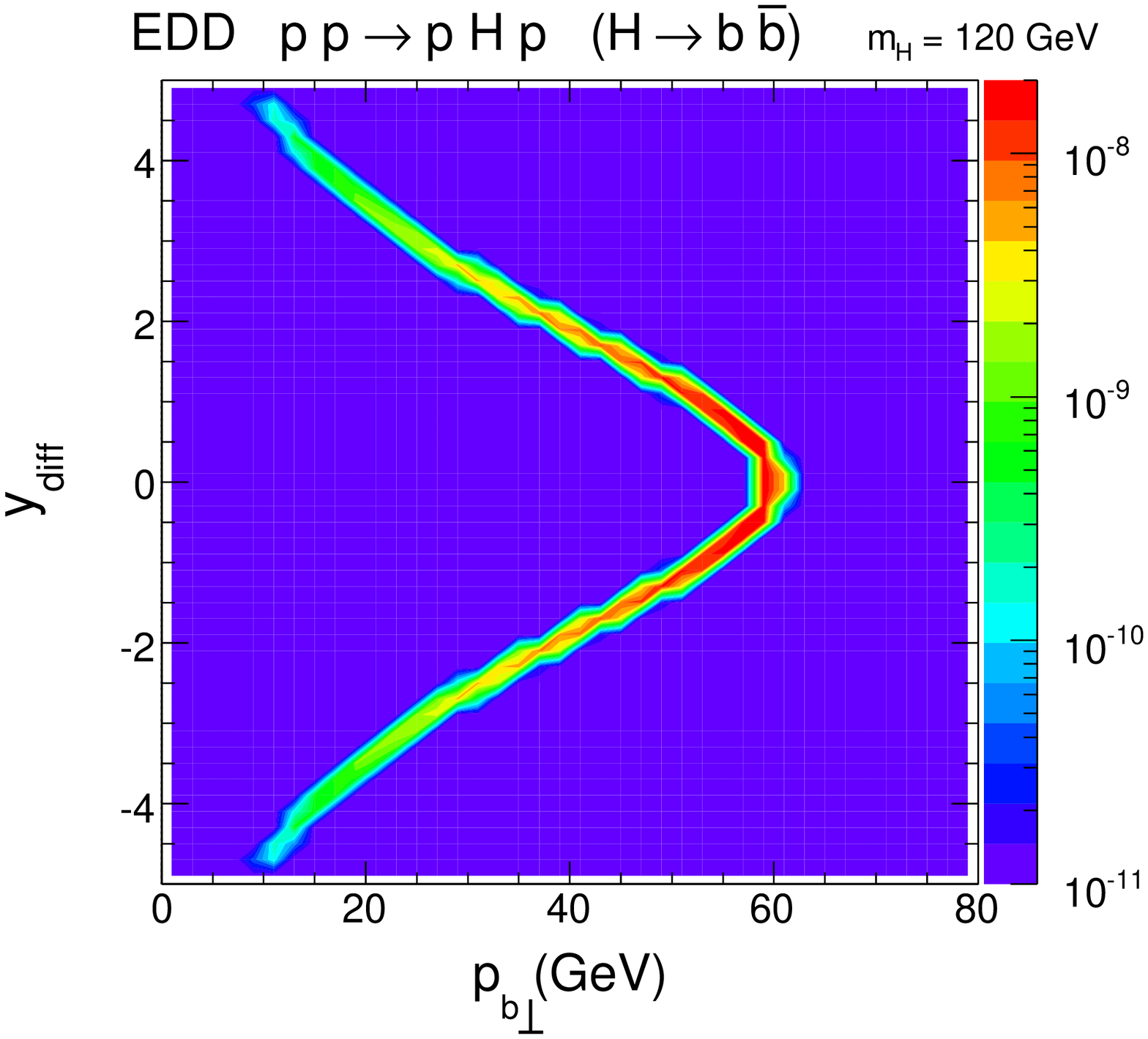}
\end{minipage}
   \caption{Two-dimensional distribution in the quark-antiquark rapidity
   difference and quark transverse momentum.
}
\label{fig:ydiffp3t}
\end{figure}

\begin{center}
\begin{figure}[!h]
\begin{minipage}{0.328\textwidth}
 \centerline{\includegraphics[width=1.0\textwidth]{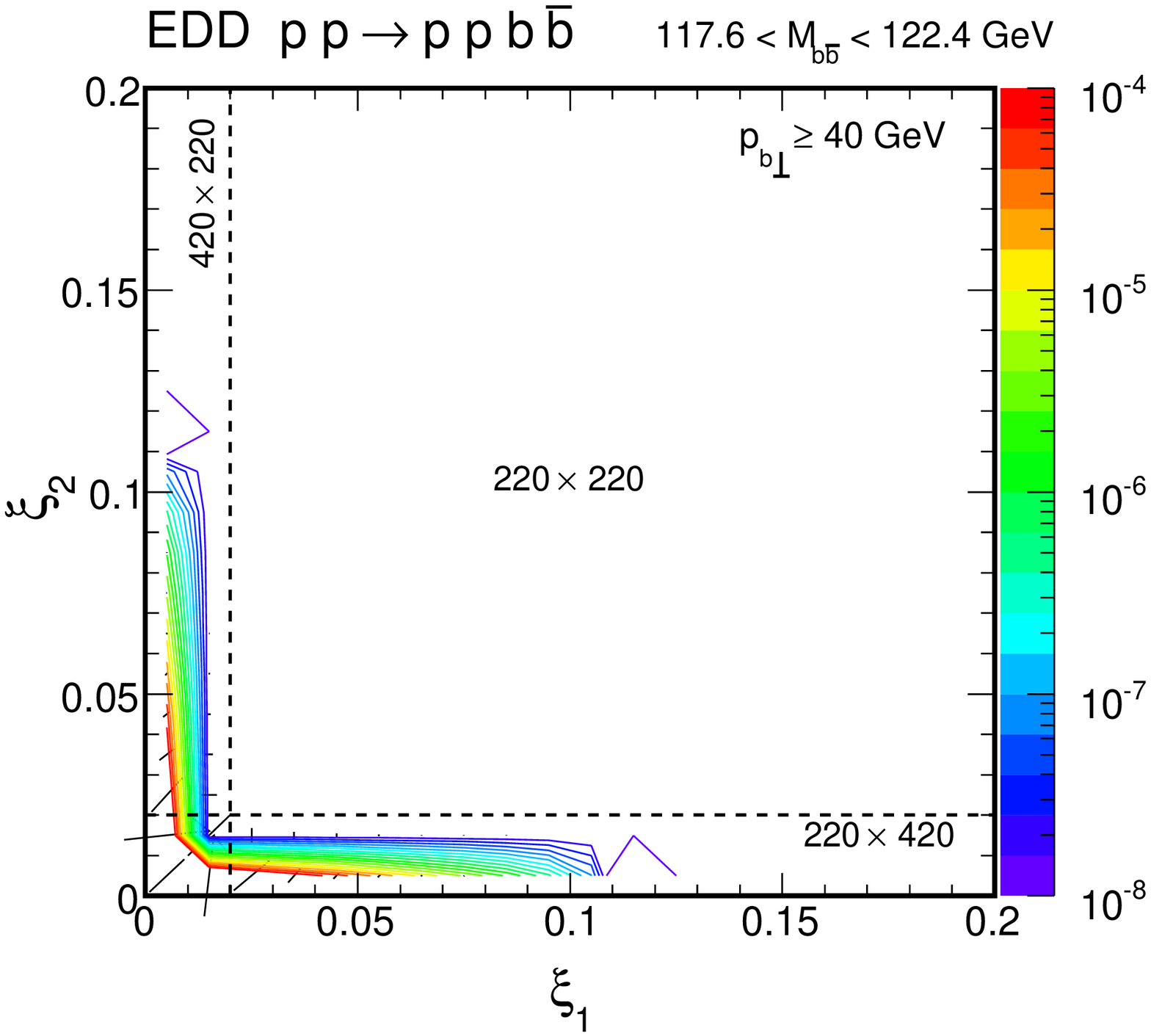}}
\end{minipage}
\begin{minipage}{0.328\textwidth}
 \centerline{\includegraphics[width=1.0\textwidth]{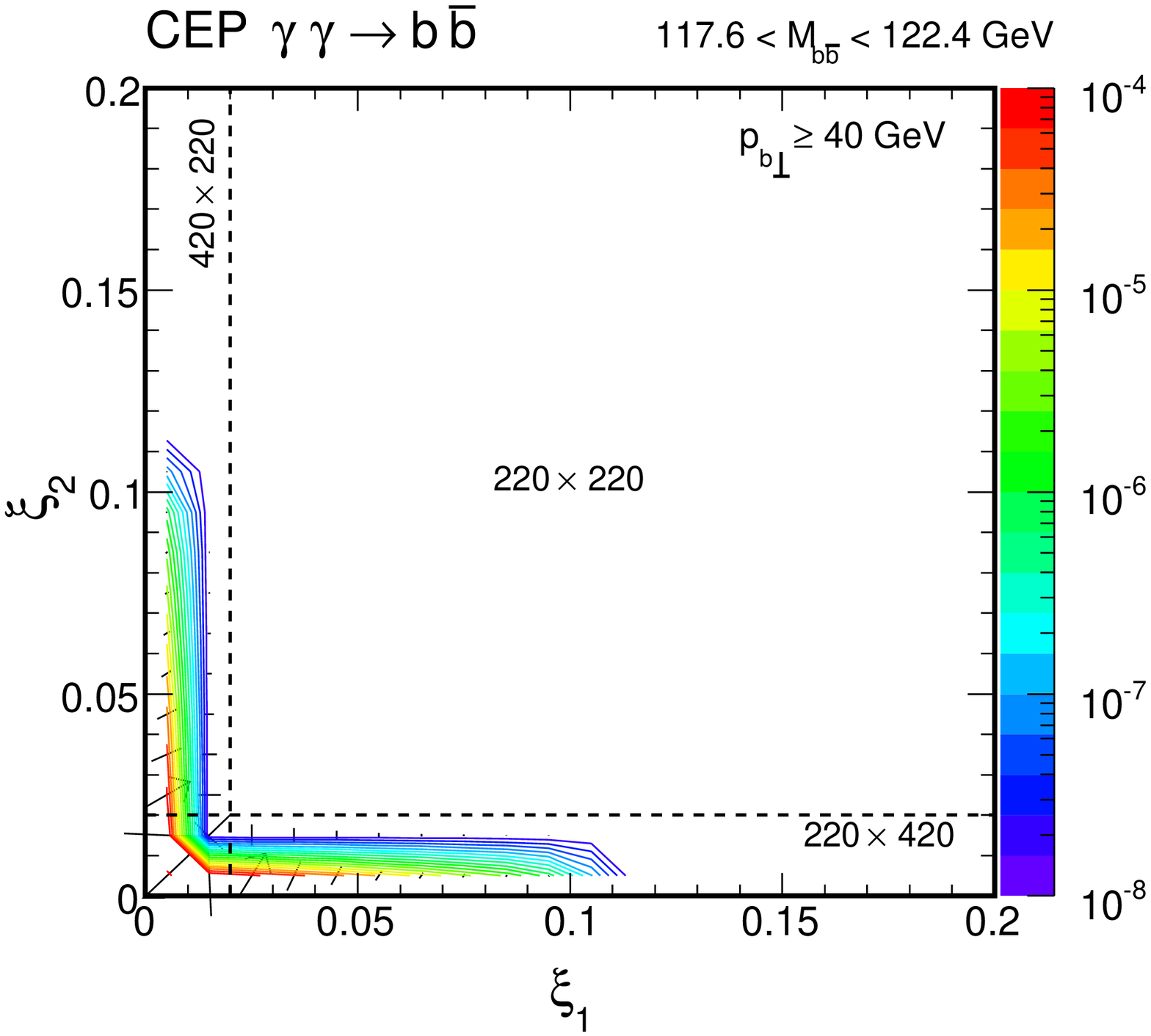}}
\end{minipage}
\begin{minipage}{0.328\textwidth}
 \centerline{\includegraphics[width=1.0\textwidth]{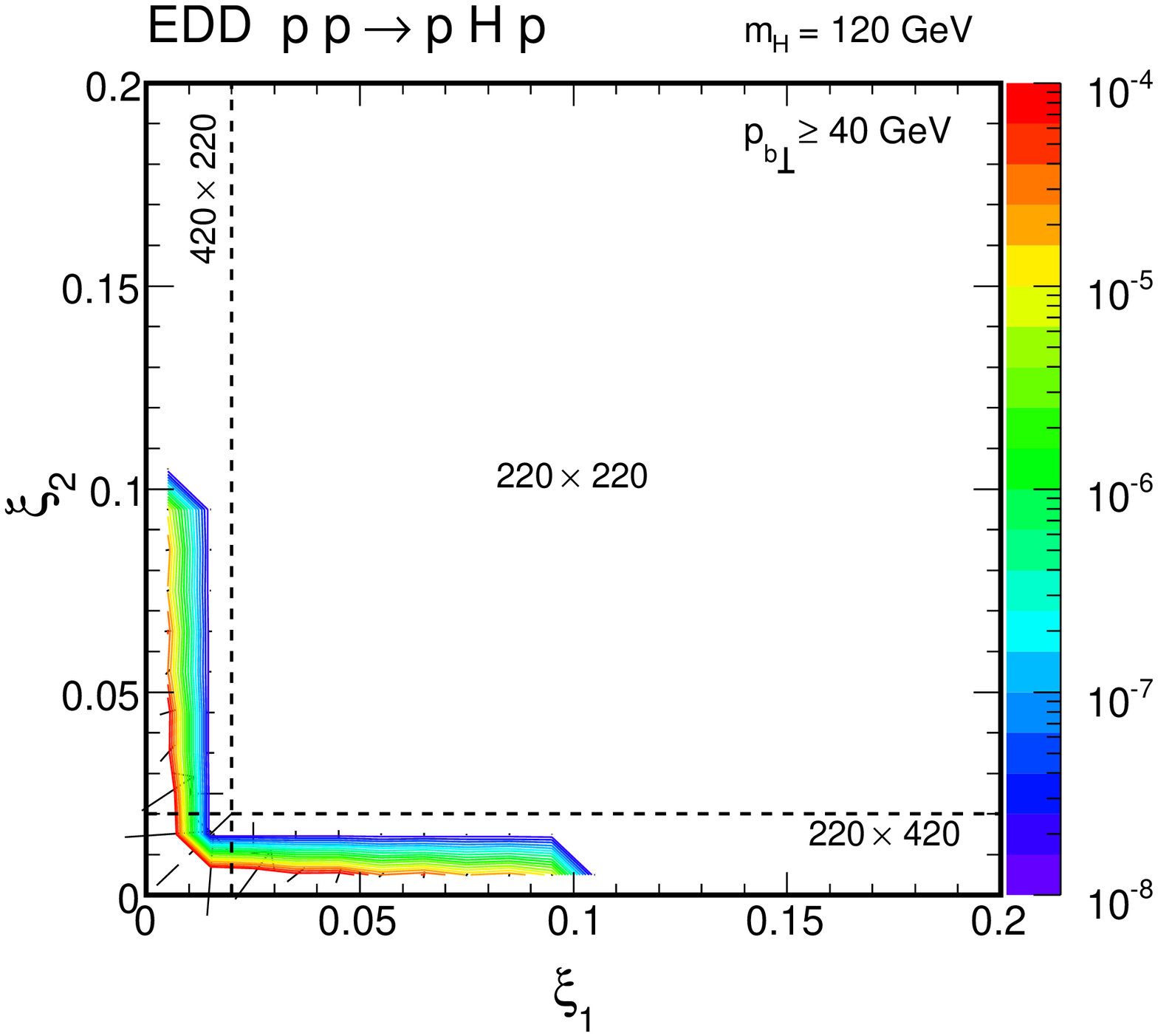}}
\end{minipage}
   \caption{
 \small The two-dimensional distributions in $\xi_{1}$ and $\xi_{2}$ for
 diffractive $b\bar b$ (left panel), $\gamma^* \gamma^* \rightarrow b \bar{b}$
 (middle panel) and CEP of Higgs (right panel).}
 \label{fig:xi1xi2}
\end{figure}
\end{center}
\vspace{-12.5mm}

The equivalence of the cuts in rapidity and transverse momentum can
be even better understood in the two-dimensional distribution in
$b$-quark transverse momentum and the difference between quark and
antiquark rapidities. Fig.~\ref{fig:ydiffp3t} shows that for a fixed narrow
interval of the $b \bar b$ invariant mass the rapidity difference
between $b$ and $\bar b$ and the transverse momentum of $b$ or $\bar
b$ are strongly correlated. This is of purely kinematical origin (see Eq.~(\ref{Mqq})) and
demonstrates that imposing cuts on one of the two variables is
equivalent and completely sufficient.

Above we have considered proton transverse momentum cuts. Can other
outgoing proton variables be useful to improve the
signal-to-background ratio? In Fig.~\ref{fig:xi1xi2} we show the
distribution in proton longitudinal momentum fraction losts $\xi_1$
and $\xi_2$ defined as:
\begin{equation}
\xi_1 = 1 - x_F(\text{proton}_1)\;, \;\;\; \xi_2 =
1-x_F(\text{proton}_2) \; , \label{xi1_xi2}
\end{equation}
where $x_F$'s are the Feynman variables of outgoing protons 1 or 2.
Only slightly different distributions for exclusive Higgs production
(right panel), $b \bar b$ EDD (left panel) and QED $b \bar b$ (middle panel)
continua can be seen. Imposing cuts on $\xi_1
> \xi_{cut}$ or $\xi_2 > \xi_{cut}$ could slightly improve the
signal-to-background ratio but at the expense of severe
deteriorating the statistics. In addition these cuts are quite correlated
with the cuts on $b$-quark rapidities. By the dotted horizontal and
vertical lines we have also marked limitations of the detectors
planned at 220 and 420 meters from the collision point by the ATLAS
and CMS collaborations. One can see that a pair of the 220 $m$
detectors is not sufficient to measure Higgs boson. Both detectors
on both sides are needed to measure most of the yield.

\subsection{Other backgrounds}

Our analysis in the previous subsection has been concentrated on the irreducible background
only. Other contributions, although, in principle, reducible, can in
practice be also rather troublesome \cite{Pilkington}. The cross
section for the gluonic dijets was found to be much larger than that
for the $b \bar b$ jets \cite{Cudell:2008gv}. If the gluon jets are
misidentified as $b$-jets, which was estimated to be 1.3\% for the
ATLAS detector, they contribute to the Higgs background. In
Ref.~\cite{Pilkington} the authors discuss in addition pile up events
when the measured protons are not related to the exclusive Higgs
production. Table 2 in their analysis presents detailed results for
the issue. Inclusive double-pomeron processes
\cite{Pomeron-Pomeron-Higgs} can also contribute to the
background. Further analyses, especially for the Standard Model
Higgs boson production, seem to be necessary to understand whether
the Higgs boson can be identified in the exclusive production,
perhaps not only in the $b \bar b$ decay channel. The present parton
level analysis should be supplemented in the future by an additional
analysis of $b \bar b$ jets by including a model of hadronization.
Then standard jet algorithms could be imposed and the quality of the
$b$ and $\bar b$ kinematical reconstruction could be studied in
detail.

\subsection{Some other remarks}

We have not been interested here in the precise
estimation of the cross section but rather in understanding the
signal-to-background ratio which is of the major importance for the
upcoming Higgs boson searches in exclusive mode at the LHC. Consequently, we have
presented results with only one UGDF. This ratio is practically the
same for other UGDFs. The absorption effects have been included here
in a simple multiplicative form. They are expected to be the same
both for the signal and the background, and thus are not affecting
the ratio under consideration. The same gap survival factor has been
used in both cases.

As was mentioned above, in the current analysis we do not take into
account the next-to-leading order QCD corrections in hard subprocess
parts in both the $b{\bar b}$ background and Higgs CEP. Calculations
of such corrections in the hard subprocess $g^*g^*\to q{\bar
q}$ within the $k_{\perp}$-factorization approach are rather
cumbersome, and we postpone them for our future studies.

We have already analyzed the sensitivity of the results on the
choice of UGDF. Different PDFs used to calculate UGDFs are
defined in different range of factorization scales (gluon transverse
momenta squared), some like CTEQ and MRST only for higher scales
($q_{\perp,min}^2 >$ 1 GeV$^2$), some like GRV and GJR for lower values
($q_{\perp,min}^2 >$ 0.4 GeV$^2$).

\begin{figure}[!h]
\begin{minipage}{0.47\textwidth}
 \centerline{\includegraphics[width=1.0\textwidth]{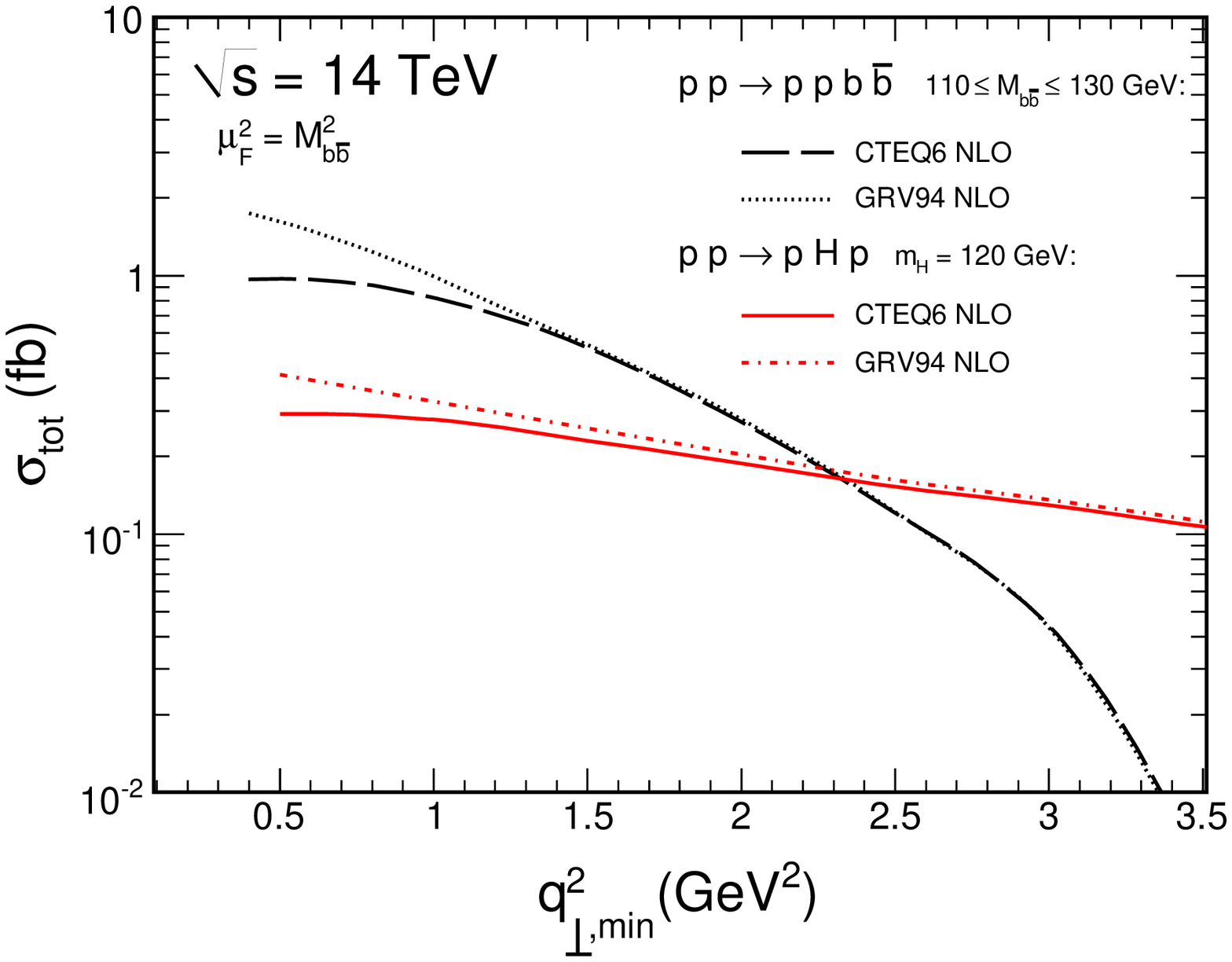}}
\end{minipage}
\hspace{0.5cm}
\begin{minipage}{0.47\textwidth}
 \centerline{\includegraphics[width=1.0\textwidth]{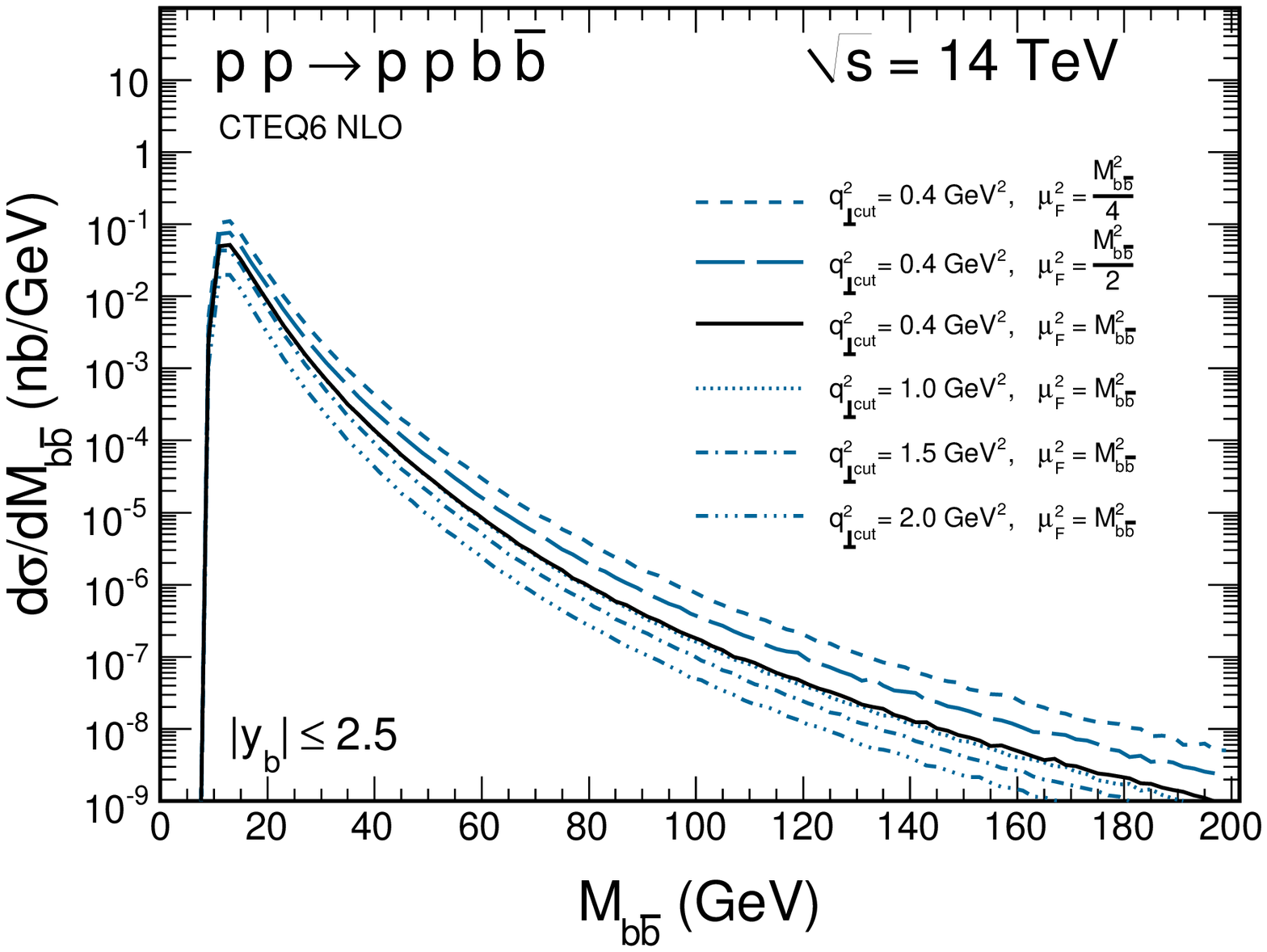}}
\end{minipage}
   \caption{
\small Total $b{\bar b}$ and Higgs CEP cross section as a function
of the lowest cut on the screening gluon transverse momentum scale (left panel)
and the invariant mass distribution of the $b{\bar b}$ CEP for
different values of the cut and different factorization scales (right panel).
}
 \label{fig:q0t2}
\end{figure}

Let us analyze how important are the low gluon transverse
momenta in evaluation of the cross section. In Fig.~\ref{fig:q0t2}
we show how the total and differential in $M_{b\bar b}$ cross
sections depend on the lowest value of the screening gluon
transverse momentum squared used in evaluating the corresponding
amplitude. There is much stronger dependence
of the background than of the signal. This is caused by the
specificity of matrix elements and the different three-body and four-body
kinematics. It is interesting to note that at high lowest limit ($
> $ 1 GeV$^2$) the cross section for different gluon distributions
coincide. This shows that the differences of the cross section
between different UGDFs come mainly from the region of relatively
small values of the screening gluon transverse momenta. There is a
stronger sensitivity on $q_{\perp,min}^2$ for larger values of
$b\bar b$ invariant mass (see the right panel in
Fig.~\ref{fig:q0t2}).

\begin{figure}[!h]
\begin{minipage}{0.5\textwidth}
\includegraphics[width=1.0\textwidth]{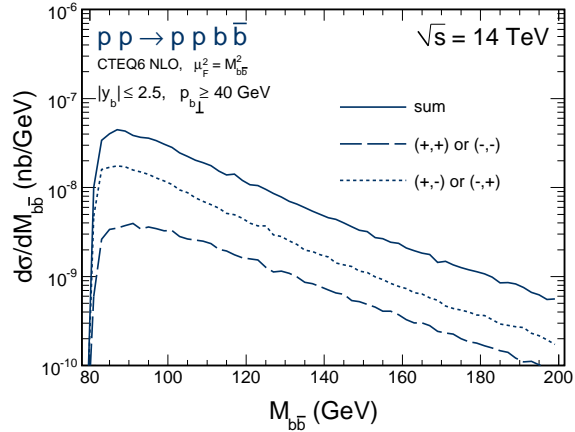}
\end{minipage}
   \caption{
 \small Invariant mass distributions of EDD $b{\bar b}$ pair production for different (anti)quark
 helicities with the lower cut ($p_{b\perp}\geq$ 40 GeV)
on both quark and antiquark jets transverse momenta.}
 \label{fig:jzrule}
\end{figure}

In Fig.~\ref{fig:jzrule} we show the invariant mass distributions of
exclusive $b{\bar b}$ pair production for
$\lambda_b\lambda_{\bar b}=++$ and $+-$
(anti)quark helicity contributions with the realistic lower cut
$p_{b\perp}\geq$ 40 GeV (so the high-$p_{\perp}$ limit is concerned) on
both quark and antiquark jets transverse momenta. The $+-$
contribution clearly dominates, however the $++$ contribution is
not negligible, especially for very large invariant masses of the
$b{\bar b}$ pair. Decreasing the cut-off on $p_{b\perp}$ relatively
enlarges the $++$ contribution, making it important for
low-$p_{\perp}$ jets production.

\section{Conclusions}

We have derived leading-order formula for the amplitude for
EDD production of heavy quarks in the
$k_{\perp}$-factorization approach. This formula takes into account
both gluon virtualities (transverse momenta) as well as the quark
masses neglected in earlier works in the literature. We have shown
that corresponding $g^*g^*\to q{\bar q}$ vertex is gauge invariant.
We have also discussed purely QED double-photon component.

Using the $2 \to 4$ diffractive amplitude we have calculated
differential cross section for $c {\bar c}$ and $b {\bar b}$ central
exclusive production (CEP) in (anti)quark rapidities, quark and
proton transverse momenta, transverse momentum of the $q{\bar q}$
pair and in azimuthal angles between outgoing protons and quark
dijets in the whole four-body phase space for the nominal LHC energy
$\sqrt{s}$ = 14 TeV. Large cross sections have been found in
contrast to previous expectations in the literature.

We have also discussed how the cross sections depends on quark
masses. While at low quark-antiquark invariant masses the cross
section for light quarks $(u,\,d,\,s)$ is considerably larger than
for heavy quarks $(c,\,b)$ at large invariant masses the situation
reverses. For instance, at invariant mass $\sim$120 GeV (relevant
for Higgs searches) it is the $b {\bar b}$ contribution which
dominates. Since experimentally one can misidentify the other
non-$b$ quark jets as $b$-jets, our calculation shows that this is
not so dangerous provided that the misidentification probability is
not too high. The gluonic jets seems in this context more difficult
because of much larger cross section \cite{Pilkington}.

We have also calculated differential distributions for exclusive
Higgs production as well as for $b$ and $\bar b$ quarks (antiquarks)
from the decay of the Higgs boson. We have used,
for the first time in exclusive Higgs case, the vertex function
which is consistently with the $k_{\perp}$-factorization approach i.e. takes
into account the gluon virtualities in the hard subprocess vertex. We
have discussed the role of the off-shell effects. In contrast to the
exclusive $\chi_c$ production, the off-shell effects for Higgs boson
are rather small and can be neglected given other sizeable
theoretical uncertainties.

The $b{\bar b}$  EDD and QED continua constitute an irreducible background to
the exclusive Higgs boson production. We have discussed in detail
how to improve the signal-to-background ratio by imposing cuts in
quark rapidities, proton transverse momenta, longitudinal momentum
fraction of outgoing protons. The analysis in the $(y_{b},\,y_{\bar
b})$-space is very useful to separate the two contributions as there
they are located in quite different parts of this space. An optimal
two-dimensional cut was proposed and the corresponding invariant
mass distribution of the signal and background was presented.

\section{Acknowledgments}

Useful discussions and helpful correspondence with Mike Albrow,
Sergey Baranov, Rikard Enberg, Gunnar Ingelman,
Igor Ivanov, Valery Khoze, Risto Orava, Andy Pilkington, Christophe
Royon, Mikhail Ryskin and Oleg Teryaev are
gratefully acknowledged. This study was partially supported by the
Carl Trygger Foundation and by the polish grant of MNiSW N N202
249235.


\end{document}